# Polyhedral Metal Nanoparticles with Cubic Lattice: Theory of Structural Properties

## Klaus E. Hermann

Inorg. Chem. Dept., Fritz-Haber-Institut der Max-Planck-Gesellschaft,
Faradayweg 4-6, 14195 Berlin, Germany.

## Abstract

We examine the structure of compact metal nanoparticles (NPs) forming polyhedral sections of face centered (fcc) and body centered (bcc) cubic lattices, which are confined by facets characterized by highly dense {100}, {110}, and {111} monolayers. Together with the constraint that the NPs exhibit the same point symmetry as the ideal cubic lattice, i.e. $O_h$, different types of generic NPs serve for the definition of general compact polyhedral cubic NPs. Their structural properties, such as shape, size, and surface facets, can be described by only three integer valued polyhedral NP parameters $N$, $M$, $K$. Corresponding analytical details are discussed with visualization of characteristic examples. While the overall NP shapes are quite similar between the different cubic lattice types, structural fine details differ. In particular, monolayer planes of adjacent NP facets can join at corners and edges which are not occupied by atoms of the ideal lattice. This gives rise to microfacets and narrow facet strips depending on the lattice type. The discussion illustrates the complexity of seemingly simple nanoparticles in a quantitative account. The geometric relationships of the model particles can also be used to classify shapes and estimate sizes of real compact metal nanoparticles observed by experiment.

# I. Introduction

Nanoparticles of many sizes, shapes, and composition have become the target of a large number of recent experimental and theoretical studies. This is due to their exciting physical and chemical properties [1, 2] which can deviate from those of corresponding bulk material. Here we mention only important applications in medicine [3] or in catalytic chemistry where metal nanoparticles have become ubiquitous [4, 5].

Physical and chemical properties of real metal nanoparticles (NPs), observed by experiment, are intimately connected with their size and shape, since in an NP the individual atoms are exposed to different local environments. Atoms close to the particle surface experience fewer neighbors compared with those inside the particle volume, which influences their interatomic binding. While this occurs for any particle size, the relative number of surface atoms compared with those inside the NP volume decreases with particle size. This is expected to affect their structure and physical behavior in its size dependence.

Structural properties of metal NPs with only a few atoms do not reflect those of corresponding bulk crystal sections in many cases and there are no general guidelines as to interatomic distances, angles, or as to symmetry. This is illustrated by theoretical studies on silver NPs up to $Ag_{12}$ [6] where equilibrium structures are found to deviate substantially from those of local sections of face centered cubic bulk silver. Further, very small NPs offer different stable isomers with varying shape and structure [6]. Larger compact metal NPs can also exhibit symmetry properties which are not compatible with those of bulk crystals. As examples, many alkaline earth and transition metal NPs (e.g. nickel, cobalt) in gas phase with up to 5000 atoms [7, 8] are believed to form compact particles of icosahedral symmetry $I_h$. This includes 5-fold rotational axes which cannot appear in perfect bulk crystals. However, their NP structure can be described by the concept of polyhedral atom shell filling, which yields preferred NP sizes connected with so-called magic numbers of atoms [8, 9].

Many metal NPs have been shown in experiments to exhibit internal cubic $O_h$ symmetry, which can be associated with compact sections of cubic bulk crystal structures, both face- and body-centered cubic, or can be approximated accordingly [10]. Examples are aluminum and indium NPs between 1000 and 10000 atoms [7]. They are suggested to form compact polyhedral particles of internal face centered cubic structure where, for energetic reasons, confining facets describe sections of densest (low Miller index) {$hkl$} monolayers. Amongst these, cuboctahedral shapes enclosed by both triangular {111} and square {100} facets, have been discussed [7]. Also other high-symmetry structures, representing bulk crystal sections, have been proposed for compact metal NPs. For example, octahedral NPs have been mentioned in the literature [11, 7, 8]. Finally, metal NPs of $O_h$ symmetry described by sections of body-centered cubic bulk crystals have been reported [11, 7, 12].

In general, real metal particles of larger size, examined by experiment [13], can assume overall shapes, which are polyhedral exhibiting flat facets, or spherical showing extended step and kink regions, or a mixture of both. Crystalline particles may be of global symmetry, reminiscent of the internal crystal lattice or may be irregularly shaped. Clearly, these properties are determined by energetics where surface and volume contributions of the particle compete. For example, polyhedral NPs may be favored energetically over spherical species, since the surface energy of large flat facet areas is assumed to be lower than that of curved areas with steps and kinks separating small facets. On the other hand, the total surface area of a spherical NP is smaller than that of a polyhedral NP of the same size, which may minimize the surface energy and favor a spherical particle. Further, atom relaxation near corners and edges as well as reconstruction of facet surfaces influences the energetics and can contribute to particle stability [14, 15]. In addition, environmental constraints, such as binding to a solid surface or contact with external gases, can determine details of the particle shape. A quantitative theoretical account of all energetic aspects of real metal NPs, going beyond a simple Wulff construction [16], is extremely complicated and requires a large number of approximations to become tractable and computationally feasible. As to local structure details, model particles derived from single crystal sections can serve as a basis for theoretical analyses [14, 15]. Here structure studies applying crystallographic methods will be helpful for any total energy calculations. In this spirit, we consider geometric details of compact nanoparticles, forming ideal lattice sections. This allows us to describe and classify structural properties, which can also help to understand structure details of real metal nanoparticles observed in experiment.

In the present work, extending previous theoretical analyses [17, 18], we examine theoretical nanoparticles forming polyhedral sections of the ideal face centered (fcc) and body centered (bcc) cubic lattice, relevant for metal NPs. In addition, NPs with simple cubic (sc) lattice and very large NPs are included for completeness. All particles are assumed to be confined by facets describing finite sections of highly dense monolayers, characterized by Miller indexed {*hkl*} families, {100}, {110}, and {111}. Together with the constraint that the NPs exhibit the same point symmetry as the ideal cubic lattice, i.e. $O_h$, there are different types of generic NPs. These can serve for the definition of general polyhedral NPs, describing models of cubic particles as intersections of corresponding generic NPs. Their structural properties, such as shape, size, and surface facets, are shown to be described by only three integer valued parameters $N$, $M$, $K$ (polyhedral NP parameters). Corresponding analytical results are discussed in detail with visualization of characteristic examples. While the overall NP shapes are quite similar between the different cubic lattice types, structural fine details differ. In particular, monolayer planes of adjacent NP facets can join at corners and edges which are not occupied by atoms of the ideal lattice. As a result, NP corners are capped and edges flattened, leading to microfacets and narrow facet strips, where the effect depends on the cubic lattice type. This illustrates the complexity of seemingly simple model nanoparticles in a quantitative structure account. The different examples can also be used as models to describe facet geometries at real metal NPs, and as a repository for structures of compact NPs with internal cubic lattice. The work shows that the multitude of NP shapes can be described uniquely by an ($N$, $M$, $K$) notation, in contrast to limited classifications of NPs based on counting facet edge atoms [11].

All analytical results of this work have been obtained by extensive calculus based on number theory and verified by mathematical proofs of induction, not discussed in detail. In addition, visualization was essential where NP options of the Balsac software [19] were used. The paper is grouped into four independent sections, dealing with fcc, bcc, and sc lattices separately and including simplifications for very large cubic lattice NPs. The sections are structured identically and presented in parts with very similar phrasing to enable easy comparison.

Stimulating discussions with Prof. Michel Van Hove, Emeritus of Hong Kong Baptist University, and his valuable suggestions are gratefully acknowledged.

## II. Formalism and Discussion

In the following we discuss structural properties of highly symmetric nanoparticles with atom arrangements reflecting local sections of the face centered (fcc) and body centered (bcc) cubic bulk. Thus, atom positions inside the nanoparticle are given by

$$\underline{R} = n_1 \underline{R}_1 + n_2 \underline{R}_2 + n_3 \underline{R}_3 + \underline{r}_i + \underline{o}$$

where $\underline{R}_1, \underline{R}_2, \underline{R}_3$ are lattice vectors of the corresponding crystal lattice and $n_1, n_2, n_3$ are integer multiples describing the bulk periodicity. Lattice basis vectors $\underline{r}_i$ describe atom positions inside the lattice unit cell. Further, vector $\underline{o}$ denotes the lattice origin describing a high symmetric site ($O_h$ symmetry) of the cubic lattice where $\underline{o}$ is assumed to form the origin of a Cartesian coordinate system, i.e. $\underline{o} = (0, 0, 0)$. In the following, we treat nanoparticles reflecting the two different cubic lattice structures separately.

### A. Face Centered Cubic Nanoparticles

The face centered cubic (**fcc**) lattice can be defined as a non-primitive simple cubic lattice by lattice vectors $\underline{R}_1, \underline{R}_2, \underline{R}_3$ together with four lattice basis vectors $\underline{r}_1$ to $\underline{r}_4$ according to

$$\underline{R}_1 = a_o (1, 0, 0), \qquad \underline{R}_2 = a_o (0, 1, 0), \qquad \underline{R}_3 = a_o (0, 0, 1) \qquad \text{(A.1a)}$$

$$\underline{r}_1 = a_o (0, 0, 0), \quad \underline{r}_2 = a_o/2 \, (0, 1, 1), \quad \underline{r}_3 = a_o/2 \, (1, 0, 1), \quad \underline{r}_3 = a_o/2 \, (1, 1, 0) \qquad \text{(A.1b)}$$

in Cartesian coordinates, where $a_o$ is the lattice constant. The three densest monolayer families {$hkl$} of the fcc lattice are described by six {100} netplanes (square mesh), twelve {110} netplanes (rectangular mesh), and eight {111} netplanes (hexagonal mesh, highest atom density) Distances between adjacent parallel netplanes are given by

$$d_{\{100\}} = a_o/2, \qquad d_{\{110\}} = a_o/(2\sqrt{2}), \qquad d_{\{111\}} = a_o/\sqrt{3} \qquad \text{(A.2)}$$

The point symmetry of the fcc lattice is characterized by $O_h$ with high symmetry centers at all atom sites and at the void centers of each elementary cell.

Compact fcc nanoparticles (NPs) are confined by finite sections of monolayers (facets) whose structure is described by different ($hkl$) netplanes. For NPs of central $O_h$ symmetry this includes all members of corresponding {$hkl$} families. As an example, we mention the {111} family with its eight netplane orientations (±1 ±1 ±1). Thus, surfaces of fcc NPs with $O_h$ symmetry are described by facets with orientations of different {$hkl$} families (denoted {$hkl$} facets in the following). These facets are limited by edges which are determined by families of Miller index directions <$hkl$> (denoted <$hkl$> edges in the following). In addition, NP corners can be characterized by directions <*hkl*> pointing from the NP center to the corresponding corner (denoted <*hkl*> corners in the following). Some <*hkl*> corners are capped to form {*hkl*} microfacets ({111} of three atoms, {100} of four or five atoms), if the confining {*hkl*} monolayer planes join at corners which are not occupied by atoms of the ideal lattice. Finally, the symmetry of the fcc host lattice allows only atom sites or $O_h$ symmetry void sites for possible NP centers. Thus, we distinguish between atom centered and void centered fcc NPs denoted **ac** and **vc** in the following.

Assuming an fcc NP to be confined by facets of the three cubic netplane families, {100}, {110}, and {111}, its size and shape can be described by three integer parameters, *N*, *M*, *K* (polyhedral NP parameters). These refer to the distances $D_{\{100\}}$, $D_{\{110\}}$, $D_{\{111\}}$ (NP diameters) between parallel facets of a given netplane family, expressed by multiples of corresponding netplane distances where

$$D_{\{100\}} = 2N\, d_{\{100\}}, \qquad D_{\{110\}} = 2M\, d_{\{110\}}, \qquad D_{\{111\}} = K\, d_{\{111\}} \qquad (A.3)$$

with $d_{\{hkl\}}$ according to (A.2). Thus, in the most general case fcc NPs can be denoted **fcc(*N*, *M*, *K*)**. If a facet type does not appear in the NP (or shows only as a very small microfacet), the corresponding parameter value *N*, *M*, or *K* may be ignored and is replaced by a minus sign. As an example, an fcc NP with only {100} and {111} facets is denoted fcc(*N*, -, *K*). These notations will be applied in the following. Further, we introduce auxiliary parameters *g*, *h*, *h'* referring to parity of *N*, *M*, and *K* with

$$\begin{aligned} g &= 0 \quad (\text{ac};\ K\ \text{even}), & &= 1 \quad (\text{vc};\ K\ \text{odd}) & &(A.4) \\ h &= 0 \quad (N + K\ \text{even}), & &= 1 \quad (N + K\ \text{odd}) & &(A.5) \\ h' &= 0 \quad (M + K\ \text{even}), & &= 1 \quad (M + K\ \text{odd}) & &(A.6) \end{aligned}$$

which help to simplify many algebraic expressions throughout Sec. A.

## A.1. Generic fcc(*N*, -, -), (-, *M*, -), and (-, -, *K*) Nanoparticles

Generic fcc nanoparticles (NPs) of $O_h$ symmetry are confined by facets with orientations of only one netplane family {*hkl*} (except for very small microfacets, see below). Here we focus on {100}, {110}, and {111} facets derived from highly dense monolayers of the fcc lattice with {111} representing the densest. These allow three different generic NP types.

(a) **Generic cubic** fcc NPs, denoted **fcc(*N*, -, -)** are confined by all six {100} monolayers with distances $D_{\{100\}} = 2N\, d_{\{100\}}$ between parallel confining monolayers. This yields six {100} facets and possibly eight {111} facets, see Fig. A.1.

The **{100} facets** for ac, *N* even or vc, *N* odd, are square with four <100> edges of length *N* $a_o$ while for ac, *N* odd or vc, *N* even, they are octagonal (capped square) with alternating edges, four <100> of length $(N - 1) a_o$ and four <110> of length $a_o/\sqrt{2}$.

The **{111} facets** are microfacets of three atoms and appear only for ac, *N* odd or vc, *N* even. They are triangular with three <110> edges of length $a_o/\sqrt{2}$.

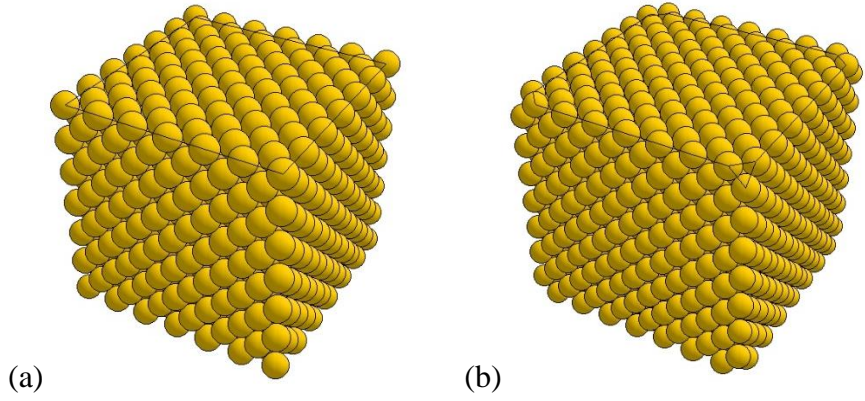

(a)  (b)

**Figure A.1.** Atom ball models of atom centered generic cubic NPs, (a) fcc(6, -, -,) and (b) fcc(7, -, -,). Black lines sketch the square and octagonal {100} facets as well as triangular {111} microfacets.

The total number of NP atoms, $N_{vol}(N, -, -)$, and the number of facet atoms, $N_{facet}(N, -, -)$, (outer polyhedral shell), are given with (A.5) by

$$N_{vol}(N, -, -) = [(2N + 1)^3 + 1]/2 - h \tag{A.7}$$
$$N_{facet}(N, -, -) = 12N^2 + 2(1 - h) \tag{A.8}$$

The largest distance from the NP center to its surface along <*hkl*> directions, $s_{<hkl>}$, is given with (A.5) by

$$s_{<100>}(N, -, -) = N\, d_{\{100\}} \tag{A.9a}$$
$$s_{<110>}(N, -, -) = 2N\, d_{\{110\}} \tag{A.9b}$$
$$s_{<111>}(N, -, -) = (3N - h)/2\, d_{\{111\}} \tag{A.9c}$$

with $d_{\{hkl\}}$ according to (A.2). These quantities will be used in Secs. A.2.

**(b)** **Generic rhombohedral** fcc NPs, denoted **fcc(-, *M*, -)**, are confined by all twelve {110} monolayers with distances $D_{\{110\}} = 2M\, d_{\{110\}}$ between parallel confining monolayers. This yields twelve {110} facets as well as possibly six smaller {100} and eight {111} facets, see Figs. A.2, A.3.

The **{100} facets** are microfacets of five atoms and appear only for ac, *M* odd or vc, *M* even NPs. They are square with four <100> edges of length $a_o$.

The **{110} facets** are rhombic, hexagonal, or octagonal with two <100> edges of length $n\, a_o$, two <110> edges of length $m\, a_o/\sqrt{2}$, and four <111> edges of length $k\, \sqrt{3}a_o$.

The **{111} facets** are triangular with three <110> edges of length $m\, a_o/\sqrt{2}$.

Corresponding edge parameters $n$, $m$, $k$, depending on $M$, are given in Table A.1 where $M$ is represented by $M = 4p + x$ with integer $p$, $0 \leq x < 4$.

| Centering | $M = 4p$ | $M = 4p + 1$ | $M = 4p + 2$ | $M = 4p + 3$ |
|---|---|---|---|---|
| ac | $n = 0$<br>$m = 0$<br>$k = M/4$ | $n = 1$<br>$m = 3$<br>$k = (M - 5)/4$ | $n = 0$<br>$m = 2$<br>$k = (M - 2)/4$ | $n = 1$<br>$m = 1$<br>$k = (M - 3)/4$ |
| vc | $n = 1$<br>$m = 2$<br>$k = (M - 4)/4$ | $n = 0$<br>$m = 1$<br>$k = (M - 1)/4$ | $n = 1$<br>$m = 0$<br>$k = (M - 2)/4$ | $n = 0$<br>$m = 3$<br>$k = (M - 3)/4$ |

**Table A.1.** Edge parameters $n$, $m$, $k$ of {100}, and {110} and {111} facets of fcc(-, $M$, -) NPs, see text.

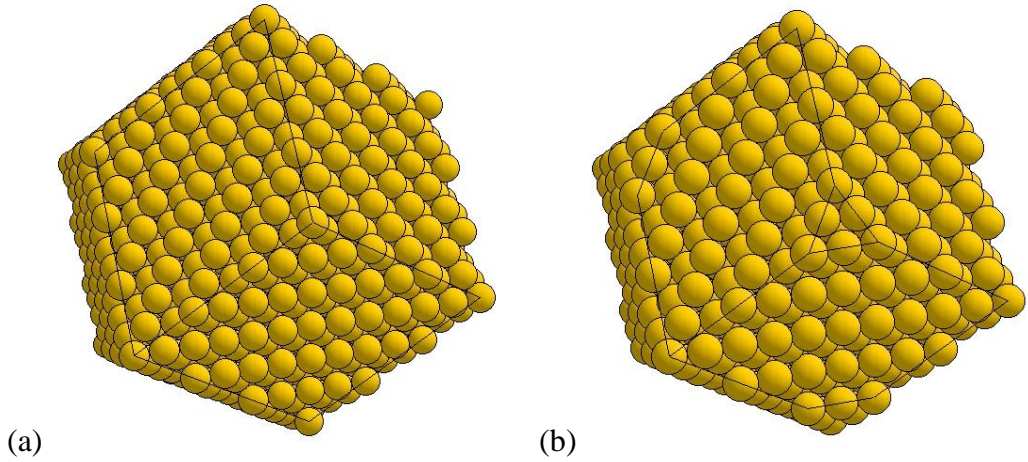

(a)                      (b)

**Figure A.2.** Atom ball models of atom centered generic rhombohedral NPs for $M$ even, (a) fcc(-, 12, -) and (b) fcc(-, 10, -). Black lines sketch the (capped) rhombic {110} and triangular {111} facets.

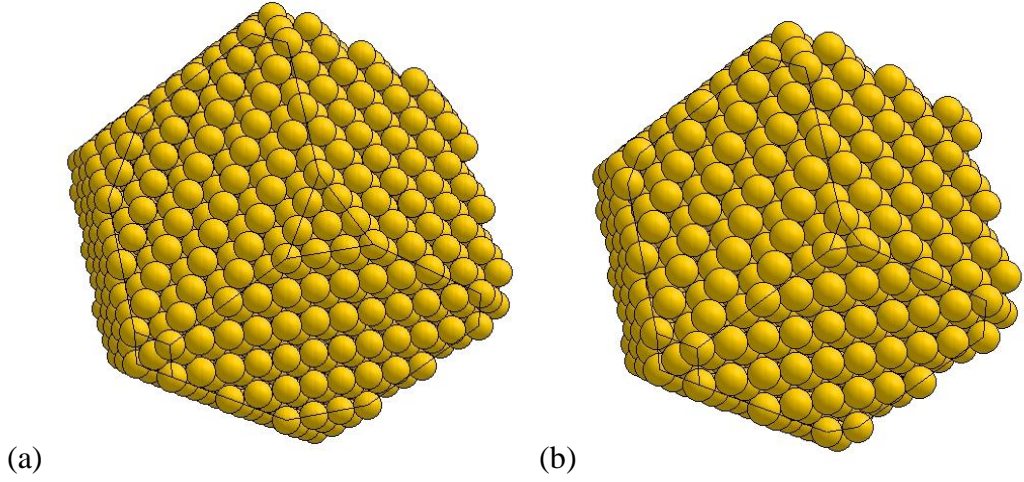

(a)                      (b)

**Figure A.3.** Atom ball models of atom centered generic rhombohedral NPs for $M$ odd, (a) fcc(-, 13, -) and (b) fcc(-, 11, -). Black lines sketch the (capped) rhombic {110} and triangular {111} facets.

The total number of NP atoms, $N_{vol}(-, M, -)$, and the number of facet atoms, $N_{facet}(-, M, -)$, (outer polyhedral shell) are given by

$$N_{vol}(-, M, -) = (2M^3 + 3M^2 + 2M + b)/2 \qquad (A.10)$$

$$N_{facet}(-, M, -) = 3M^2 + c \qquad (A.11)$$

using constants $b$, $c$ from the following Table.

| Centering | $M = 4p$ | $M = 4p + 1$ | $M = 4p + 2$ | $M = 4p + 3$ |
|---|---|---|---|---|
| ac | $b = 2$<br>$c = 2$ | $b = -5$<br>$c = 11$ | $b = 6$<br>$c = 6$ | $b = -1$<br>$c = 3$ |
| vc | $b = 0$<br>$c = 6$ | $b = 5$<br>$c = 3$ | $b = -4$<br>$c = 2$ | $b = 1$<br>$c = 11$ |

The largest distance from the NP center to its surface along $\langle hkl \rangle$ directions, $s_{\langle hkl \rangle}$, is given by

$$s_{\langle 100 \rangle}(-, M, -) = M\, d_{\{100\}} \qquad (\text{ac}, M \text{ even}; \text{vc}, M \text{ odd}) \qquad (A.12a)$$

$$\qquad\qquad\qquad = (M - 1)\, d_{\{100\}} \qquad (\text{ac}, M \text{ odd}; \text{vc}, M \text{ even}) \qquad (A.12b)$$

$$s_{\langle 110 \rangle}(-, M, -) = M\, d_{\{110\}} \qquad (A.12c)$$

$$s_{\langle 111 \rangle}(-, M, -) = 3M/4\, d_{\{111\}} \qquad (\text{ac}, M = 4p; \text{vc}, M = 4p + 2) \qquad (A.12d)$$

$$\qquad\qquad\qquad = (3M - 3)/4\, d_{\{111\}} \qquad (\text{ac}, M = 4p + 1; \text{vc}, M = 4p + 3) \qquad (A.12e)$$

$$\qquad\qquad\qquad = (3M - 2)/4\, d_{\{111\}} \qquad (\text{ac}, M = 4p + 2; \text{vc}, M = 4p) \qquad (A.12f)$$

$$\qquad\qquad\qquad = (3M - 1)/4\, d_{\{111\}} \qquad (\text{ac}, M = 4p + 3; \text{vc}, M = 4p + 1) \qquad (A.12g)$$

with $d_{\{hkl\}}$ according to (A.2). These quantities will be used in Secs. A.2.

**(c)** **Generic octahedral** fcc NPs, denoted **fcc(-, -, K)**, are confined by all eight {111} monolayers with distances $D_{\{111\}} = K\, d_{\{111\}}$ between parallel confining monolayers (ac NPs for $K$ even, vc NPs for $K$ odd). This yields eight {111} facets, see Fig. A.4, where ac and vc NPs are structurally identical. All **{111} facets** are triangular with three <110> edges of length $K\, a_o/\sqrt{2}$.

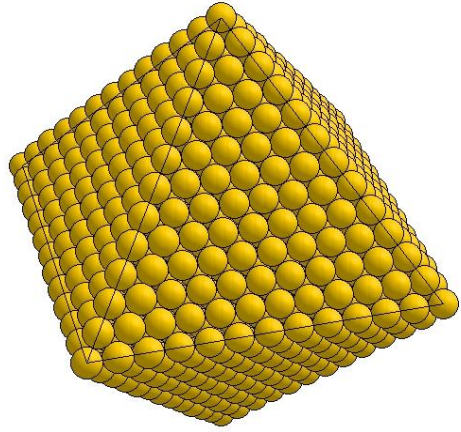

**Figure A.4.** Atom ball model of an atom centered generic octahedral NP, fcc(-, -, 12). Black lines sketch the triangular {111} facets.

The total number of NP atoms, $N_{vol}(-, -, K)$, and the number of facet atoms, $N_{facet}(-, -, K)$, (outer polyhedral shell), are given by

$$N_{vol}(-, -, K) = (K + 1)\,[2\,(K + 1)^2 + 1]/3 \tag{A.13}$$
$$N_{facet}(-, -, K) = 4\,K^2 + 2 \tag{A.14}$$

The largest distance from the NP center to its surface along <hkl> directions, $s_{<hkl>}$, is given by

$$s_{<100>}(-, -, K) = K\, d_{\{100\}} \tag{A.15a}$$
$$s_{<110>}(-, -, K) = K\, d_{\{110\}} \tag{A.15b}$$
$$s_{<111>}(-, -, K) = K/2\, d_{\{111\}} \tag{A.15c}$$

with $d_{\{hkl\}}$ according to (A.2). These quantities will be used in Secs. A.2.

Table T.1 of the Supplement collects types, constraints, and shapes of all generic fcc NPs.

## A.2. Non-generic fcc Nanoparticles

Non-generic fcc nanoparticles of $O_h$ symmetry show facets with orientations of more than one {*hkl*} netplane family. This can be considered as combining confinements of corresponding generic NPs, discussed in Sec. A.1, sharing their symmetry center, atom centered (**ac**) or high symmetry void centered (**vc**). Thus, non-generic NPs are mutual intersections of more than one generic NP, where one NP cuts corners and edges from the other(s) to form additional facets. Here we discuss non-generic NPs **fcc(*N*, *M*, *K*),** which combine constraints of up to three generic NPs, cubic fcc(*N*, -, -), rhombohedral fcc(-, *M*, -), and octahedral fcc(-, -, *K*). Thus, they allow {100}, {110}, as well as {111} facets. Clearly, the corresponding polyhedral parameters *N*, *M*, *K* depend on each other and determine the overall NP shape. In particular, ac NPs always require *K* even, while vc NPs require *K* odd. Further, if a participating generic NP encloses another participant it will not contribute to the overall NP shape. Thus, the respective {*hkl*} facets will not appear at the surface of the non-generic NP. In the following, we consider the three types of non-generic NPs, which combine constraints due to two generic NPs (Secs. A.2.1-3), before we discuss the most general case of fcc(*N*, *M*, *K*) NPs in Sec. A.2.4.

### A.2.1. Truncated fcc(*N*, *M*, -) Nanoparticles

Non-generic **cubo-rhombic** NPs, denoted **fcc(*N*, *M*, -)**, are confined by facets of the two generic NPs, fcc(*N*, -, -) (cubic) and fcc(-, *M*, -) (rhombohedral), see Figs. A.5, A.6. If the edges of the cubic NP fcc(*N*, -, -) lie inside the rhombohedral NP fcc(-, *M*, -), the resulting combination fcc(*N*, *M*, -) will be generic cubic. This requires

$$s_{<110>}(N, -, -) \leq s_{<110>}(-, M, -) \quad (A.16)$$

and with (A.9), (A.12), leads to

$$2N \leq M \quad (A.17)$$

for both ac and vc NPs. On the other hand, if the corners of the rhombohedral NP fcc(-, *M*, -) lie inside the cubic NP fcc(*N*, -, -), the resulting combination fcc(*N*, *M*, -) will be generic rhombohedral. This requires

$$s_{<100>}(-, M, -) \leq s_{<100>}(N, -, -) \quad (A.18)$$

and with (A.9), (A.12), (A.6), leads to

$$N \geq (M - h') \quad (A.19)$$

Thus, the two generic NPs intersect, to yield an NP fcc(*N*, *M*, -) with both {100} and {110} facets (apart from {111} microfacets), only for *N*, *M* values, where with (A.6)

$$N + h' < M < 2N \tag{A.20}$$

In contrast, fcc(*N*, *M*, -) is generic cubic for larger *M* according to (A.17) and generic rhombohedral for smaller *M* according to (A.19). Further, generic cubic and rhombohedral fcc NPs can be described by fcc(*N*, *M*, -) where with (A.6)

$$\text{fcc}(N, -, -) = \text{fcc}(N, M = 2N, -) \qquad \text{(cubic)} \tag{A.21a}$$

$$\text{fcc}(-, M, -) = \text{fcc}(N = M - h', M, -) \qquad \text{(rhombohedral)} \tag{A.21b}$$

The surfaces of cubo-rhombic fcc(*N*, *M*, -) NPs exhibit six {100} facets, twelve {110} facets, and eight smaller {111} facets, see Figs. A.5, A.6.

The **{100} facets** for ac, *N* even or vc, *N* odd, are square with four <100> edges of length $n\, a_o$ while for ac, *N* odd or vc, *N* even, they are octagonal (capped square) with alternating edges, four <100> of length $(n - 2)\, a_o$ and four <110> of length $a_o/\sqrt{2}$.

The **{110} facets** are octagonal (hexagonal) with two <100> edges of length $n\, a_o$, two <110> edges of length $m\, a_o/\sqrt{2}$, and four <111> edges of length $k\, \sqrt{3} a_o$.

The **{111} facets** are triangular with three <110> edges of length $m\, a_o/\sqrt{2}$.

Corresponding edge parameters *n*, *m*, *k*, depending on *N*, *M*, are given in Table A.2 where *M* is represented by $M = 4p + x$ with integer *p*, $0 \leq x < 4$.

| Centering | $M = 4p$ | $M = 4p + 1$ | $M = 4p + 2$ | $M = 4p + 3$ |
|---|---|---|---|---|
| ac<br>*N* even | $n = M - N$<br>$m = 0$<br>$k = (2N - M)/4$ | $n = M - N$<br>$m = 3$<br>$k = (2N - M - 3)/4$ | $n = M - N$<br>$m = 2$<br>$k = (2N - M - 2)/4$ | $n = M - N$<br>$m = 1$<br>$k = (2N - M - 1)/4$ |
| ac<br>*N* odd | $n = M - N + 1$<br>+ ext<br>$m = 0$<br>$k = (2N - M - 2)/4$ | $n = M - N + 1$<br>+ ext<br>$m = 3$<br>$k = (2N - M - 5)/4$ | $n = M - N + 1$<br>+ ext<br>$m = 2$<br>$k = (2N - M - 4)/4$ | $n = M - N + 1$<br>+ ext<br>$m = 1$<br>$k = (2N - M - 3)/4$ |
| vc<br>*N* odd | $n = M - N$<br>$m = 2$<br>$k = (2N - M - 2)/4$ | $n = M - N$<br>$m = 1$<br>$k = (2N - M - 1)/4$ | $n = M - N$<br>$m = 0$<br>$k = (2N - M)/4$ | $n = M - N$<br>$m = 3$<br>$k = (2N - M - 3)/4$ |
| vc<br>*N* even | $n = M - N + 1$<br>+ ext<br>$m = 2$<br>$k = (2N - M - 4)/4$ | $n = M - N + 1$<br>+ ext<br>$m = 1$<br>$k = (2N - M - 3)/4$ | $n = M - N + 1$<br>+ ext<br>$m = 0$<br>$k = (2N - M - 2)/4$ | $n = M - N + 1$<br>+ ext<br>$m = 3$<br>$k = (2N - M - 5)/4$ |

**Table A.2.** Edge parameters *n*, *m*, *k* of {100}, and {110} and {111} facets of fcc(*N*, *M*, -) NPs, see text. Values $m = 0$ result in hexagonal rather than octagonal facets. Further, "+ ext" indicates that each {110} facet is extended by two atom rows of length $(M - N - 1)\, a_o$ along <100>.

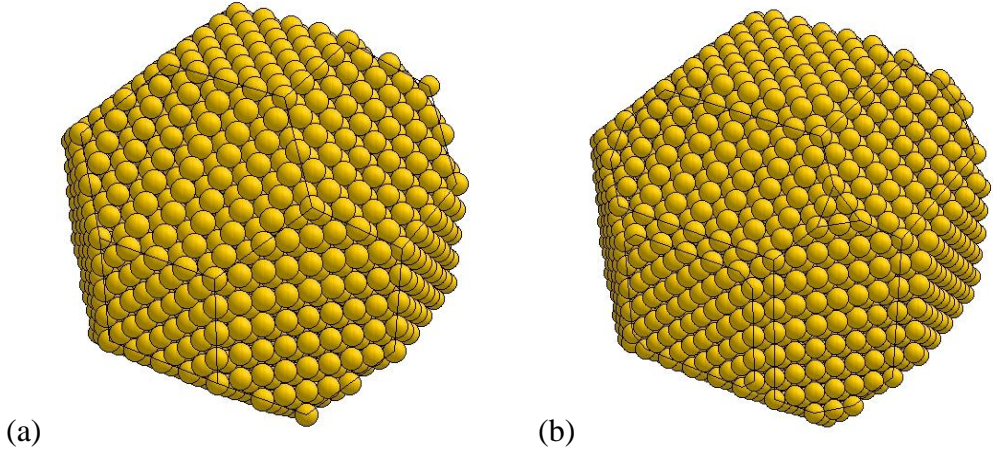

**Figure A.5.** Atom ball models of atom centered cubo-rhombic NPs for $M$ even, (a) fcc(12, 16, -) and (b) fcc(13, 18, -). Black lines sketch the (capped) square {100}, (capped) hexagonal {110} and triangular {111} facets.

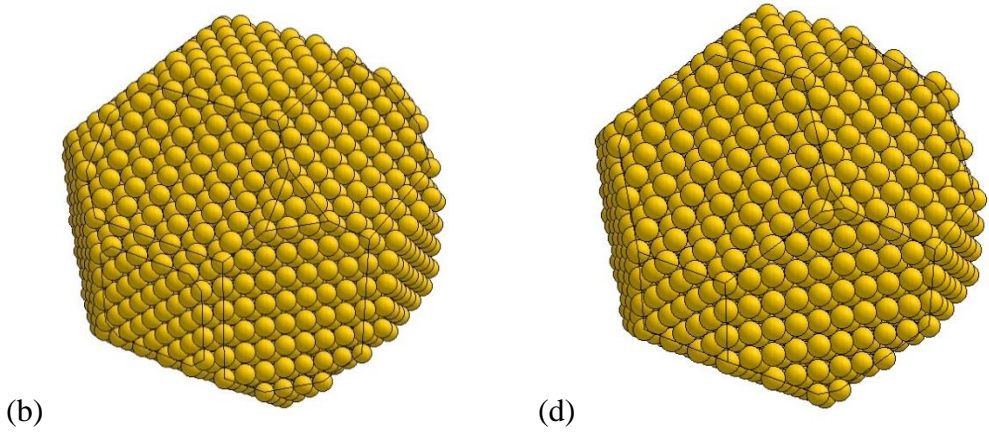

**Figure A.6.** Atom ball models of atom centered cubo-rhombic NPs, for $M$ odd, (b) fcc(13, 17, -) and (d) fcc(12, 15, -). Black lines sketch the (capped) square {100}, (capped) hexagonal {110} and triangular {111} facets.

The total number of NP atoms, $N_{vol}(N, M, -)$, and the number of facet atoms, $N_{facet}(N, M, -)$, (outer polyhedral shell) are given with (A.10), (A.11) by

$$N_{vol}(N, M, -) = N_{vol}(-, M, -) - (M - N)[4(M - N)^2 - 1] - a \quad (A.22)$$

$$\begin{aligned} a &= 0 & (N + M \text{ even}) \\ &= 3 & (N + M \text{ odd: ac}, M \text{ even; vc}, M \text{ odd}) \\ &= -3 & (N + M \text{ odd: ac}, M \text{ odd; vc}, M \text{ even}) \end{aligned}$$

$$N_{facet}(N, M, -) = N_{facet}(-, M, -) - c \quad (A.23)$$

$$c = 0 \quad (\text{ac}, N \text{ even; vc}, N \text{ odd}), \quad = 6 \quad (\text{ac}, N \text{ odd; vc}, N \text{ even})$$

A classification of fcc(*N*, *M*, -) NP types for any *N*, *M* combination is given by Table T.2 of the Supplement.

## A.2.2. Truncated fcc(*N*, -, *K*) Nanoparticles

Non-generic **cubo-octahedral** NPs, denoted **fcc(*N*, -, *K*)**, are confined by facets of the two generic NPs, fcc(*N*, -, -) (cubic) and fcc(-, -, *K*) (octahedral), see Figs. A.7, A.8. If the (capped) corners of the cubic NP fcc(*N*, -, -) lie inside the octahedral NP fcc(-, -, *K*), the resulting combination fcc(*N*, -, *K*) will be generic cubic. This requires

$$s_{<111>}(N, -, -) \leq s_{<111>}(-, -, K) \tag{A.24}$$

and with (A.9), (A.15), (A.5), leads to

$$3N \leq K + h \tag{A.25}$$

On the other hand, if the corners of the octahedral NP fcc(-, -, *K*) lie inside the cubic NP fcc(*N*, -, -), the resulting combination fcc(*N*, -, *K*) will be generic octahedral. This requires

$$s_{<100>}(-, -, K) \leq s_{<100>}(N, -, -) \tag{A.26}$$

and with (A.9), (A.15), leads to

$$N \geq K \tag{A.27}$$

Thus, the two generic NPs intersect, to yield an NP fcc(*N*, -, *K*) with both {100} and {111} facets, only for *N*, *K* values, where with (A.5)

$$N < K < 3N - h \tag{A.28}$$

In contrast, fcc(*N*, -, *K*) is generic cubic for larger *K* according to (A.25) and generic octahedral for smaller *K* according to (A.27). Further, generic cubic and octahedral fcc NPs can be described by fcc(*N*, -, *K*) where with (A.5)

$$\text{fcc}(N, -, -) = \text{fcc}(N, -, K = 3N - h) \quad \text{(cubic)} \tag{A.29a}$$
$$\text{fcc}(-, -, K) = \text{fcc}(N = K, -, K) \quad \text{(octahedral)} \tag{A.29b}$$

The surfaces of cubo-octahedral NPs fcc(*N*, -, *K*) exhibit six {100} and eight {111} facets, see Figs. A.7, A.8. Amongst the intersecting species according to (A.28) we can distinguish between **truncated octahedral** NPs where *K* < 2*N* and **truncated cubic** NPs for *K* > 2*N*, with **cuboctahedral** NPs for *K* = 2*N* separating. This will be discussed in the following.

**Truncated octahedral** NPs ($K < 2N$), Fig. A.7a, can be characterized by their facets as follows.

The **{100} facets** are square with four <110> edges of length $(K - N) a_o/\sqrt{2}$.

The **{111} facets** are hexagonal with <110> edges of alternating lengths $(K - N) a_o/\sqrt{2}$ and $(2N - K) a_o/\sqrt{2}$.

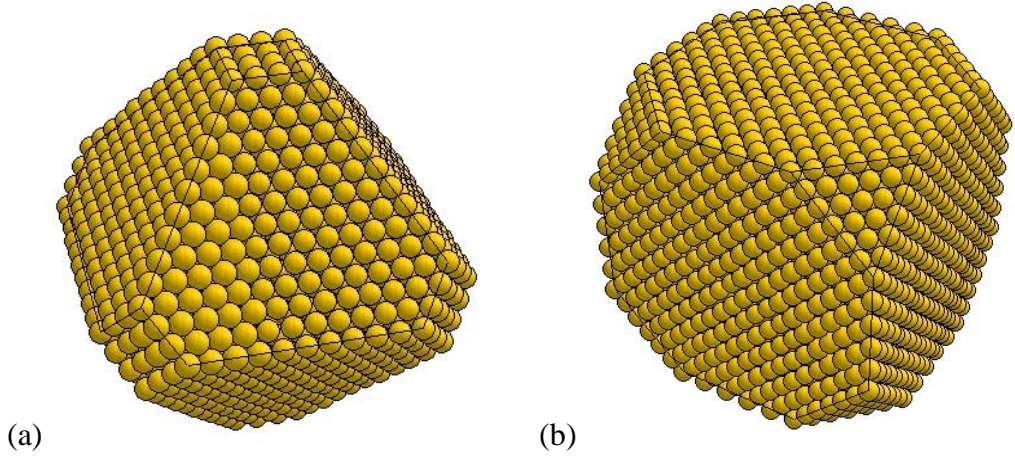

**Figure A.7.** Atom ball models of cubo-octahedral NPs, (a) ac fcc(13, -, 16) (truncated octahedral) and (b) vc fcc(11, -, 27) (truncated cubic). Black lines sketch the square/octagonal {100} and hexagonal/triangular {111} facets.

The total number of NP atoms, $N_{vol}(N, -, K)$, and the number of facet atoms, $N_{facet}(N, -, K)$, (outer polyhedral shell) are given with (A.13), (A.14) by

$$N_{vol}(N, -, K) = N_{vol}(-, -, K) - H (H + 1) (2H + 1), \qquad H = K - N \qquad (A.30)$$

$$N_{facet}(N, -, K) = N_{facet}(-, -, K) - 6 (K - N)^2 \qquad (A.31)$$

**Truncated cubic** NPs ($K > 2N$), Fig. A.7b, can be characterized by their facets as follows.

The **{100} facets** are octagonal with alternating edges, four <100> of length $(K - 2N) a_o$ and <110> of length $(3N - K) a_o/\sqrt{2}$.

The **{111} facets** are triangular with <110> edges of length $(3N - K) a_o/\sqrt{2}$.

The total number of NP atoms, $N_{vol}(N, -, K)$, for ac and vc NPs and the number of facet atoms, $N_{facet}(N, -, K)$, (outer polyhedral shell) are given with (A.7), (A.8), (A.5) by

$$N_{vol}(N, -, K) = N_{vol}(N, -, -) - H (H + 2) (2H - 1)/3 + h \qquad H = 3N - K \qquad (A.32)$$

$$N_{facet}(N, -, K) = N_{facet}(N, -, -) - 2H^2 + 2h \qquad (A.33)$$

There are fcc NPs which can be assigned to both truncated cubic and truncated octahedral type, the **cuboctahedral** NPs fcc($N$, -, $K$), defined by $K = 2N$. These NPs exist only as atom centered species since $K$ must be even. They exhibit six {100} and eight {111} facets, see Fig. A.8. All **{100} facets** are square with four <110> edges of length $N\, a_o/\sqrt{2}$ while all **{111} facets** are triangular with three <110> edges of length $N\, a_o/\sqrt{2}$ shared with those of the {100} facets.

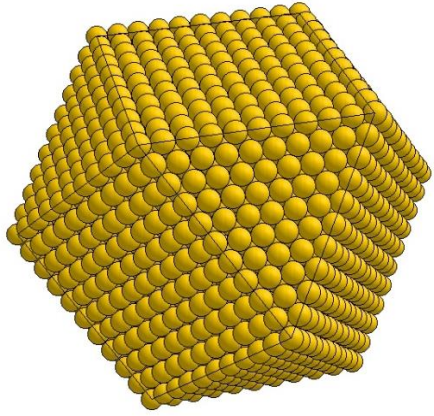

**Figure A.8.** Atom ball model of an atom centered cuboctahedral fcc(10, -, 20). Black lines sketch the triangular {111} and octagonal/square {100} facets.

A classification of fcc($N$, -, $K$) NP types for any $N$, $K$ combination is given by Table T.3 of the Supplement.

## A.2.3. Truncated fcc(-, $M$, $K$) Nanoparticles

Non-generic **rhombo-octahedral** NPs, denoted **fcc(-, $M$, $K$)**, are confined by facets of the two generic NPs, fcc(-, $M$, -) (rhombohedral) and fcc(-, -, $K$) (octahedral), see Figs. A.9, A.10. If the corners of the rhombohedral NP fcc(-, $M$, -) lie inside the octahedral NP fcc(-, -, $K$), the resulting combination fcc(-, $M$, $K$) will be generic rhombohedral. This requires

$$s_{<111>}(-, M, -) \leq s_{<111>}(-, -, K) \tag{A.34}$$

and with (A.12), (A.15), leads to

| | | |
|---|---|---|
| $3M \leq 2K$ | (ac, $M = 4p$; vc, $M = 4p + 2$) | (A.35a) |
| $3M \leq 2K + 3$ | (ac, $M = 4p + 1$; vc, $M = 4p + 3$) | (A.35b) |
| $3M \leq 2K + 2$ | (ac, $M = 4p + 2$; vc, $M = 4p$) | (A.35c) |
| $3M \leq 2K + 1$ | (ac, $M = 4p + 3$; vc, $M = 4p + 1$) | (A.35d) |

On the other hand, if the corners of the octahedral NP fcc(-, -, $K$) lie inside the rhombohedral NP fcc(-, $M$, -), the resulting combination fcc(-, $M$, $K$) will be generic octahedral. This requires

$$s_{<100>}(-, -, K) \leq s_{<100>}(-, M, -) \tag{A.36}$$

and with (A.12), (A.15), (A.6), leads to

$$K \leq M - h' \tag{A.37}$$

Thus, the two generic NPs intersect, to yield an NP fcc(-, $M$, $K$) with both {110} and {111} facets (apart from small {100} facets), only for $M$, $K$ values, where with (A.4)

$$2M - 2g \quad < \; 2K \; < \; 3M - 2g \qquad (M = 4p) \tag{A.38a}$$
$$2M - 2(1 - g) \; < \; 2K \; < \; 3M - 1 - 2(1 - g) \qquad (M = 4p + 1) \tag{A.38b}$$
$$2M - 2g \quad < \; 2K \; < \; 3M - 2(1 - g) \qquad (M = 4p + 2) \tag{A.38c}$$
$$2M - 2(1 - g) \; < \; 2K \; < \; 3M - 1 - 2g \qquad (M = 4p + 3) \tag{A.38d}$$

In contrast, fcc(-, $M$, $K$) is generic rhombohedral for larger $K$ according to (A.35) and generic octahedral for smaller $K$ according to (A.37). Further, generic rhombohedral and octahedral fcc NPs can be described by fcc(-, $M$, $K$) where with (A.4)

$$\begin{aligned}
\text{fcc}(-, M, -) &= \text{fcc}(-, M, K = 3M/2) && (\text{rhombohedral}, \; M = 4p + 2g) & \text{(A.39a)} \\
&= \text{fcc}(-, M, K = (3M - 3)/2) && ( \text{''} \quad M = 4p + 1 + 2g) & \text{(A.39b)} \\
&= \text{fcc}(-, M, K = (3M - 2)/2) && ( \text{''} \quad M = 4p + 2 - 2g) & \text{(A.39c)} \\
&= \text{fcc}(-, M, K = (3M - 1)/2) && ( \text{''} \quad M = 4p + 3 - 2g) & \text{(A.39d)}
\end{aligned}$$

$$\text{fcc}(-, -, K) = \text{fcc}(-, M = K, K) \qquad (\text{octahedral}) \tag{A.40}$$

The surfaces of rhombo-octahedral NPs fcc(-, $M$, $K$) exhibit twelve {110}, eight {111} facets, and six possible {100} facets, see Figs. A.9, A.10.

The **{100} facets** are microfacets of five atoms and appear only for $K + M$ odd. They are square with <100> edges of length $a_o$.

The **{110} facets** for $K + M$ even, are hexagonal (capped rhombic) with four <111> edges of length $(K - M)/2 \sqrt{3} a_o$ and two <110> edges of $(3M - 2K) a_o/\sqrt{2}$. For $K + M$ odd, the facets are octagonal with four <111> edges of length $(K - M - 1)/2 \sqrt{3} a_o$ and two <110> edges of $(3M - 2K) a_o/\sqrt{2}$.

The **{111} facets** are triangular with three <110> edges of length $(3M - 2K) a_o/\sqrt{2}$.

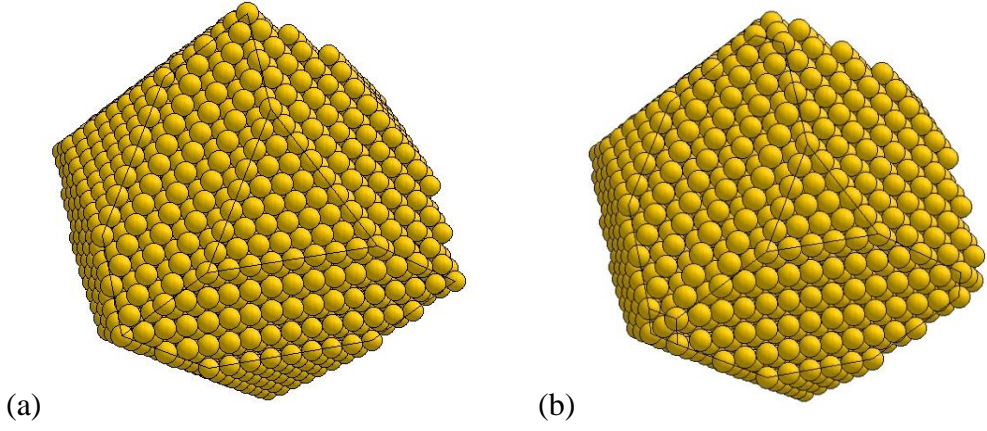

**Figure A.9.** Atom ball models of atom centered rhombo-octahedral NPs, (a) fcc(-, 16, 20) and (b) fcc(-, 15, 20). Black lines sketch the hexagonal/octagonal {110}, triangular {111}, and small square {100} facets.

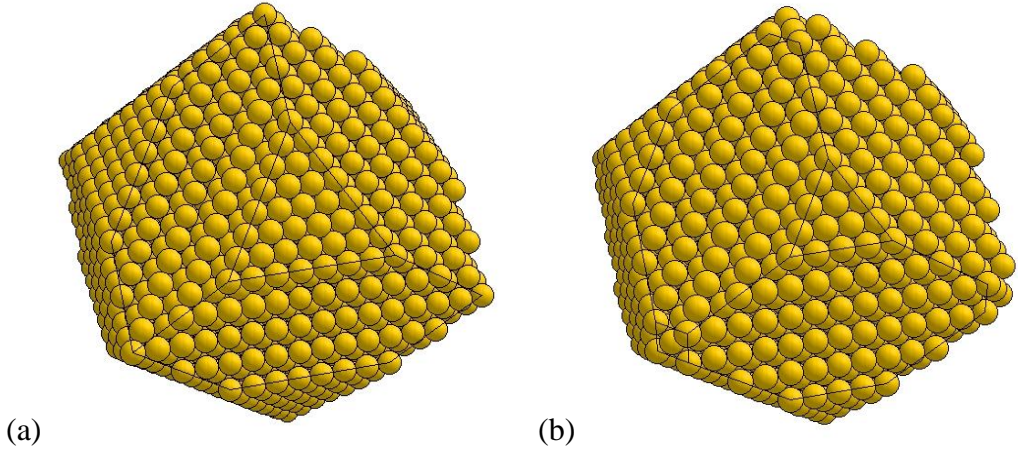

**Figure A.10.** Atom ball models of void centered rhombo-octahedral NPs, (a) fcc(-, 15, 19) and (b) fcc(-, 14, 19). Black lines sketch the hexagonal/octagonal {110}, triangular {111}, and small square {100} facets.

The total number of NP atoms, $N_{vol}(\text{-}, M, K)$, and the number of facet atoms, $N_{facet}(\text{-}, M, K)$, (outer polyhedral shell) are given with (A.6) by

$$N_{vol}(\text{-}, M, K) = (2M^3 + 3M^2 + 2M)/2 - H(2H^2 - 3H - 8)/6 + 1 - 3h' \qquad (A.41)$$

$$N_{facet}(\text{-}, M, K) = 3M^2 + H^2 + 2 \qquad H = 3M - 2K \qquad (A.42)$$

A classification of fcc(-, $M$, $K$) NP types for any $M$, $K$ combination is given by Table T.4 of the Supplement where parameter $K_a$ inside the table is defined by

$$K_a(N, M) = 3M/2 \qquad \text{(ac, } M = 4p\text{; vc, } M = 4p + 2\text{)} \qquad (A.43a)$$
$$= (3M - 3)/2 \qquad \text{(ac, } M = 4p + 1\text{; vc, } M = 4p + 3\text{)} \qquad (A.43b)$$
$$= (3M - 2)/2 \qquad \text{(ac, } M = 4p + 2\text{; vc, } M = 4p\text{)} \qquad (A.43c)$$
$$= (3M - 1)/2 \qquad \text{(ac, } M = 4p + 3\text{; vc, } M = 4p + 1\text{)} \qquad (A.43d)$$

## A.2.4. General fcc($N$, $M$, $K$) Nanoparticles

Non-generic **cubo-rhombo-octahedral** NPs, denoted **fcc($N$, $M$, $K$)**, are confined by facets of all three generic NPs, fcc($N$, -, -) (cubic), fcc(-, $M$, -) (rhombohedral), and fcc(-, -, $K$) (octahedral), see Fig. A.11. Thus, they can show {100}, {110}, and {111} facets. A general discussion of these NPs requires results for generic and non-generic NPs, see Secs. A.1, A.2.1-3, as will be detailed in the following.

First, we consider the general notation for generic NPs discussed in Sec. A.1. Cubic NPs fcc($N$, -, -) are surrounded by smallest rhombohedral NPs fcc(-, $M$, -) if $M = 2N$ and by smallest octahedral NPs fcc(-, -, $K$) if $K = 3N - h$, see (A.17), (A.25). This allows a notation

$$\text{fcc}(N, -, -) = \text{fcc}(N, M = 2N, K = 3N - h) \qquad (A.44)$$

Rhombohedral NPs fcc(-, $M$, -) are surrounded by smallest cubic NPs fcc($N$, -, -) if $N = M - h'$ and by smallest octahedral NPs fcc(-, -, $K$) if $K = K_a$, see (A.19), (A.35). This yields with (A.43), (A.4)

$$\text{fcc}(-, M, -) = \text{fcc}(N = M - h', M, K_a) \qquad (A.45)$$

Octahedral NPs fcc(-, -, $K$) are surrounded by smallest cubic NPs fcc($N$, -, -) if $N = K$ and by smallest rhombohedral NPs fcc(-, $M$, -) if $M = K$, see (A.27), (A.37). This yields

$$\text{fcc}(-, -, K) = \text{fcc}(N = K, M = K, K) \qquad (A.46)$$

General notations for non-generic fcc NPs with two facet types, discussed in Secs. A.2.1-3, are obtained by analogous arguments. Cubo-rhombic NPs fcc($N$, $M$, -) are surrounded by smallest octahedral NPs fcc(-, -, $K$) if $K = K_a$ with $K_a$ defined by (A.43). This allows a notation

$$\text{fcc}(N, M, -) = \text{fcc}(N, M, K = K_a) \qquad (A.47)$$

Cubo-octahedral NPs fcc($N$, -, $K$) are surrounded by smallest rhombohedral NPs fcc(-, $M$, -) if $M = M_a$ with

$$M_a(N, K) = \min(K, 2N) \qquad (A.48)$$

yielding

$$\text{fcc}(N, -, K) = \text{fcc}(N, M = M_a, K) \tag{A.49}$$

Rhombo-octahedral NPs fcc(-, $M$, $K$) are surrounded by smallest cubic NPs fcc($N$, -, -) if $N = N_a$ with

$$N_a(M, K) = M - h' \tag{A.50}$$

yielding

$$\text{fcc}(-, M, K) = \text{fcc}(N = N_a, M, K) \tag{A.51}$$

In the most general case of an fcc($N$, $M$, $K$) NP with {100}, {110}, and {111} facets, we start from a cubo-rhombic NP, fcc($N$, $M$, -) with its constraints $N + h' \leq M \leq 2N$. Then we add constraints of a generic octahedral NP, fcc(-, -, $K$), to yield the cubo-rhombo-octahedral NP fcc($N$, $M$, $K$). This requires, according to the discussion above, $K$ values below $K_a$. Here we can distinguish four different ranges of parameter $K$, defined by separating values $K_a \geq K_b \geq K_c$, where with $K_a$ from (A.43)

$$K_b(N, M) = 2M - N - h \tag{A.52}$$
$$K_c(N, M) = M - h' \tag{A.53}$$

The ranges are illustrated in Fig. A.11 for the atom centered cubo-rhombic NP fcc(20, 26, 38).

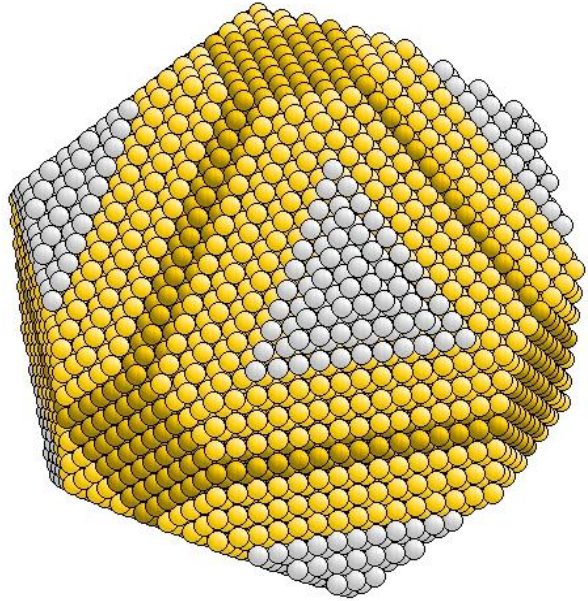

**Figure A.11.** Atom ball model of an atom centered cubo-rhombic NP, fcc(20, 26, 38) ($K = K_a$, all atom balls), with its cubo-rhombo-octahedral NP components, fcc(20, 26, 32) ($K = K_b$, dark and light yellow balls), and fcc(20, 26, 26) ($K = K_c$, dark yellow balls), see text.

**Outer *K* range** of fcc(*N*, *M*, *K*) where with (A.43)

$$K \geq K_a \tag{A.54}$$

For these *K* values the NP becomes cubo-rhombic and exhibits only small triangular {111} facets of 1, 3, 6, or 10 atoms depending on *M*, see Sec. A.2.1. It is structurally identical with fcc(*N*, *M*, $K_a$) as discussed above and in Sec. A.2.1.

**Upper central *K* range** of fcc(*N*, *M*, *K*) where with (A.43), (A.52)

$$K_b \leq K \leq K_a \tag{A.55}$$

For these *K* values the initial fcc(*N*, *M*, $K_a$) NP is capped at its <111> corners forming eight larger triangular {111} facets. Altogether, these NPs exhibit six {100} facets, twelve {110} facets, and eight {111} facets, see Fig. A.12.

The **{100} facets** for *N* + *K* even, are square with four <100> edges of length (*M* - *N*) $a_o$.
For *N* + *K* odd, the facets are octagonal with alternating edges, four <100> of length (*M* - *N* - 1) $a_o$ and four <110> of length $a_o/\sqrt{2}$.

The **{110} facets** are octagonal or rectangular (*K* = $K_b$) with two <110> edges of length (3*M* - 2*K*) $a_o/\sqrt{2}$, two <100> edges of (*M* - *N* + *h*) $a_o$, and four <111> edges of (*K* + *N* - 2*M* - *h*)/2 $\sqrt{3} a_o$ with (A.5).

The **{111} facets** are triangular with three <110> edges of length (3*M* - 2*K*) $a_o/\sqrt{2}$.

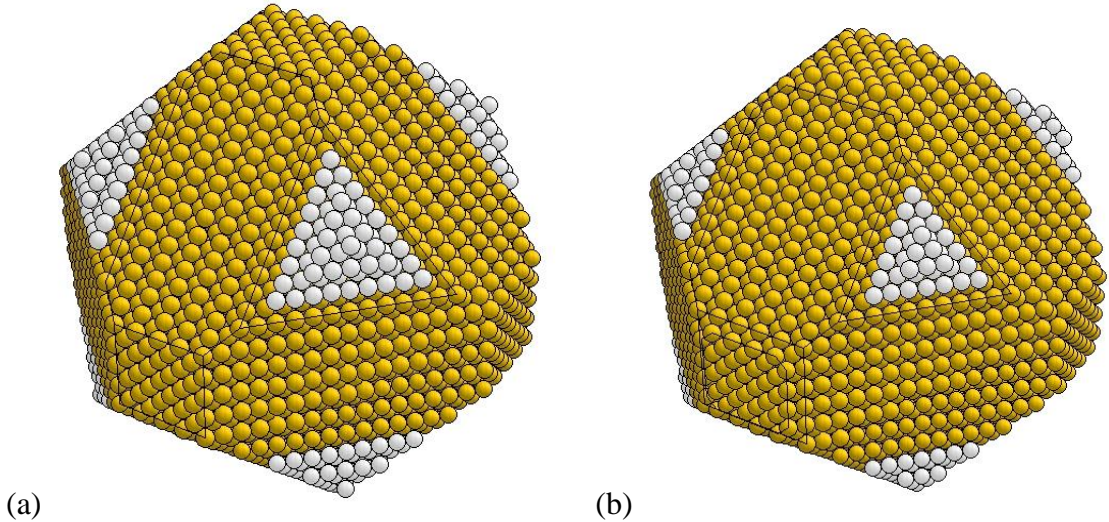

(a)    (b)

**Figure A.12.** Atom ball model of cubo-rhombo-octahedral NPs, (a) atom centered fcc(20, 24, 30) and (b) void centered fcc(20, 24, 31), see text. Black lines sketch the square/octagonal {100}, octagonal {110}, and triangular {111} facets.

The NP structures are illustrated in Fig. A.12 for the ac NP fcc(20, 24, 30) ($K_a = 36$, $K_b = 28$) and the vc NP fcc(20, 24, 31) ($K_a = 36$, $K_b = 27$), both shown by yellow atom balls, where white balls above the {111} facets are added to yield the corresponding cubo-rhombic fcc($N$, $M$, $K_a$) NP.

The total number of NP atoms, $N_{vol}(N, M, K)$, and the number of facet atoms, $N_{facet}(N, M, K)$, (outer polyhedral shell) are given with (A.22), (A.23) by

$$N_{vol}(N, M, K) = N_{vol}(N, M, -) - (2 H^3 - 3 H^2 + 8 H)/6 + b \qquad (A.56)$$

$$N_{facet}(N, M, K) = N_{facet}(N, M, -) + H^2 - c \qquad H = 3M - 2K \qquad (A.57)$$

using constants $b$, $c$ from the following Table where $M$ is represented by $M = 4p + x$ with integer $p$, $0 \leq x < 4$.

| Centering | $M = 4p$ | $M = 4p + 1$ | $M = 4p + 2$ | $M = 4p + 3$ |
|---|---|---|---|---|
| ac, $K$ even | $b = 0$<br>$c = 0$ | $b = -3$<br>$c = 9$ | $b = 12$<br>$c = 4$ | $b = 9$<br>$c = 1$ |
| vc, $K$ odd | $b = 12$<br>$c = 4$ | $b = 9$<br>$c = 1$ | $b = 0$<br>$c = 0$ | $b = -3$<br>$c = 9$ |

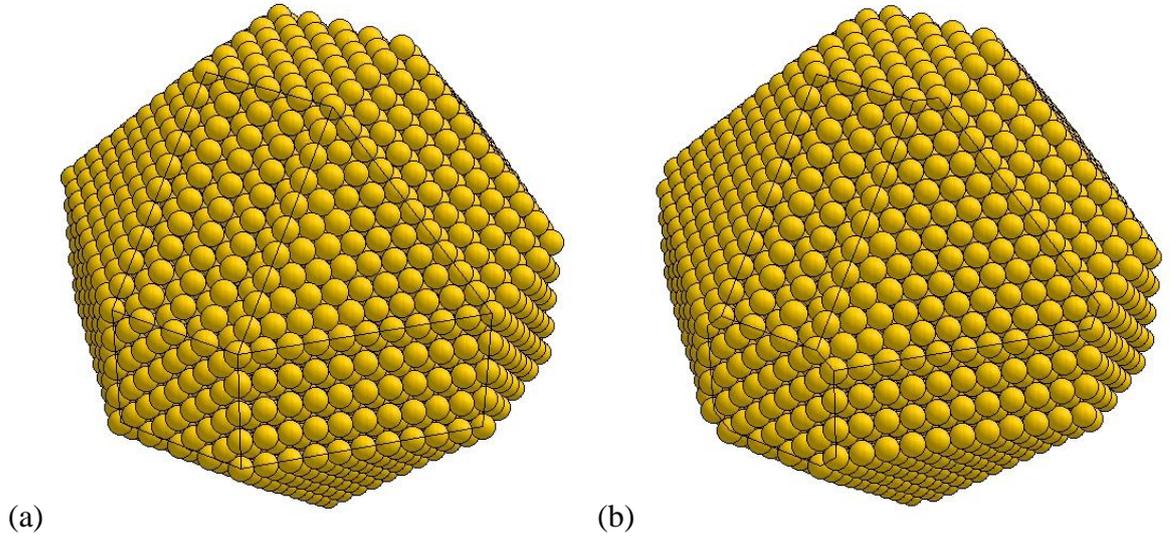

(a)          (b)

**Figure A.13.** Atom ball model of cubo-rhombo-octahedral NPs, (a) atom centered fcc(14, 18, 22), (b) void centered fcc(14, 18, 21). Black lines sketch the square {100}, rectangular {110}, and triangular {111} facets.

For $K = K_b$, the fcc($N$, $M$, $K$) NP assumes a particular shape. Its six **{100} facets** are square/octahedral with alternating edges, four <100> of length $(M - N - h) a_o$ and four

<110> of length $h\,a_o/\sqrt{2}$. Its twelve **{110} facets** are rectangular with two <110> edges of length $(2N - M)\,a_o/\sqrt{2}$ and two <100> edges of $(M - N - h)\,a_o$. Finally, its eight **{111} facets** are triangular/hexagonal with alternating edges, three <110> of length $(2N - M)\,a_o/\sqrt{2}$ and three <110> of length $h\,a_o/\sqrt{2}$. In all cases, $h$ is given by (A.5). The NP structures are illustrated in Fig. A.13 for (a) fcc(14, 18, 22) ($K_b = 22$) and (b) fcc(14, 18, 21) ($K_b = 21$).

**Lower central $K$ range** of fcc($N$, $M$, $K$) where with (A.52), (A.53)

$$K_c \leq K \leq K_b \tag{A.58}$$

For these $K$ values the capping of the initial fcc($N$, $M$, $K_b$) along the <111> directions is continued to yield eight hexagonal {111} facets. As before, these NPs exhibit six {100} facets, twelve {110} facets, and eight {111} facets, see Fig. A.14.

The **{100} facets** are octagonal with alternating edges, four <100> of length $(K - M)\,a_o$ and four <110> of length $(K_b - K)\,a_o/\sqrt{2}$.

The **{110} facets** are rectangular with two <110> edges of length $(2N - M)\,a_o/\sqrt{2}$ and two <100> edges of length $(K - M)\,a_o$.

The **{111} facets** are hexagonal with <110> edges of alternating lengths $(K_b - K)\,a_o/\sqrt{2}$ and $(2N - M)\,a_o/\sqrt{2}$.

This is illustrated in Fig. A.14 for the vc NP fcc(15, 21, 23) ($K_b = 27$, $K_c = 21$, yellow atom balls) where white balls above the {111} facets are added to fcc($N$, $M$, $K$) to yield the corresponding fcc($N$, $M$, $K_b$) NP.

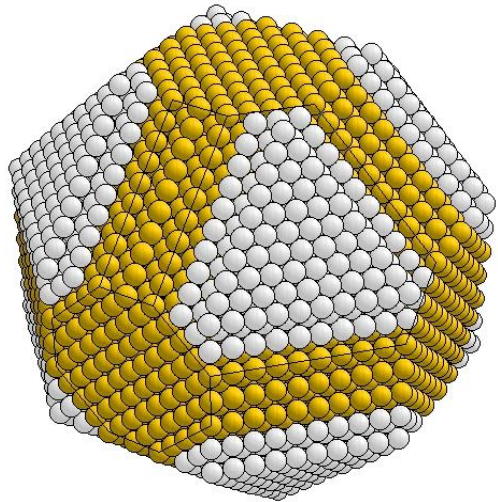

**Figure A.14.** Atom ball model of a void centered cubo-rhombo-octahedral NP, fcc(15, 21, 23), yellow balls, see text. Black lines sketch the octagonal {100}, rectangular {110}, and hexagonal {111} facets.

The total number of NP atoms, $N_{vol}(N, M, K)$, and the number of facet atoms, $N_{facet}(N, M, K)$, (outer polyhedral shell) are given with (A.56), (A.57), (A.5) by

$$N_{vol}(N, M, K) = N_{vol}(N, M, K_b)$$
$$- 2/3\, H\, \{(H+2)(2H + 12G - 1 + 6h)/2 + 3G(G - 5 + 4h)\} \quad (A.59)$$

$$N_{facet}(N, M, K) = N_{facet}(N, M, K_b) + 2H(2G - H - 2h) \quad (A.60)$$

$$H = K_b - K, \quad G = 2N - M$$

**Inner $K$ range** of fcc($N, M, K$) where with (A.53)

$$K \leq K_c \quad (A.61)$$

For these $K$ values the NP becomes cubo-octahedral and does not exhibit any {110} facets. It is structurally identical with fcc($N, M_a, K$) as discussed above and in Sec. A.2.2.

A classification of fcc($N, M, K$) NP types for any $N, M, K$ combination is given by Table T.5 of the Supplement.

## B. Body Centered Cubic Nanoparticles

The body centered cubic (**bcc**) lattice can be defined as a non-primitive simple cubic lattice by lattice vectors $\underline{R}_1$, $\underline{R}_2$, $\underline{R}_3$ together with two lattice basis vectors $\underline{r}_1$, $\underline{r}_2$ according to

$$\underline{R}_1 = a_o\,(1, 0, 0)\,, \qquad \underline{R}_2 = a_o\,(0, 1, 0)\,, \qquad \underline{R}_3 = a_o\,(0, 0, 1) \qquad \text{(B.1a)}$$

$$\underline{r}_1 = a_o\,(0, 0, 0)\,, \qquad \underline{r}_2 = a_o/2\,(1, 1, 1) \qquad \text{(B.1b)}$$

in Cartesian coordinates, where $a_o$ is the lattice constant. The two densest monolayer families $\{hkl\}$ of the bcc lattice are described by six $\{100\}$ netplanes (square mesh) and twelve $\{110\}$ netplanes (centered rectangular mesh, highest atom density). In addition, we consider the monolayer family given by eight $\{111\}$ netplanes (hexagonal mesh), which are less dense but included for comparison with the fcc and sc cases. Distances between adjacent parallel netplanes are given by

$$d_{\{100\}} = a_o/2\,, \qquad d_{\{110\}} = a_o/\sqrt{2}\,, \qquad d_{\{111\}} = a_o/(2\sqrt{3}) \qquad \text{(B.2)}$$

The point symmetry of the bcc lattice is characterized by $O_h$ with high symmetry centers at all atom sites.

Compact bcc nanoparticles (NPs) are confined by finite sections of monolayers (facets) whose structure is described by different $(hkl)$ netplanes. For NPs of central $O_h$ symmetry this includes all members of corresponding $\{hkl\}$ families. As an example, we mention the $\{110\}$ family with its twelve netplane orientations $(\pm 1\ \pm 1\ 0)$, $(\pm 1\ 0\ \pm 1)$, $(0\ \pm 1\ \pm 1)$. Thus, surfaces of bcc NPs with $O_h$ symmetry are described by facets with orientations of different $\{hkl\}$ families (denoted $\{hkl\}$ facets in the following). These facets are limited by edges which are determined by families of Miller index directions $<hkl>$ (denoted $<hkl>$ edges in the following). In addition, NP corners can be characterized by directions $<hkl>$ pointing from the NP center to the corresponding corner (denoted $\{hkl\}$ corners in the following). Finally, the symmetry of the bcc host lattice allows only atom sites for possible NP centers.

Assuming a bcc NP to be confined by facets of the three cubic netplane families, $\{100\}$, $\{110\}$, and $\{111\}$, its size and shape can be described by three integer type structure parameters, $N$, $M$, $K$ (polyhedral NP parameters). These refer to the distances $D_{\{100\}}$, $D_{\{110\}}$, $D_{\{111\}}$ (NP diameters) between parallel facets of a given netplane family, expressed by multiples of corresponding netplane distances where

$$D_{\{100\}} = 2N\,d_{\{100\}}\,, \qquad D_{\{110\}} = 2M\,d_{\{110\}}\,, \qquad D_{\{111\}} = 2K\,d_{\{111\}} \qquad \text{(B.3)}$$

with $d_{\{hkl\}}$ according to (B.2). Thus, in the most general case bcc NPs can be denoted **bcc(N, M, K)**. If a facet type does not appear in the NP (or shows only as a very small microfacet), the corresponding parameter value $N$, $M$, or $K$ may be ignored and is replaced by a minus sign. As an example, a bcc NP with only {100} and {110} facets is denoted bcc($N$, $M$, -). These notations will be used in the following. Further, we introduce auxiliary parameters **g**, **h** referring to parity of $N$ and $K$ with

$$g = 0 \quad (K \text{ even}), \qquad = 1 \quad (K \text{ odd}) \tag{B.4}$$
$$h = 0 \quad (N + K \text{ even}), \qquad = 1 \quad (N + K \text{ odd}) \tag{B.5}$$

which help to simplify many algebraic expressions throughout Sec. B.

## B.1. Generic bcc(N, -, -), (-, M, -), and (-, -, K) Nanoparticles

Generic bcc nanoparticles (NPs) of $O_h$ symmetry are confined by facets with orientations of only one {$hkl$} netplane family (except for very small microfacets, see below). Here we consider {100}, {110}, and {111} facets, derived from highly dense monolayers of the bcc lattice with {110} representing the densest. These allow three different generic NP types.

(a) **Generic cubic** bcc NPs, denoted **bcc(N, -, -)** are confined by all six {100} monolayers with distances $D_{\{100\}} = 2N\, d_{\{100\}}$ between parallel confining monolayers. This yields six {100} facets, see Fig. B.1.

The **{100} facets** are square with four <100> edges of length $N\, a_o$.

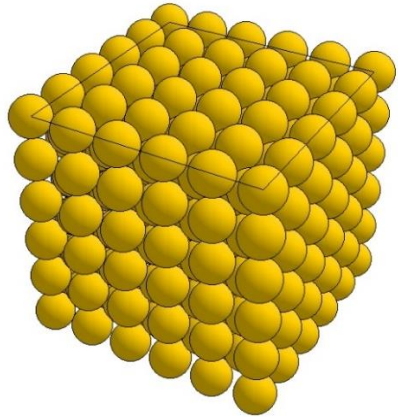

**Figure B.1.** Atom ball model of a generic cubic bcc NP, bcc(5, -, -). Black lines sketch a square {100} facet.

The total number of NP atoms, $N_{vol}(N, -, -)$, and the number of facet atoms, $N_{facet}(N, -, -)$, (outer polyhedral shell), are given by

$$N_{vol}(N, -, -) = (N + 1)^3 + N^3 \tag{B.6}$$

$$N_{facet}(N, -, -) = 6N^2 + 2 \tag{B.7}$$

The largest distance from the NP center to its surface along <hkl> directions, $s_{<hkl>}$, is given by

$$s_{<100>}(N, -, -) = N\, d_{\{100\}} \tag{B.8a}$$

$$s_{<110>}(N, -, -) = N\, d_{\{110\}} \tag{B.8b}$$

$$s_{<111>}(N, -, -) = 3N\, d_{\{111\}} \tag{B.8c}$$

with $d_{\{hkl\}}$ according to (B.2). These quantities will be used in Secs. B.2.

(b) **Generic rhombohedral** bcc NPs, denoted **bcc(-, *M*, -)**, are confined by all twelve {110} monolayers with distances $D_{\{110\}} = 2M\, d_{\{110\}}$ between parallel confining monolayers. This yields twelve {110} facets, see Fig. B.2.

The **{110} facets** are rhombic with <111> edges of length $M/2\, \sqrt{3}a_o$. Thus, the NPs can be described as rhombic dodecahedra reminding of the shape of Wigner-Seitz cells of the fcc crystal lattice [20].

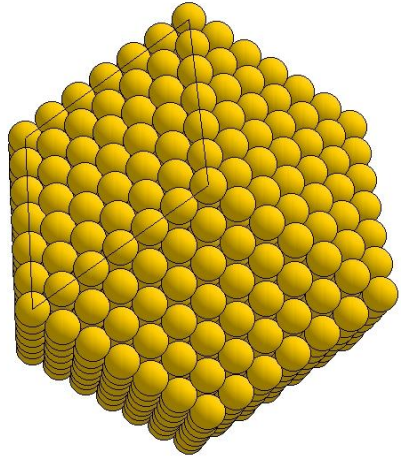

**Figure B.2.** Atom ball model of a generic cubic bcc(-, 6, -) NP. Black lines sketch a rhombic {110} facet.

The total number of NP atoms, $N_{vol}(-, M, -)$, and the number of facet atoms, $N_{facet}(-, M, -)$, (outer polyhedral shell), are given by

$$N_{vol}(-, M, -) = (2M + 1)\,[(2M + 1)^2 + 1]/2 \tag{B.9}$$

$$N_{facet}(-, M, -) = 12M^2 + 2 \tag{B.10}$$

The largest distance from the NP center to its surface along <hkl> directions, $s_{<hkl>}$, is given by

$$s_{<100>}(-, M, -) = 2M\, d_{\{100\}} \tag{B.11a}$$

$$s_{<110>}(-, M, -) = M\, d_{\{110\}} \tag{B.11b}$$

$$s_{<111>}(-, M, -) = 3M\, d_{\{111\}} \tag{B.11c}$$

with $d_{\{hkl\}}$ according to (B.2). These quantities will be used in Secs. B.2.

(c) **Generic octahedral** bcc NPs, denoted **bcc(-, -, K)**, are confined by all eight {111} monolayers with distances $D_{\{111\}} = 2K\, d_{\{111\}}$ between parallel confining monolayers. This yields eight {111} facets as well as possibly twelve {110} facets, see Fig. B.3.

The **{111} facets** are triangular with three <110> edges of length $K\, a_o/\sqrt{2}$ for $K$ even and of length $(K - 3)\, a_o/\sqrt{2}$ for $K$ odd.

The **{110} facets** appear only for $K$ odd. They form narrow hexagonal strips with two <110> edges of length $(K - 3)\, a_o/\sqrt{2}$ and four very small <111> edges of length $1/2\, \sqrt{3}a_o$.

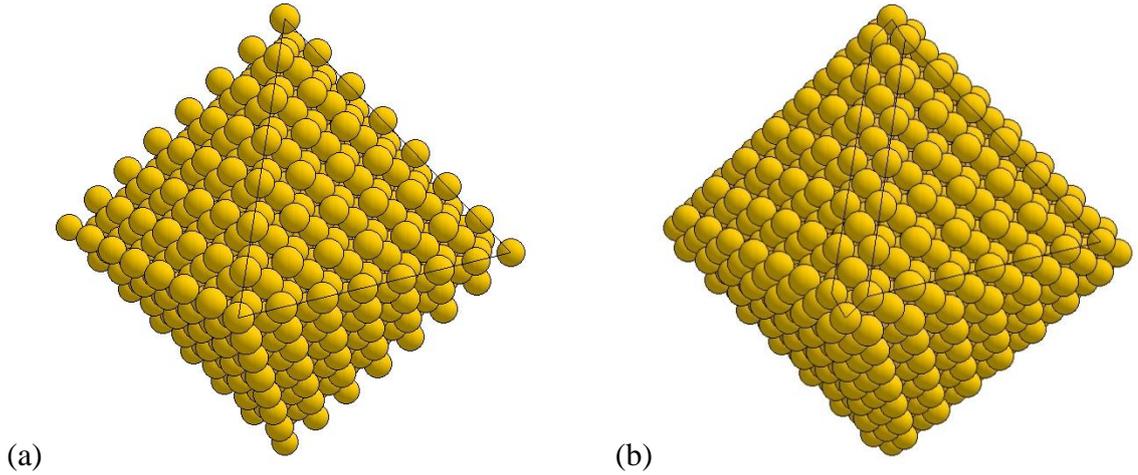

(a)          (b)

**Figure B.3.** Atom ball models of generic octahedral bcc NPs, (a) bcc(-, -, 14) and (b) bcc(-, -, 15). Black lines sketch the triangular {111} and stripped {110} facets.

The total number of NP atoms, $N_{vol}(-, -, K)$, and the number of facet atoms, $N_{facet}(-, -, K)$, (outer polyhedral shell), are given with (B.4) by

$$N_{vol}(-, -, K) = \{(K+1)\,[(K+1)^2 + 1] + K^3 + 4 - 9g\}/6 \tag{B.12}$$

$$N_{facet}(-, -, K) = K^2 + 2 - 3g \tag{B.13}$$

The largest distance from the NP center to its surface along <hkl> directions, $s_{<hkl>}$, is given with (B.4) by

$$s_{<100>}(-, -, K) = (K - g)\, d_{\{100\}} \tag{B.14a}$$

$$s_{<110>}(-, -, K) = (K - g)/2\, d_{\{110\}} \tag{B.14b}$$

$$s_{<111>}(-, -, K) = K\, d_{\{111\}} \tag{B.14c}$$

with $d_{\{hkl\}}$ according to (B.2). These quantities will be used in Secs. B.2.

Table T.6 of the Supplement collects types, constraints, and shapes of all generic bcc NPs.

## B.2. Non-generic bcc Nanoparticles

Non-generic bcc nanoparticles of $O_h$ symmetry show facets with orientations of more than one {hkl} netplane family. This can be considered as combining confinements of the corresponding generic NPs, discussed in Sec. B.1, sharing their symmetry center at an atom. Thus, non-generic NPs are mutual intersections of more than one generic NP, where one NP cuts corners and edges from the other(s), to form additional facets. Here we discuss non-generic NPs **bcc(N, M, K),** which combine constraints of up to three generic NPs, cubic bcc(N, -, -), rhombohedral bcc(-, M, -), and octahedral bcc(-, -, K). Thus, they allow {100}, {110}, as well as {111} facets. Clearly, the corresponding polyhedral parameters N, M, K depend on each other and determine the overall NP shape. In particular, if a participating generic NP encloses another participant it will not contribute to the overall NP shape. Thus, the respective {hkl} facets will not appear at the surface of the non-generic NP. In the following, we consider the three types of non-generic NPs, which combine constraints due to two generic NPs (Secs. B.2.1-3), before we discuss the most general case of bcc(N, M, K) NPs in Sec. B.2.4.

### B.2.1. Truncated bcc(N, M, -) Nanoparticles

Non-generic **cubo-rhombic** NPs, denoted **bcc(N, M, -)**, are confined by facets of the two generic NPs, bcc(N, -, -) (cubic) and bcc(-, M, -) (rhombohedral), see Fig. B.4. If the edges of the cubic NP bcc(N, -, -) lie inside the rhombohedral NP bcc(-, M, -), the resulting combination bcc(N, M, -) will be generic cubic. This requires

$$s_{<110>}(N, -, -) \leq s_{<110>}(-, M, -) \tag{B.15}$$

and with (B.8), (B.11), leads to

$$N \leq M \tag{B.16}$$

On the other hand, if the corners of the rhombohedral NP bcc(-, $M$, -) lie inside the cubic NP bcc($N$, -, -), the resulting combination bcc($N$, $M$, -) will be generic rhombohedral. This requires

$$s_{<100>}(-, M, -) \leq s_{<100>}(N, -, -) \tag{B.17}$$

and with (B.8), (B.11), leads to

$$N \geq 2M \tag{B.18}$$

Thus, the two generic NPs intersect, to yield an NP bcc($N$, $M$, -) with both {100} and {110} facets, only for $N$, $M$ values with

$$M < N < 2M \tag{B.19}$$

In contrast, bcc($N$, $M$, -) is generic cubic for smaller $N$ according to (B.16) and generic rhombohedral for larger $N$ according to (B.18). Further, generic cubic and rhombohedral bcc NPs can be described by bcc($N$, $M$, -) where

$$\text{bcc}(N, -, -) = \text{bcc}(N, M = N, -) \quad \text{(cubic)} \tag{B.20a}$$
$$\text{bcc}(-, M, -) = \text{bcc}(N = 2M, M, -) \quad \text{(rhombohedral)} \tag{B.20b}$$

The surfaces of cubo-rhombic NPs bcc($N$, $M$, -) exhibit six {100} facets and twelve {110} facets, see Fig. B.4.

**The {100} facets** are square with four <100> edges of length $(2M - N) a_o$.

**The {110} facets** are hexagonal with four <111> edges of length $(N - M)/2 \sqrt{3} a_o$ and two <100> edges of length $(2M - N) a_o$.

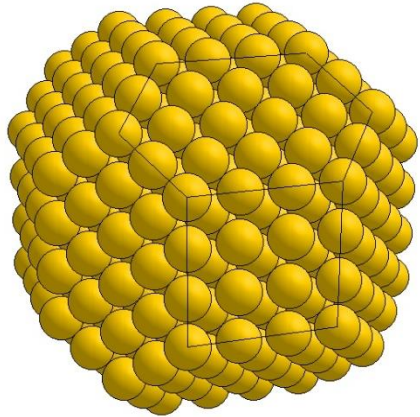

**Figure B.4.** Atom ball model of the cubo-rhombic NP bcc(7, 5, -). Black lines sketch the square {100} and hexagonal {110} facets.

The total number of NP atoms, $N_{vol}(N, M, -)$, and the number of facet atoms, $N_{facet}(N, M, -)$, (outer polyhedral shell) are given with (B.9), (B.10) by

$$N_{vol}(N, M, -) = N_{vol}(-, M, -) - H(H + 1)(2H + 1), \quad H = 2M - N \tag{B.21}$$

$$N_{facet}(N, M, -) = N_{facet}(-, M, -) - 6(2M - N)^2 \tag{B.22}$$

A classification of bcc(*N*, *M*, -) NP types for any *N*, *M* combination is given by Table T.7 of the Supplement.

### B.2.2. Truncated bcc(*N*, -, *K*) Nanoparticles

Non-generic **cubo-octahedral** NPs, denoted **bcc(*N*, -, *K*)**, are confined by facets of the two generic NPs, bcc(*N*, -, -) (cubic) and bcc(-, -, *K*) (octahedral), see Figs. B.5, B.6, B.7. If the corners of the cubic NP bcc(*N*, -, -) lie inside the octahedral NP bcc(-, -, *K*), the resulting combination bcc(*N*, -, *K*) will be generic cubic. This requires

$$s_{<111>}(N, -, -) \leq s_{<111>}(-, -, K) \tag{B.23}$$

and with (B.8), (B.14), leads to

$$3N \leq K \tag{B.24}$$

On the other hand, if the corners of the octahedral NP bcc(-, -, *K*) lie inside the cubic NP bcc(*N*, -, -), the resulting combination bcc(*N*, -, *K*) will be generic octahedral. This requires

$$s_{<100>}(-, -, K) \leq s_{<100>}(N, -, -) \tag{B.25}$$

and with (B.8), (B.14), leads to

$$N \geq K - g \tag{B.26}$$

Thus, the two generic NPs intersect, to yield an NP bcc(*N*, -, *K*) with both {100} and {111} facets (apart from {110} microstrips), only for *N*, *K* values with

$$N + g < K < 3N \tag{B.27}$$

In contrast, bcc(*N*, -, *K*) is generic cubic for larger *K* according to (B.24) and generic octahedral for smaller *K* according to (B.26). Further, generic cubic and octahedral bcc NPs can be described by bcc(*N*, -, *K*) where

$$\text{bcc}(N, -, -) = \text{bcc}(N, -, K = 3N) \quad \text{(cubic)} \tag{B.28a}$$

$$\text{bcc}(-, -, K) = \text{bcc}(N = K - g, -, K) \quad \text{(octahedral)} \tag{B.28b}$$

The surfaces of cubo-octahedral NPs bcc(*N*, -, *K*) exhibit six {100}, twelve {110}, and eight {111} facets, see Figs. B.5, B.6, B.7. Amongst the intersecting species according to (B.27) we can distinguish between **truncated octahedral** NPs where *K* < 2*N* and **truncated cubic** NPs for *K* > 2*N*, with **cuboctahedral** NPs for *K* = 2*N* separating. This will be discussed in the following.

**Truncated octahedral** NPs ($K < 2N$), Figs. B.5, B.6, can be characterized by their facets as follows.

> The **{100} facets** for $N$ even, are square with four <110> edges of length $(K - N)/2 \sqrt{2}a_o$ (with $K$ even) or $(K - N - 1)/2 \sqrt{2}a_o$ (with $K$ odd). For $N$ odd, the facets are octagonal (capped square) with alternating edges, four <100> of length $a_o$ and four <110> of length $(K - N - 3)/2 \sqrt{2}a_o$ (with $K$ even) or $(K - N - 2)/2 \sqrt{2}a_o$ (with $K$ odd).
>
> The **{111} facets** are hexagonal with <110> edges of alternating lengths $(K - N + b)/2 \sqrt{2}a_o$ and $(2N - K + c)/2 \sqrt{2}a_o$ where constants $b$, $c$ are given in the following Table.

|  |  | $b$ | $c$ |
|---|---|---|---|
| $N$ even | $K$ even | 0 | 0 |
|  | $K$ odd | -1 | -1 |
| $N$ odd | $K$ even | 1 | -2 |
|  | $K$ odd | -2 | 1 |

> The **{110} facets** appear only for $K$ odd and form narrow strips. For $N$ even, they are hexagonal with two <110> edges of lengths $(2N - K - 1)/2 \sqrt{2}a_o$ and four <111> edges of length $1/2 \sqrt{3}a_o$. For $N$ odd, the facets are rectangular with two <110> edges of lengths $(2N - K + 1)/2 \sqrt{2}a_o$ and two <100> edges of length $a_o$.

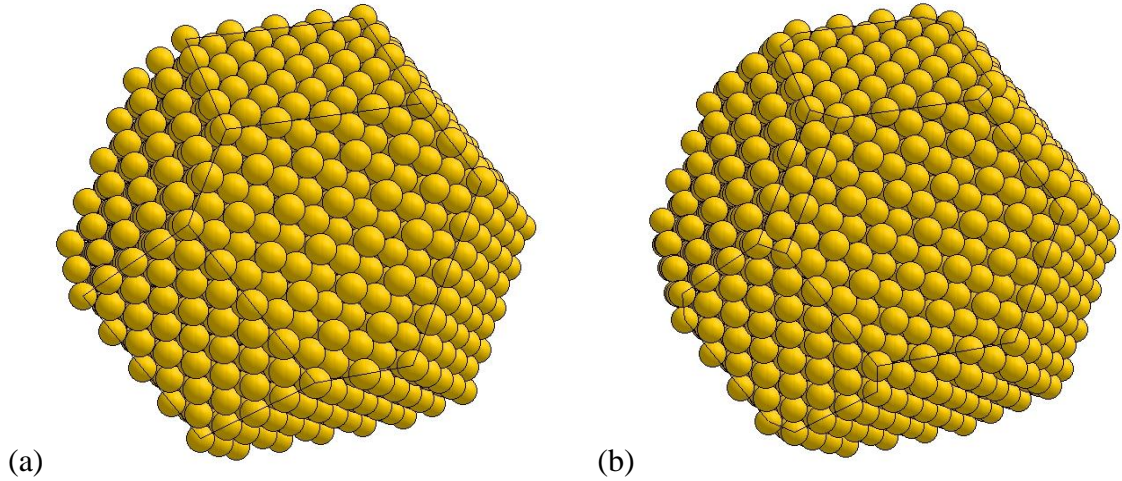

(a)      (b)

**Figure B.5.** Atom ball models of cubo-octahedral bcc NPs of truncated octahedral type, (a) bcc(12, -, 20) and (b) bcc(13, -, 21). Black lines sketch the square/octagonal {100}, hexagonal {111}, and connecting {110} facets, see text.

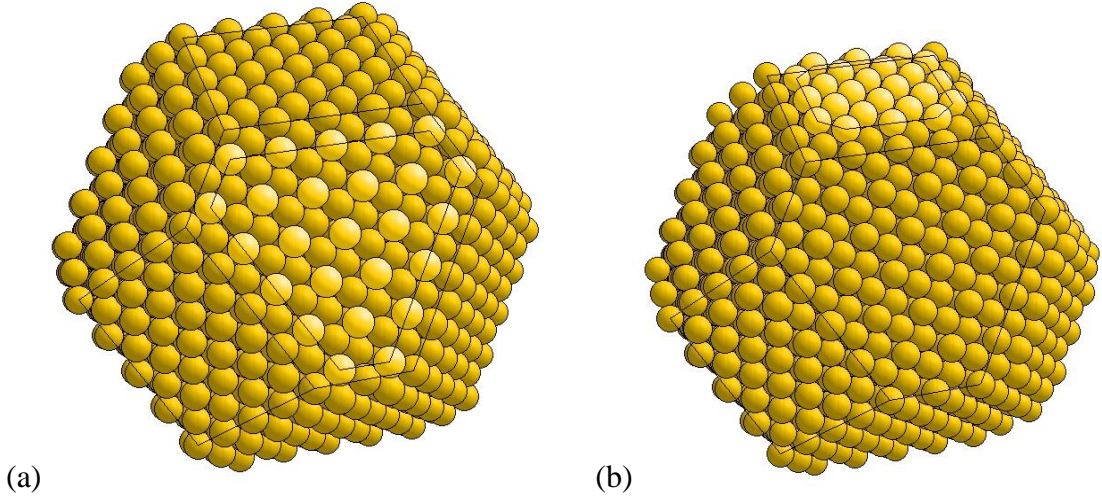

**Figure B.6.** Atom ball models of cubo-octahedral bcc NPs of truncated octahedral type, (a) bcc(12, -, 21) and (b) bcc(13, -, 20). Black lines sketch the square/octagonal {100} and hexagonal {111} facets. The light color balls indicate one (a) {111} and (b) {100} facet, see text.

The total number of NP atoms, $N_{vol}(N, -, K)$, and the number of facet atoms, $N_{facet}(N, -, K)$, (outer polyhedral shell) are given with (B.12), (B.13), (B.4), (B.5) by

$$N_{vol}(N, -, K) = N_{vol}(-, -, K) - H(H^2 - 1) - 3h(H + 1 - 2g), \qquad H = K - N \qquad (B.29)$$

$$N_{facet}(N, -, K) = N_{facet}(-, -, K) + 6h(2g - 1) \qquad (B.30)$$

**Truncated cubic** NPs ($K > 2N$), Fig. B.7, can be characterized by their facets as follows.

The **{100} facets** are octagonal with alternating edges, four <110> of length $(3N - K + h)/2 \sqrt{2}a_o$ and four <100> of length $(K - 2N - h) a_o$ with (B.5).

The **{111} facets** are triangular with three <110> edges of length $(3N - K)/2 \sqrt{2}a_o$ (with $N + K$ even) or of length $(3N - K - 3)/2 \sqrt{2}a_o$ (with $N + K$ odd).

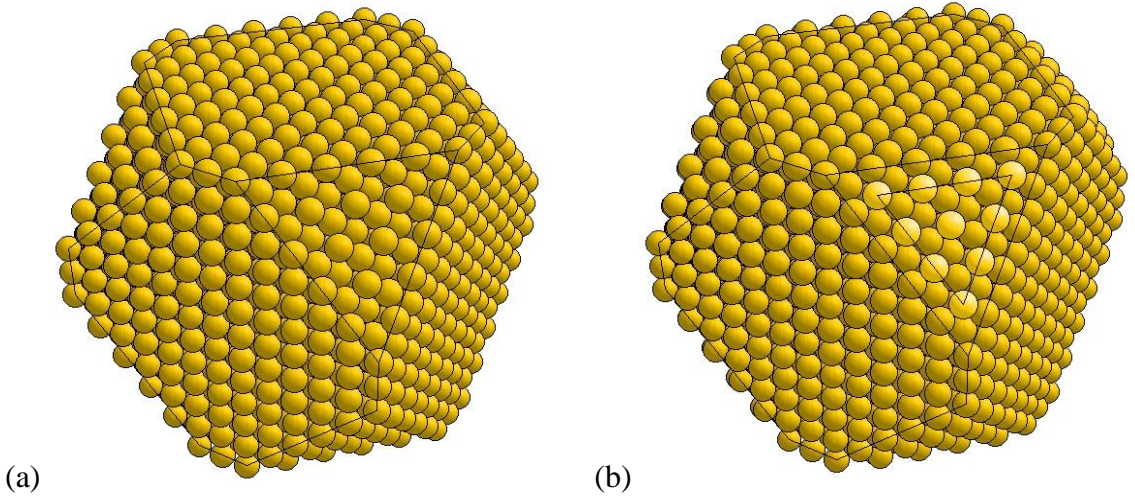

**Figure B.7.** Atom ball models of cubo-octahedral bcc NPs of truncated cubic type, (a) bcc(12, -, 26) and (b) bcc(12, -, 27). Black lines sketch the octagonal {100} and triangular {111} facets. The light color balls indicate one {111} facet, see text.

The total number of NP atoms, $N_{vol}(N, -, K)$, and the number of facet atoms, $N_{facet}(N, -, K)$, (outer polyhedral shell) are given with (B.6), (B.7), (B.5) by

$$N_{vol}(N, -, K) = N_{vol}(N, -, -) - (H + 1)(H^2 + 2H + 9h)/3, \quad H = 3N - K \quad \text{(B.31)}$$

$$N_{facet}(N, -, K) = N_{facet}(N, -, -) - 2(K - 3N)^2 - 6h \quad \text{(B.32)}$$

There are bcc NPs which can be assigned to both truncated cubic and truncated octahedral type, the **cuboctahedral** NPs bcc($N$, -, $K$), defined by $K = 2N$. These NPs exhibit six {100}, eight {111}, and twelve possible {110} facets, see Fig. B.8.

**The {100} facets** are square with four <110> edges of length $N/2 \sqrt{2}a_o$ if $N$ even, while for $N$ odd the facets are octagonal (capped square) with alternating edges, four <110> of length $(N - 3)/2 \sqrt{2}a_o$ and four <100> of length $a_o$.

**The {110} facets** appear only for $N$ odd. They are hexagonal with two <100> edges of length $a_o$ and four <111> edges of length $1/2 \sqrt{3}a_o$.

**The {111} facets** are triangular with <110> edges of length $N/2 \sqrt{2}a_o$ if $N$ even and of length $(N - 3)/2 \sqrt{2}a_o$ if $N$ odd.

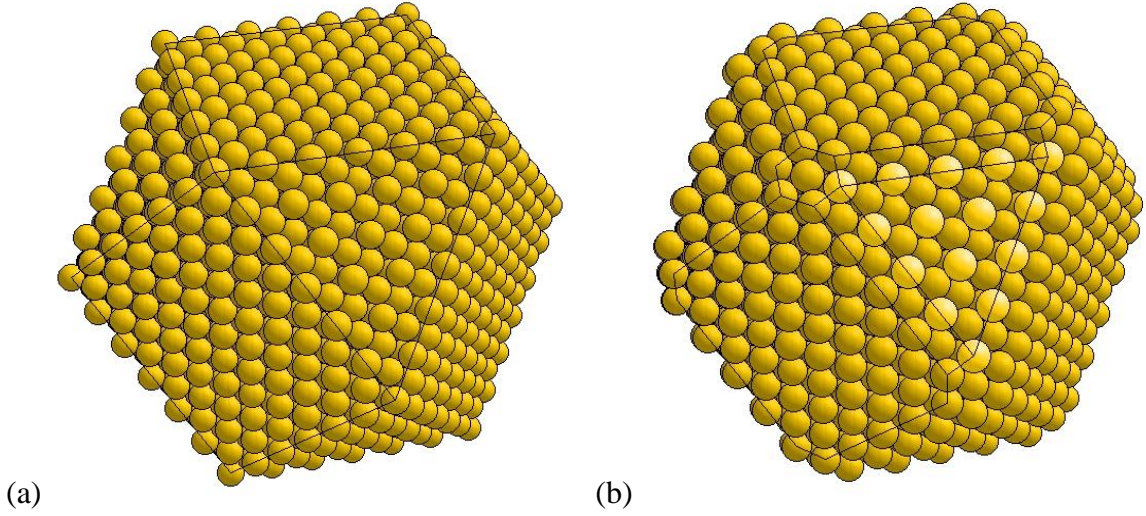

(a) (b)

**Figure B.8.** Atom ball models of cuboctahedral bcc NPs, (a) bcc(12, -, 24) and (b) bcc(11, -, 22). Black lines sketch the square/octagonal {100} and triangular {111} facets with connecting hexagonal {110} facets and {112} strips. The light color balls indicate one {111} facet, see text.

A classification of bcc($N$, -, $K$) NP types for any $N$, $K$ combination is given by Table T.8 of the Supplement.

### B.2.3. Truncated bcc(-, $M$, $K$) Nanoparticles

Non-generic **rhombo-octahedral** NPs, denoted **bcc(-, $M$, $K$)**, are confined by facets of the two generic NPs, bcc(-, $M$, -) (rhombohedral) and bcc(-, -, $K$) (octahedral), see Fig. B.9. If the corners of the rhombohedral NP bcc(-, $M$, -) lie inside the octahedral NP bcc(-, -, $K$), the resulting combination bcc(-, $M$, $K$) will be generic rhombohedral. This requires

$$s_{<111>}(-, M, -) \leq s_{<111>}(-, -, K) \tag{B.33}$$

and with (B.11), (B.14), leads to

$$3M \leq K \tag{B.34}$$

On the other hand, if the corners of the octahedral NP bcc(-, -, $K$) lie inside the rhombohedral NP bcc(-, $M$, -), the resulting combination bcc(-, $M$, $K$) will be generic octahedral. This requires

$$s_{<100>}(-, -, K) \leq s_{<100>}(-, M, -) \tag{B.35}$$

and with (B.11), (B.14), leads to

$$2M \geq K - g \tag{B.36}$$

Thus, the two generic NPs intersect, to yield an NP bcc(-, $M$, $K$) with both {110} and {111} facets, only for $M$, $K$ values with

$$2M + g < K < 3M \tag{B.37}$$

In contrast, bcc(-, $M$, $K$) is generic rhombohedral for larger $K$ according to (B.34) and generic octahedral for smaller $K$ according to (B.36). Further, generic rhombohedral and octahedral bcc NPs can be described by bcc(-, $M$, $K$) where

| | | |
|---|---|---|
| bcc(-, $M$, -) = bcc(-, $M$, $K = 3M$) | (rhombohedral) | (B.38a) |
| bcc(-, -, $K$) = bcc(-, $M = K/2$, $K$) | (octahedral, $K$ even) | (B.38b) |
| bcc(-, -, $K$) = bcc(-, $M = (K - 1)/2$, $K$) | (octahedral, $K$ odd) | (B.38c) |

The surfaces of rhombo-octahedral NPs bcc(-, $M$, $K$) exhibit twelve {110} and eight {111} facets, see Fig. B.9.

The **{110} facets** are hexagonal with four <111> edges of length $(K - 2M)/2 \sqrt{3}a_o$ and two <110> edges of length $(3M - K) \sqrt{2}a_o$.

The **{111} facets** are triangular with three <110> edges of length $(3M - K) \sqrt{2}a_o$.

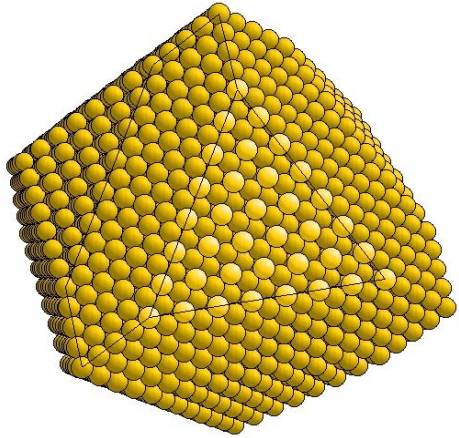

**Figure B.9.** Atom ball models of the rhombo-octahedral NP bcc(-, 11, 26). One {111} facet is emphasized by atom balls of light color. Black lines sketch the hexagonal {110} and triangular {111} facets.

The total number of NP atoms, $N_{vol}(\textbf{-, M, K})$, and the number of facet atoms, $N_{facet}(\textbf{-, M, K})$, (outer polyhedral shell) are given with (B.9), (B.10) by

$$N_{vol}(-, M, K) = N_{vol}(-, M, -) - 4H (H + 1) (H + 2)/3 , \qquad H = 3M - K \tag{B.39}$$

$$N_{facet}(-, M, K) = N_{facet}(-, M, -) - 8 (3M - K)^2 \tag{B.40}$$

A classification of bcc(-, *M*, *K*) NP types for any *M*, *K* combination is given by Table T.9 of the Supplement.

### B.2.4. General bcc(*N*, *M*, *K*) Nanoparticles

Non-generic **cubo-rhombo-octahedral** NPs, denoted **bcc(*N*, *M*, *K*)**, are confined by facets of all three generic NPs, bcc(*N*, -, -) (cubic), bcc(-, *M*, -) (rhombohedral), and bcc(-, -, *K*) (octahedral), see Fig. B.10. Thus, they can show {100}, {110}, and {111} facets. A general discussion of these NPs requires results for generic and non-generic NPs, see Secs. B. 1, B.2.1-3, as will be detailed in the following.

First, we consider the general notation for generic NPs discussed in Sec. B.1. Cubic NPs bcc(*N*, -, -) are surrounded by smallest rhombohedral NPs bcc(-, *M*, -) if *M* = *N* and by smallest octahedral NPs bcc(-, -, *K*) if *K* = 3*N*, see (B.16), (B.24). This allows a notation

$$\text{bcc}(N, -, -) = \text{bcc}(N, M = N, K = 3N) \tag{B.41}$$

Rhombohedral NPs bcc(-, *M*, -) are surrounded by smallest cubic NPs bcc(*N*, -, -) if *N* = 2*M* and by smallest octahedral NPs bcc(-, -, *K*) if *K* = 3*M*, see (B.18), (B.34). This yields

$$\text{bcc}(-, M, -) = \text{bcc}(N = 2M, M, K = 3M) \tag{B.42}$$

Octahedral NPs bcc(-, -, *K*) are surrounded by smallest cubic NPs bcc(*N*, -, -) if *N* = *K* - *g* and by smallest rhombohedral NPs bcc(-, *M*, -) if *M* = (*K* - *g*)/2, see (B.26), (B.36). This yields

$$\text{bcc}(-, -, K) = \text{bcc}(N = K - g, M = (K - g)/2, K) \tag{B.43}$$

General notations for non-generic bcc NPs with two facet types, discussed in Secs. B.2.1-3, are obtained by analogous arguments. Cubo-rhombic NPs bcc(*N*, *M*, -) are surrounded by smallest octahedral NPs bcc(-, -, *K*) if *K* = *K*$_a$ with

$$K_a(N, M) = \min(3N, 3M) = 3M \tag{B.44}$$

This allows a notation

$$\text{bcc}(N, M, -) = \text{bcc}(N, M, K = K_a) \tag{B.45}$$

Cubo-octahedral NPs bcc(*N*, -, *K*) are surrounded by smallest rhombohedral NPs bcc(-, *M*, -) if *M* = *M*$_a$ with

$$M_a(N, K) = \min((K - g)/2, N) \tag{B.46}$$

yielding

$$\text{bcc}(N, -, K) = \text{bcc}(N, M = M_a, K) \tag{B.47}$$

Rhombo-octahedral NPs bcc(-, $M$, $K$) are surrounded by smallest cubic NPs bcc($N$, -, -) if $N = N_a$ with

$$N_a(M, K) = \min(2M, K) = 2M \tag{B.48}$$

yielding

$$\text{bcc}(-, M, K) = \text{bcc}(N = N_a, M, K) \tag{B.49}$$

In the most general case of a bcc($N$, $M$, $K$) NP with {100}, {110}, and {111} facets, we start from a cubo-rhombic NP, bcc($N$, $M$, -) with its constraints $M \leq N \leq 2M$. Then we add constraints of a generic octahedral NP, bcc(-, -, $K$), to yield the cubo-rhombo-octahedral NP bcc($N$, $M$, $K$). This requires, according to the discussion above, $K$ values below $K_a$. Here we can distinguish four different ranges of parameter $K$, defined by separating values $K_a \geq K_b \geq K_c$, where with $K_a$ from (B.44)

$$K_b(N, M) = 4M - N \tag{B.50}$$
$$K_c(N, M) = 2M \quad (N \text{ even}) \tag{B.51a}$$
$$= 2M + 1 \quad (N \text{ odd}) \tag{B.51b}$$

The ranges are illustrated in Fig. B.10 for the cubo-rhombic NP bcc(18, 12, 36).

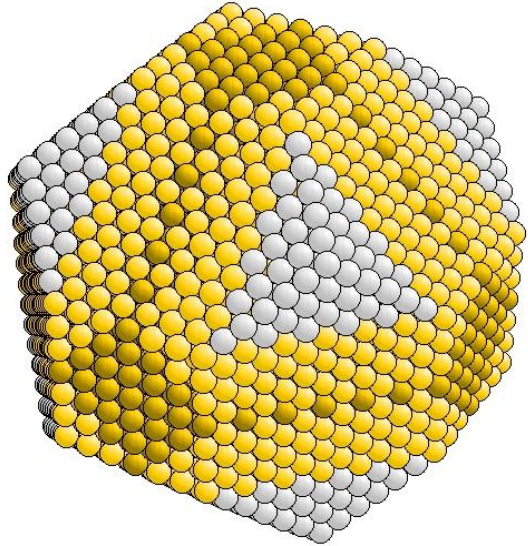

**Figure B.10.** Atom ball model of a cubo-rhombic NP, bcc(18, 12, 36) ($K = K_a$, all atom balls), with its cubo-rhombo-octahedral NP components, bcc(18, 12, 30) ($K = K_b$, dark and light yellow balls), and bcc(18, 12, 24) ($K = K_c$, dark yellow balls), see text.

**Outer $K$ range** of bcc($N$, $M$, $K$) where with (B.44)

$$K \geq K_a \tag{B.52}$$

For these $K$ values the NP becomes cubo-rhombic and does not exhibit any {111} facets (except for microfacets with three atoms). It is structurally identical with bcc($N$, $M$, $K_a$) as discussed above and in Sec. B.2.1.

**Upper central $K$ range** of bcc($N$, $M$, $K$) where with (B.50), (B.51)

$$K_b \leq K \leq K_a \tag{B.53}$$

For these $K$ values the initial bcc($N$, $M$, $K_a$) NP is capped at its <111> corners creating eight triangular {111} facets. Altogether, these NPs exhibit six {100} facets, twelve {110} facets, and eight {111} facets, see Fig. B.11.

The **{100} facets** are square with four <100> edges of length $(2M - N)\,a_o$.

The **{110} facets** are octagonal or rectangular ($K = K_b$) with two <110> edges of length $(3M - K)\,\sqrt{2}a_o$, two <100> of length $(2M - N)\,a_o$, and two <111> of length $(K + N - 4M)/2\,\sqrt{3}a_o$.

The **{111} facets** are triangular with three <110> edges of length $(3M - K)\,\sqrt{2}a_o$.

The NP structure is illustrated in Fig. B.11 for the NP bcc(18, 10, 26) ($K_a = 30$, $K_b = 22$, yellow atom balls), where white balls above the {111} facets are added to yield the corresponding cubo-rhombic bcc($N$, $M$, $K_a$) NP.

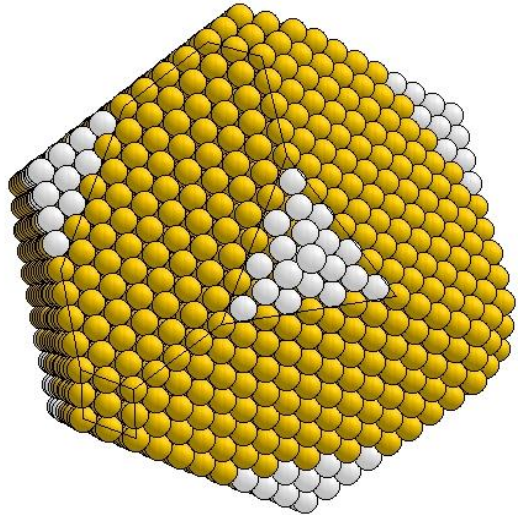

**Figure B.11.** Atom ball model of a cubo-rhombo-octahedral bcc NP, bcc(18, 10, 26), see text. Black lines sketch the square {100}, octagonal {110}, and triangular {111} facets.

The total number of NP atoms, $N_{vol}(N, M, K)$, and the number of facet atoms, $N_{facet}(N, M, K)$, (outer polyhedral shell) are given with (B.21), (B.22) by

$$N_{vol}(N, M, K) = N_{vol}(N, M, \text{-}) - 4H(H+1)(H+2)/3, \quad H = 3M - K \quad \text{(B.54)}$$
$$N_{facet}(N, M, K) = N_{facet}(N, M, \text{-}) - 8H^2 \quad \text{(B.55)}$$

For $K = K_b$, the bcc($N, M, K$) NP assumes a particular shape where its twelve **{110} facets** are rectangular with two edges of length $(N - M)\sqrt{2}a_o$ and of $(2M - N)a_o$ while the **{100}** and **{111} facets** are square and triangular as described before. This is illustrated in Fig. B.12 for the NP bcc(12, 8, 20) ($K_b = 20$).

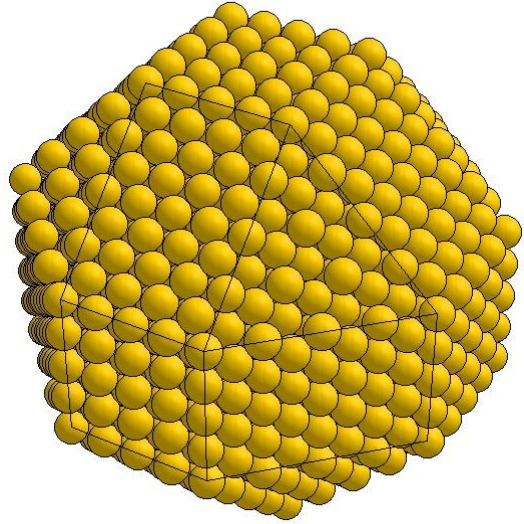

**Figure B.12.** Atom ball model of a cubo-rhombo-octahedral bcc NP, void centered bcc(12, 8, 20). Black lines sketch the square {100}, rectangular {110}, and triangular {111} facets.

**Lower central $K$ range** of bcc($N, M, K$) where with (B.51)

$$K_c \leq K \leq K_b \quad \text{(B.56)}$$

For these $K$ values the capping of the initial bcc($N, M, K_b$) along the <111> directions is continued to yield eight hexagonal {111} facets. As before, these NPs exhibit six {100} facets, twelve {110} facets, and eight {111} facets, see Fig. B.13.

The **{100} facets** are octagonal with alternating edges, four <100> of length $(K - 2M)a_o$ and four <110> of length $(4M - N - K)/2\sqrt{2}a_o$.

The **{110} facets** are rectangular with two <110> edges of length $(N - M)\sqrt{2}a_o$ and two <100> edges of length $(K - 2M)a_o$.

The **{111} facets** are hexagonal with <110> edges of alternating lengths

$(4M - N - K)/2 \sqrt{2} a_o$ and $(N - M) \sqrt{2} a_o$.

The NP structure is illustrated in Fig. B.13 for the NP bcc(18, 12, 26) ($K_b = 30$, $K_c = 24$, yellow atom balls) where white balls above the {111} facets are added to bcc($N, M, K$) to yield the corresponding bcc($N, M, K_b$) NP.

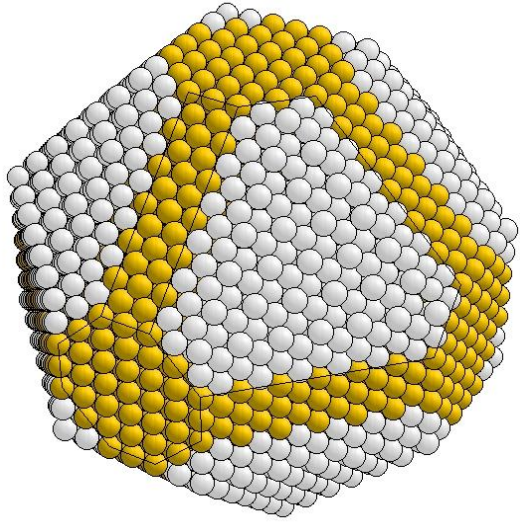

**Figure B.13.** Atom ball model of a cubo-rhombo-octahedral bcc NP, atom centered bcc(18, 12, 26), see text. Black lines sketch the octagonal {100}, rectangular {110}, and hexagonal/triangular {111} facets.

The total number of NP atoms, $N_{vol}(N, M, K)$, and the number of facet atoms, $N_{facet}(N, M, K)$, (outer polyhedral shell) are given with (B.54), (B.55), (B.5) by

$$N_{vol}(N, M, K) = N_{vol}(N, M, K_b) -$$
$$- H \{(H + 2)(H + 12G + 1)/3 + 4 G^2 + 3h\} - 3h \quad (B.57)$$

$$N_{facet}(N, M, K) = N_{facet}(N, M, K_b) - 2H(H + 8G) - 6h \quad (B.58)$$

$$H = 4M - N - K, \quad G = N - M$$

**Inner $K$ range** of bcc($N, M, K$) where with (B.51)

$$K \leq K_c \quad (B.59)$$

For these $K$ values the NP becomes cubo-octahedral and does not exhibit {110} facets (except for possible microstrips). It is structurally identical with bcc($N, M_a, K$) as discussed above and in Sec. B.2.2.

A classification of bcc($N, M, K$) NP types for any $N, M, K$ combination is given by Table T.10 of the Supplement.

## C. Simple Cubic Nanoparticles

The simple cubic (**sc**) lattice is defined by lattice vectors $\underline{R}_1, \underline{R}_2, \underline{R}_3$ according to

$$\underline{R}_1 = a_o\,(1, 0, 0)\,, \qquad \underline{R}_2 = a_o\,(0, 1, 0)\,, \qquad \underline{R}_3 = a_o\,(0, 0, 1) \qquad (C.1)$$

$$\underline{r}_1 = a_o\,(0, 0, 0) \qquad (C.2)$$

in Cartesian coordinates, where $a_o$ is the lattice constant. The lattice is primitive with only one lattice basis vector $\underline{r}_1$ which may be positioned at the lattice origin. The three densest monolayer families {$hkl$} of the sc lattice are described by six {100} netplanes (square mesh, highest atom density), twelve {110} netplanes (rectangular mesh), and eight {111} netplanes (hexagonal mesh). Distances between adjacent parallel netplanes are given by

$$d_{\{100\}} = a_o\,, \qquad d_{\{110\}} = a_o/\sqrt{2}\,, \qquad d_{\{111\}} = a_o/\sqrt{3} \qquad (C.3)$$

The point symmetry of the sc lattice is characterized by $O_h$ with high symmetry centers at all atom sites and at the void centers of each elementary cell.

Compact sc nanoparticles (NPs) are confined by finite sections of monolayers (facets) whose structure is described by different netplanes ($hkl$). For NPs of central $O_h$ symmetry this includes all members of corresponding {$hkl$} families. As an example, we mention the {100} family with its six netplane orientations (±1 0 0), (0 ±1 0), (0 0 ±1). Thus, surfaces of sc NPs with $O_h$ symmetry are described by facets with orientations of different {$hkl$} families (denoted {$hkl$} facets in the following). These facets are limited by edges which are determined by families of Miller index directions <$hkl$> (denoted <$hkl$> edges in the following). In addition, NP corners can be characterized by directions <$hkl$> pointing from the NP center to the corresponding corner (denoted <$hkl$> corners in the following). Some <$hkl$> corners are capped to form {$hkl$} microfacets ({111} of three atoms, {100} of four or five atoms), if the confining {$hkl$} monolayer planes join at corners which are not occupied by atoms of the ideal lattice. Finally, the symmetry of the sc host lattice allows only atom sites or $O_h$ symmetry void sites for possible NP centers. Thus, we distinguish between atom centered and void centered sc NPs denoted **ac** and **vc** in the following.

Assuming an sc NP to be confined by facets of the three cubic netplane families, {100}, {110}, and {111}, its size and shape can be described by three integer type structure parameters, $N$, $M$, $K$ (polyhedral NP parameters). These refer to the distances $D_{\{100\}}$, $D_{\{110\}}$, $D_{\{111\}}$ (NP diameters) between parallel facets of a given netplane family, expressed by multiples of corresponding netplane distances where

$$D_{\{100\}} = N\,d_{\{100\}}\,, \qquad D_{\{110\}} = 2M\,d_{\{110\}}\,, \qquad D_{\{111\}} = K\,d_{\{111\}} \qquad (C.4)$$

with $d_{\{hkl\}}$ according to (C.3). Thus, in the most general case sc NPs can be denoted **sc(N, M, K)**. If a facet type does not appear in the NP (or shows only as a very small microfacet), the corresponding parameter value $N$, $M$, or $K$ may be ignored and is replaced by a minus sign. As an example, an sc NP with only {100} and {111} facets is denoted sc($N$, -, $K$). These notations will be used in the following. Further, we introduce auxiliary parameters **g**, **h** referring to parity of $N$, $M$, and $K$ with

$$g = 0 \quad (\text{ac}; N, K \text{ even}), \qquad = 1 \quad (\text{vc}; N, K \text{ odd}) \qquad (C.5)$$
$$h = 0 \quad (M + N \text{ even}; M + K \text{ even}), \qquad = 1 \quad (M + N \text{ odd}; M + K \text{ odd}) \qquad (C.6)$$

which help to simplify many algebraic expressions throughout Sec. C.

## C.1. Generic sc(N, -, -), (-, M, -), and (-, -, K) Nanoparticles

Generic sc nanoparticles (NPs) of $O_h$ symmetry are confined by facets with orientations of only one {$hkl$} netplane family (except for very small microfacets, see below). Here we consider {100}, {110}, and {111} facets derived from highly dense monolayers of the sc lattice with {100} representing the densest. These allow three different generic NP types.

**(a)** **Generic cubic** sc NPs, denoted **sc(N, -, -)** are confined by all six {100} monolayers with distances $D_{\{100\}} = N\, d_{\{100\}}$ between parallel confining monolayers (ac NPs for $N$ even, vc NPs for $N$ odd). This yields six {100} facets, see Fig. C.1.

The **{100} facets** are always square with four <100> edges of length $N\, a_o$.

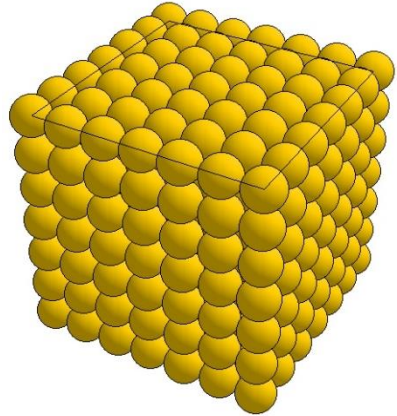

**Figure C.1.** Atom ball model of a generic atom centered NP, sc(6, -, -). Black lines sketch a square {100} facet.

The total number of NP atoms, $N_{vol}(N, -, -)$, and the number of facet atoms, $N_{facet}(N, -, -)$, (outer polyhedral shell), are given by

$$N_{vol}(N, -, -) = (N + 1)^3 \quad \text{(C.7)}$$

$$N_{facet}(N, -, -) = 6N^2 + 2 \quad \text{(C.8)}$$

The largest distance from the NP center to its surface along <hkl> directions, $s_{<hkl>}$, for <hkl> = <100>, <110>, and <111>, is given by

$$s_{<100>}(N, -, -) = N/2 \; d_{\{100\}} \quad \text{(C.9a)}$$

$$s_{<110>}(N, -, -) = N \; d_{\{110\}} \quad \text{(C.9b)}$$

$$s_{<111>}(N, -, -) = 3N/2 \; d_{\{111\}} \quad \text{(C.9c)}$$

with $d_{\{hkl\}}$ according to (C.3). These quantities will be used in Secs. C.2.

(b) **Generic rhombohedral** sc NPs, denoted **sc(-, M, -)** are confined by all twelve {110} monolayers with distances $D_{\{110\}} = 2M \; d_{\{110\}}$ between parallel confining monolayers. This yields twelve {110} facets as well as possibly six smaller {100} and eight {111} facets, see Fig. C.2, C.3.

The **{100} facets** are microfacets of four atoms and appear only for vc NPs. They are square with four <100> edges of length $a_o$.

The **{110} facets** are rhombic, hexagonal, or octagonal with two <100> edges of length $n \; a_o$, two <110> edges of length $m \; a_o/\sqrt{2}$, and four <111> edges of length $k \; \sqrt{3}a_o$.

Corresponding edge parameters $n$, $m$, $k$ are given in Table C.1.

| Centering | $M$ even | $M$ odd |
|---|---|---|
| ac | $n = 0$<br>$m = 0$<br>$k = M/2$ | $n = 0$<br>$m = 2$<br>$k = (M - 1)/2$ |
| vc | $n = 1$<br>$m = 2$<br>$k = (M - 2)/2$ | $n = 1$<br>$m = 0$<br>$k = (M - 1)/2$ |

**Table C.1.** Edge parameters $n$, $m$, $k$ of {100}, and {110} and {111} facets of sc(-, $M$, -) NPs, see text.

The **{111} facets** are microfacets of three atoms and appear only for ac, $M$ odd or for vc, $M$ even NPs. They are triangular with three <110> edges of length 2 $a_o/\sqrt{2}$.

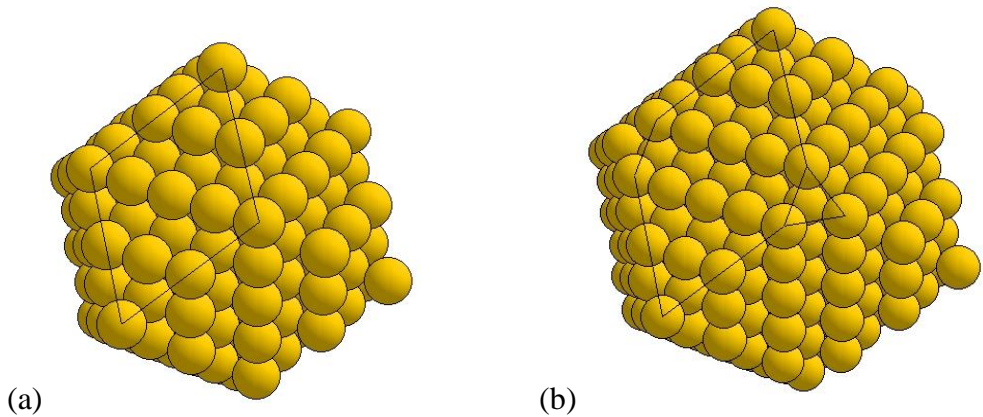

(a)                (b)

**Figure C.2.** Atom ball models of generic rhombohedral atom centered NPs, (a) sc(-, 4, -) and (b) sc(-, 5, -). Black lines sketch the (capped) rhombic {110} and triangular {111} facets.

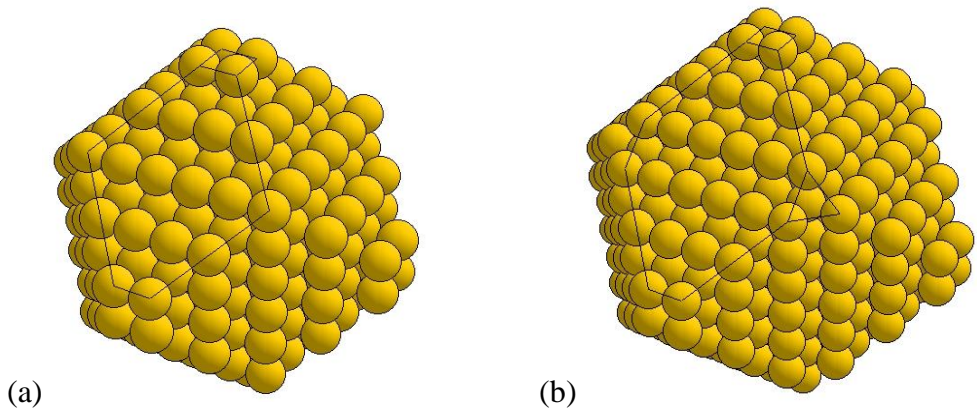

(a)                (b)

**Figure C.3.** Atom ball models of generic rhombohedral void centered NPs, (a) sc(-, 5, -) and (b) sc(-, 6, -). Black lines sketch the capped rhombic {110}, square {100}, and triangular {111} facets.

The total number of NP atoms, $N_{vol}(-, M, -)$, and the number of facet atoms, $N_{facet}(-, M, -)$, (outer polyhedral shell), are given with (C.6) by

$$N_{vol}(-, M, -) = M(2M^2 + 3M + 2) + 1 - h \qquad (C.10)$$
$$N_{facet}(-, M, -) = 6M^2 + 2(1 - h) \qquad (C.11)$$

The largest distance from the NP center to its surface along $\langle hkl \rangle$ directions, $s_{\langle hkl \rangle}$, is given with (C.5), (C.6) by

$$s_{\langle 100 \rangle}(-, M, -) = (2M - g)/2\, d_{\{100\}} \qquad (C.12a)$$
$$s_{\langle 110 \rangle}(-, M, -) = M\, d_{\{110\}} \qquad (C.12b)$$
$$s_{\langle 111 \rangle}(-, M, -) = (3M - h)/2\, d_{\{111\}} \qquad (C.12c)$$

with $d_{\{hkl\}}$ according to (C.3). These quantities will be used in Secs. C.2.

**(c)** **Generic octahedral** sc NPs, denoted **sc(-, -, K)**, are confined by all eight {111} monolayers with distances $D_{\{111\}} = K\, d_{\{111\}}$ between parallel confining monolayers (ac NPs for $K$ even, vc NPs for $K$ odd). This yields eight {111} facets as well as possibly six smaller {100} and twelve {110} facets, see Fig. C.4.

The **{100} facets** are microfacets of four atoms and appear only for $K$ odd. They are square with four <100> edges of length $a_o$.

The **{111} facets** are triangular with three <110> edges of length $K\, a_o/\sqrt{2}$ for $K$ even and of length $(K - 3)\, a_o/\sqrt{2}$ for $K$ odd.

The **{110} facets** appear only for $K$ odd. They form narrow rectangular strips with two <100> edges of length $a_o$ and two <110> edges of length $(K - 3)\, a_o/\sqrt{2}$.

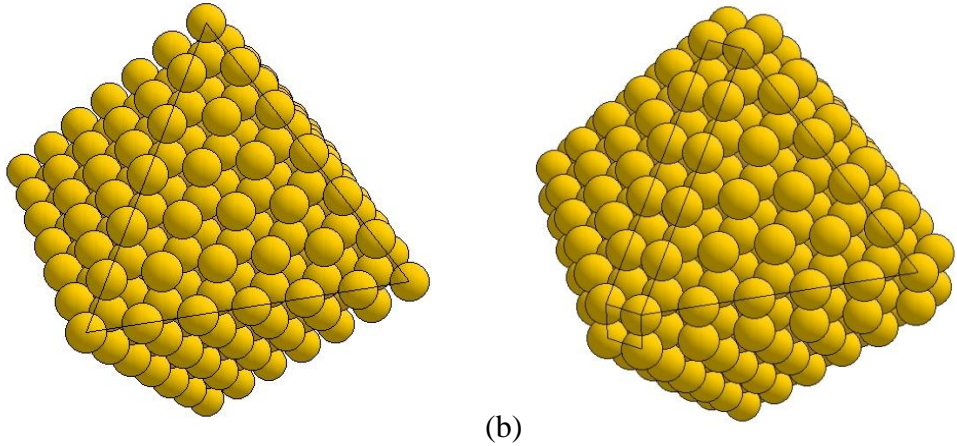

(a)              (b)

**Figure C.4.** Atom ball models of generic octahedral NPs, (a) atom centered sc(-, -, 12) and (b) void centered sc(-, -, 13). Black lines sketch the triangular {111}, stripped {110}, and square {100} facets.

The total number of NP atoms, $N_{vol}(\text{-, -, }K)$, and the number of facet atoms, $N_{facet}(\text{-, -, }K)$, (outer polyhedral shell), are given with (C.5) by

$$N_{vol}(\text{-, -, }K) = (K+1)\,[(K+1)^2 + 5 - 9g]/6 \qquad (C.13)$$

$$N_{facet}(\text{-, -, }K) = K^2 + 2 - 3g \qquad (C.14)$$

The largest distance from the NP center to its surface along $\langle hkl \rangle$ directions, $s_{\langle hkl \rangle}$, is given with (C.5) by

$$s_{\langle 100 \rangle}(\text{-, -, }K) = (K - 2g)/2 \; d_{\{100\}} \qquad (C.15a)$$

$$s_{\langle 110 \rangle}(\text{-, -, }K) = (K - g)/2 \; d_{\{110\}} \qquad (C.15b)$$

$$s_{\langle 111 \rangle}(\text{-, -, }K) = K/2 \; d_{\{111\}} \qquad (C.15c)$$

with $d_{\{hkl\}}$ according to (C.3). These quantities will be used in Secs. C.2.

Table T.11 of the Supplement collects types, constraints, and shapes of all generic sc NPs.

## C.2. Non-generic sc Nanoparticles

Non-generic sc nanoparticles of $O_h$ symmetry show facets with orientations of more than one $\{hkl\}$ netplane family. This can be considered as combining confinements of the corresponding generic NPs, discussed in Sec. C.1, sharing their symmetry center, atom centered (**ac**) or high symmetry void centered (**vc**). Thus, non-generic NPs are mutual intersections of more than one generic NP, where one NP cuts corners and edges from the other(s) to form additional facets. Here we discuss non-generic NPs **sc(*N*, *M*, *K*),** which combine constraints of up to three generic NPs, cubic sc(*N*, -, -), rhombohedral sc(-, *M*, -), and octahedral sc(-, -, *K*). Thus, they allow $\{100\}$, $\{110\}$, as well as $\{111\}$ facets. Clearly, the corresponding polyhedral parameters *N*, *M*, *K* depend on each other and determine the overall NP shape. In particular, *N*, *K* can be restricted to both even or both odd values, where ac NPs require *N*, *K* even, while vc NPs require *N*, *K* odd. Further, if a participating generic NP encloses another participant it will not contribute to the overall NP shape. Thus, the respective $\{hkl\}$ facets will not appear at the surface of the non-generic NP. In the following, we consider the three types of non-generic NPs, which combine constraints due to two generic NPs (Secs. C.2.1-3), before we discuss the most general case of sc(*N*, *M*, *K*) NPs in Sec. C.2.4.

## C.2.1. Truncated sc(*N*, *M*, -) Nanoparticles

Non-generic **cubo-rhombic** NPs, denoted **sc(*N*, *M*, -)**, are confined by facets of the two generic NPs, sc(*N*, -, -) (cubic) and sc(-, *M*, -) (rhombohedral), see Fig. C.5. If the edges of the cubic NP sc(*N*, -, -) lie inside the rhombohedral NP sc(-, *M*, -), the resulting combination sc(*N*, *M*, -) will be generic cubic. This requires

$$s_{<110>}(N, -, -) \leq s_{<110>}(-, M, -) \tag{C.16}$$

and with (C.9), (C.12), leads to

$$N \leq M \tag{C.17}$$

for both ac and vc NPs. On the other hand, if the corners of the rhombohedral NP sc(-, *M*, -) lie inside the cubic NP sc(*N*, -, -), the resulting combination sc(*N*, *M*, -) will be generic rhombohedral. This requires

$$s_{<100>}(-, M, -) \leq s_{<100>}(N, -, -) \tag{C.18}$$

and with (C.9), (C.12), (C.5), leads to

$$N \geq 2M - g \tag{C.19}$$

Thus, the two generic NPs intersect, to yield an NP sc(*N*, *M*, -) with both {100} and {110} facets (apart from {111} microfacets), only for *N*, *M* values where with (C.5)

$$M < N < 2M - g \tag{C.20}$$

In contrast, sc(*N*, *M*, -) is generic cubic for smaller *N* according to (C.17) and generic rhombohedral for larger *N* according to (C.19). Further, generic cubic and rhombohedral sc NPs can be described by sc(*N*, *M*, -) where with (C.5)

$$\text{sc}(N, -, -) = \text{sc}(N, M = N, -) \qquad \text{(cubic)} \tag{C.21a}$$

$$\text{sc}(-, M, -) = \text{sc}(N = 2M - g, M, -) \qquad \text{(rhombohedral)} \tag{C.21b}$$

The surfaces of cubo-rhombic NPs sc(*N*, *M*, -) exhibit six {100} facets, twelve {110} facets, and eight smaller {111} facets, see Fig. C.5.

The **{100} facets** are square with four <100> edges of length $(2M - N) a_o$.

The **{110} facets** for $(N + M)$ even, are hexagonal with four <111> edges of length $(N - M)/2 \sqrt{3} a_o$ and two <100> edges of length $(2M - N) a_o$. For $(N + M)$ odd, the facets are octagonal (capped hexagonal) with four <111> edges of length $(N - M - 1)/2 \sqrt{3} a_o$, two <100> edges of length $(2M - N) a_o$, and two <110> edges of length $\sqrt{2} a_o$.

The **{111} facets** are microfacets of three atoms and appear only for (N + M) odd. They are triangular with three <110> edges of length $\sqrt{2}a_o$.

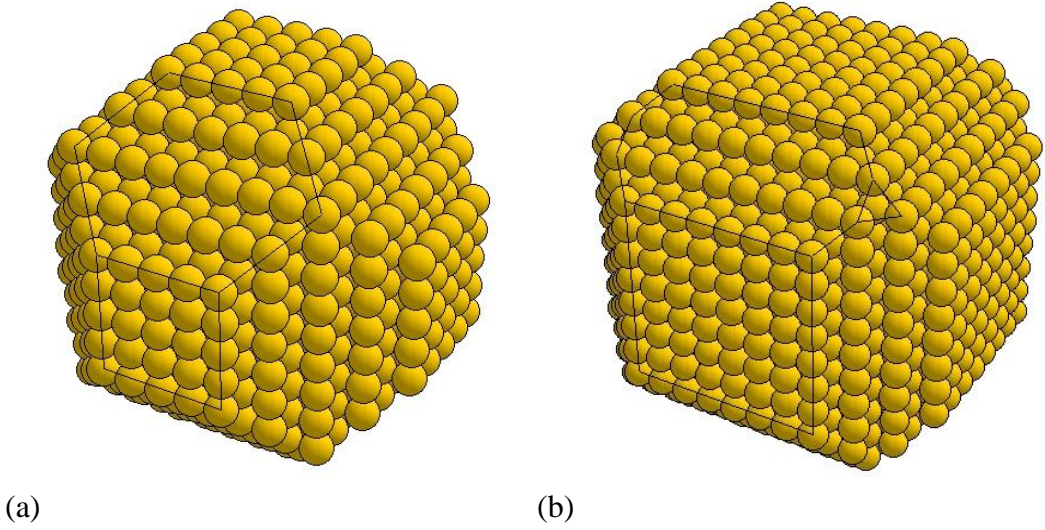

(a)              (b)

**Figure C.5.** Atom ball models of cubo-rhombic NPs, (a) atom centered sc(12, 8, -) and (b) void centered sc(13, 10, -). Black lines sketch the square {100}, (capped) hexagonal {110}, and triangular {111} facets.

The total number of NP atoms, $N_{vol}(N, M, -)$, and the number of facet atoms, $N_{facet}(N, M, -)$, (outer polyhedral shell) are given with (C.10), (C.11) by

$$N_{vol}(N, M, -) = N_{vol}(-, M, -) - H(H^2 - 1), \qquad H = 2M - N \qquad (C.22)$$
$$N_{facet}(N, M, -) = N_{facet}(-, M, -) \qquad (C.23)$$

A classification of sc(N, M, -) NP types for any N, M combination is given by Table T.12 of the Supplement.

### C.2.2. Truncated sc(N, -, K) Nanoparticles

Non-generic **cubo-octahedral** NPs, denoted **sc(N, -, K)**, are confined by facets of the two generic NPs, sc(N, -, -) (cubic) and sc(-, -, K) (octahedral), see Figs. C.6, C.7. If the corners of the cubic NP sc(N, -, -) lie inside the octahedral NP sc(-, -, K), the resulting combination sc(N, -, K) will be generic cubic. This requires

$$s_{<111>}(N, -, -) \leq s_{<111>}(-, -, K) \qquad (C.24)$$

and with (C.9), (C.15), leads to

$$3N \leq K \qquad (C.25)$$

On the other hand, if the corners of the octahedral NP sc(-, -, $K$) lie inside the cubic NP sc($N$, -, -), the resulting combination sc($N$, -, $K$) will be generic octahedral. This requires

$$s_{<100>}(-, -, K) \leq s_{<100>}(N, -, -) \tag{C.26}$$

and with (C.9), (C.15), (C.5), leads to

$$N \geq K - 2g \tag{C.27}$$

Thus, the two generic NPs intersect, to yield an NP sc($N$, -, $K$) with both {100} and {111} facets (apart from {110} microstrips), only for $N$, $K$ values, where with (C.5)

$$N + 2g < K < 3N \tag{C.28}$$

In contrast, sc($N$, -, $K$) is generic cubic for larger $K$ according to (C.25) and generic octahedral for smaller $K$ according to (C.27). Further, generic cubic and octahedral sc NPs can be described by sc($N$, -, $K$) where with (C.5)

$$\text{sc}(N, -, -) = \text{sc}(N, -, K = 3N) \quad \text{(cubic)} \tag{C.29a}$$
$$\text{sc}(-, -, K) = \text{sc}(N = K - 2g, -, K) \quad \text{(octahedral)} \tag{C.29b}$$

The surfaces of cubo-octahedral NPs sc($N$, -, $K$) exhibit six {100}, twelve {110}, and eight {111} facets, see Figs. C.6, C.7. Amongst the intersecting species according to (C.28) we can distinguish between **truncated octahedral** NPs where $K < 2N$ and **truncated cubic** NPs for $K > 2N$, with **cuboctahedral** NPs for $K = 2N$ separating. This will be discussed in the following.

**Truncated octahedral** NPs ($K < 2N$), Fig. C.6, can be characterized by their facets as follows.

- The **{100} facets** for $N$, $K$ even, are square with four <110> edges of length $(K - N)/2 \sqrt{2}a_o$. For $N$, $K$ odd, the facets are octagonal (capped square) with alternating edges, four <110> of length $(K - N - 2)/2 \sqrt{2}a_o$ and four <100> of length $a_o$.
- The **{110} facets** appear only for $N$, $K$ odd. They form narrow rectangular strips with two <110> edges of length $(2N - K + 1)/2 \sqrt{2}a_o$ and two <100> edges of length $a_o$.
- The **{111} facets** are hexagonal with <110> edges of alternating lengths $(K - N)/2 \sqrt{2}a_o$ and $(2N - K)/2 \sqrt{2}a_o$ for $N$, $K$ even, while for $N$, $K$ odd, the edges are of alternating lengths $(K - N - 2)/2 \sqrt{2}a_o$ and $(2N - K + 1)/2 \sqrt{2}a_o$.

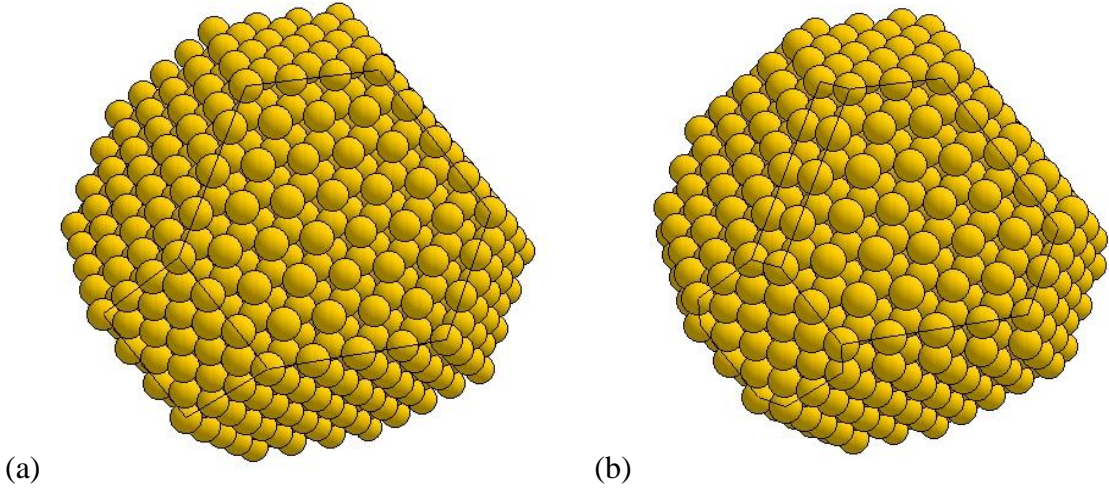

**Figure C.6.** Atom ball models of cubo-octahedral NPs of truncated octahedral type, (a) atom centered sc(14, -, 20) and (b) void centered sc(13, -, 19). Black lines sketch the square {100}, stripped {110}, and hexagonal {111} facets.

The total number of NP atoms, $N_{vol}(N, -, K)$, and the number of facet atoms, $N_{facet}(N, -, K)$, (outer polyhedral shell) are given with (C.13), (C.14), (C.5) by

$$N_{vol}(N, -, K) = N_{vol}(-, -, K) - H(H^2 + 2 - 6g)/2, \qquad H = K - N \qquad \text{(C.30)}$$

$$N_{facet}(N, -, K) = N_{facet}(-, -, K) \qquad \text{(C.31)}$$

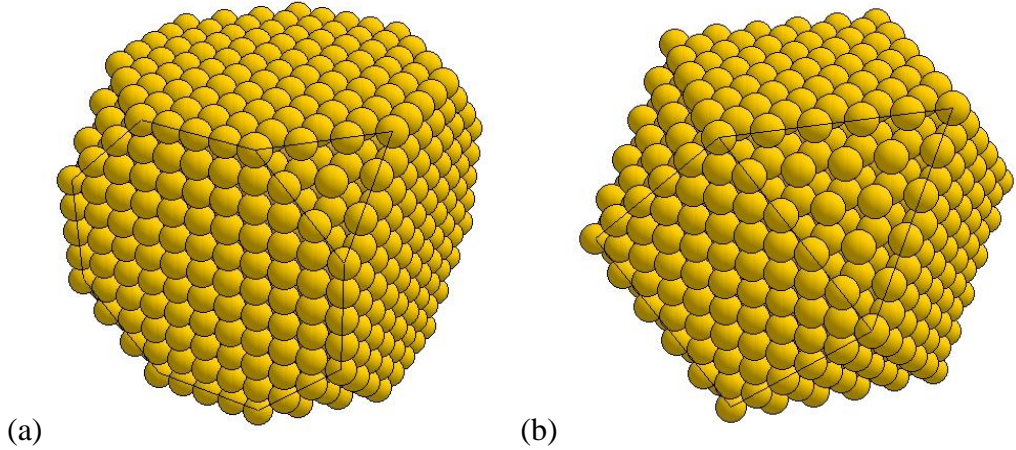

**Figure C.7.** Atom ball models of atom centered NPs, (a) cubo-octahedral sc(10, -, 24) (truncated cubic) and (b) cuboctahedral sc(10, -, 20). Black lines sketch the octagonal/square {100}, triangular {111} facets.

**Truncated cubic** NPs ($K > 2N$), Fig. 7, can be characterized by their facets as follows.

The **{100} facets** are octagonal with alternating edges, four <100> of length $(K - 2N) a_o$ and four <110> of length $(3N - K)/2 \sqrt{2} a_o$.

The **{111} facets** are triangular with <110> edges of length $(3N - K)/2 \sqrt{2} a_o$.

The total number of NP atoms, $N_{vol}(N, -, K)$, and the number of facet atoms, $N_{facet}(N, -, K)$, (outer polyhedral shell) are given with (C.7), (C.8) by

$$N_{vol}(N, -, K) = N_{vol}(N, -, -) - H (H + 2) (H + 4)/6 \tag{C.32}$$

$$N_{facet}(N, -, K) = N_{facet}(N, -, -) - 2H^2, \qquad H = 3N - K \tag{C.33}$$

There are sc NPs which can be assigned to both truncated cubic and truncated octahedral type, the **cuboctahedral** NPs sc($N, -, K$), defined by $K = 2N$. These NPs exist only as atom centered species since both $N$ and $K$ must be even. They exhibit six {100} and eight {111} facets, see Fig. C.7b. All **{100} facets** are square with four <110> edges of length $N/2 \sqrt{2} a_o$ while all **{111} facets** are triangular with three <110> edges of length $N/2 \sqrt{2} a_o$ shared with those of the {100} facets.

A classification of sc($N, -, K$) NP types for any $N$, $K$ combination is given by Table T.13 of the Supplement.

### C.2.3. Truncated sc(-, $M$, $K$) Nanoparticles

Non-generic **rhombo-octahedral** NPs, denoted **sc(-, $M$, $K$)**, are confined by facets of the two generic NPs, sc(-, $M$, -) (rhombohedral) and sc(-, -, $K$) (octahedral), see Fig. C.8. If the corners of the rhombohedral NP sc(-, $M$, -) lie inside the octahedral NP sc(-, -, $K$), the resulting combination sc(-, $M$, $K$) will be generic rhombohedral. This requires

$$s_{<111>}(-, M, -) \leq s_{<111>}(-, -, K) \tag{C.34}$$

and with (C.12), (C.15), (C.6), leads to

$$3M \leq K + h \tag{C.35}$$

On the other hand, if the corners of the octahedral NP sc(-, -, $K$) lie inside the rhombohedral NP sc(-, $M$, -), the resulting combination sc(-, $M$, $K$) will be generic octahedral. This requires

$$s_{<100>}(-, -, K) \leq s_{<100>}(-, M, -) \tag{C.36}$$

and with (C.12), (C.15), (C.5), leads to

$$2M \geq K - g \tag{C.37}$$

Thus, the two generic NPs intersect, to yield an NP sc(-, $M$, $K$) with both {110} and {111} facets (apart from {100} microfacets), only for $M$, $K$ values, where with (C.5), (C.6)

$$2M + g < K < 3M - h \tag{C.38}$$

In contrast, sc(-, $M$, $K$) is generic rhombohedral for larger $K$ according to (C.35) and generic octahedral for smaller $K$ according to (C.37). Further, generic rhombohedral and octahedral sc NPs can be described by sc(-, $M$, $K$) where with (C.5), (C.6)

$$\text{sc}(-, M, -) = \text{sc}(-, M, K = 3M - h) \qquad \text{(rhombohedral)} \tag{C.39a}$$

$$\text{sc}(-, -, K) = \text{sc}(-, M = (K - g)/2, K) \qquad \text{(octahedral)} \tag{C.39b}$$

The surfaces of rhombo-octahedral NPs sc(-, $M$, $K$) exhibit twelve {110} and eight {111} facets with six possible {100} microfacets, see Fig. C.8.

The **{100} facets** are microfacets of four atoms and appear only for $K$ odd. They are square with <100> edges of length $a_o$.

The **{110} facets** for $K$ even, are hexagonal with four <111> edges of length $(K - 2M)/2 \sqrt{3}a_o$ and two <110> edges of length $(3M - K) \sqrt{2}a_o$. For $K$ odd, the facets are octagonal with four <111> edges of length $(K - 2M - 1)/2 \sqrt{3}a_o$, two <100> edges of length $a_o$, and two <110> edges of length $(3M - K) \sqrt{2}a_o$.

The **{111} facets** are triangular with three <110> edges of length $(3M - K) \sqrt{2}a_o$.

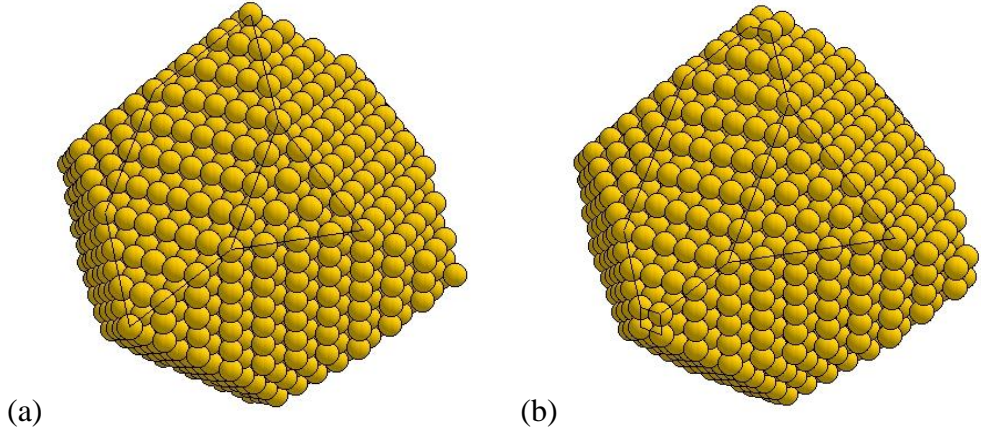

(a) (b)

**Figure C.8.** Atom ball models of rhombo-octahedral NPs, (a) atom centered sc(-, 10, 26) and (b) void centered sc(-, 10, 25). Black lines sketch the hexagonal/octagonal {110}, triangular {111}, and square {100} facets.

The total number of NP atoms, $N_{vol}$(-, $M$, $K$), and the number of facet atoms, $N_{facet}$(-, $M$, $K$), (outer polyhedral shell) are given with (C.10), (C.11), (C.6) by

$$N_{vol}(-, M, K) = N_{vol}(-, M, -) - H(H+2)(2H-1)/3 + h, \qquad H = 3M - K \qquad (C.40)$$

$$N_{facet}(-, M, K) = N_{facet}(-, M, -) - 2H^2 + 2h \qquad (C.41)$$

A classification of sc(-, $M$, $K$) NP types for any $M$, $K$ combination is given by Table T.14 of the Supplement.

### C.2.4. General sc($N$, $M$, $K$) Nanoparticles

Non-generic **cubo-rhombo-octahedral** NPs, denoted **sc($N$, $M$, $K$)**, are confined by facets of all three generic NPs, sc($N$, -, -) (cubic), sc(-, $M$, -) (rhombohedral), and sc(-, -, $K$) (octahedral), see Fig. C.9. Thus, they can show {100}, {110}, and {111} facets. A general discussion of these NPs requires results for generic and non-generic NPs, see Secs. C.1, C.2.1-3, as will be detailed in the following.

First, we consider the general notation for generic NPs discussed in Sec. C.1. Cubic NPs sc($N$, -, -) are surrounded by smallest rhombohedral NPs sc(-, $M$, -) if $M = N$ and by smallest octahedral NPs sc(-, -, $K$) if $K = 3N$, see (C.17), (C.25). This allows a notation

$$\text{sc}(N, -, -) = \text{sc}(N, M = N, K = 3N) \qquad (C.42)$$

Rhombohedral NPs sc(-, $M$, -) are surrounded by smallest cubic NPs sc($N$, -, -) if $N = 2M - g$ and by smallest octahedral NPs sc(-, -, $K$) if $K = 3M - h$, see (C.19), (C.35). This yields with (C.5), (C.6)

$$\text{sc}(-, M, -) = \text{sc}(N = 2M - g, M, K = 3M - h) \qquad (C.43)$$

Octahedral NPs sc(-, -, $K$) are surrounded by smallest cubic NPs sc($N$, -, -) if $N = K - 2g$ and by smallest rhombohedral NPs sc(-, $M$, -) if $M = (K - g)/2$, see (C.27), (C.37). This yields with (C.5)

$$\text{sc}(-, -, K) = \text{sc}(N = K - 2g, M = (K - g)/2, K) \qquad (C.44)$$

General notations for non-generic sc NPs with two facet types, discussed in Secs. C.2.1-3, are obtained by analogous arguments. Cubo-rhombic NPs sc($N$, $M$, -) are surrounded by smallest octahedral NPs sc(-, -, $K$) if $K = K_a$ where with (C.6)

$$K_a(N, M) = \min(3N, 3M - h) = 3M - h \qquad (C.45)$$

This allows a notation

$$\text{sc}(N, M, -) = \text{sc}(N, M, K = K_a) \qquad (C.46)$$

Cubo-octahedral NPs sc($N$, -, $K$) are surrounded by smallest rhombohedral NPs sc(-, $M$, -) if $M = M_a$ were with (C.5)

$$M_a(N, K) = \min((K - g)/2, N) \tag{C.47}$$

yielding

$$sc(N, -, K) = sc(N, M = M_a, K) \tag{C.48}$$

Rhombo-octahedral NPs $sc(-, M, K)$ are surrounded by smallest cubic NPs $sc(N, -, -)$ if $N = N_a$ were with (C.5)

$$N_a(M, K) = \min(2M - g, K - 2g) \tag{C.49}$$

yielding

$$sc(-, M, K) = sc(N = N_a, M, K) \tag{C.50}$$

In the most general case of an $sc(N, M, K)$ NP with $\{100\}$, $\{110\}$, and $\{111\}$ facets, we start from a cubo-rhombic NP, $sc(N, M, -)$ with its constraints $M \leq N \leq 2M$. Then we add constraints of a generic octahedral NP, $sc(-, -, K)$, to yield the cubo-rhombo-octahedral NP $sc(N, M, K)$. This requires, according to the discussion above, $K$ values below $K_a$. Here we can distinguish four different ranges of parameter $K$, defined by separating values $K_a \geq K_b \geq K_c$, where with $K_a$ from (C.45)

$$K_b(N, M) = 4M - N \tag{C.51}$$
$$K_c(N, M) = 2M \tag{C.52}$$

The ranges are illustrated in Fig. C.9 for the atom centered cubo-rhombic NP $sc(20, 14, 42)$.

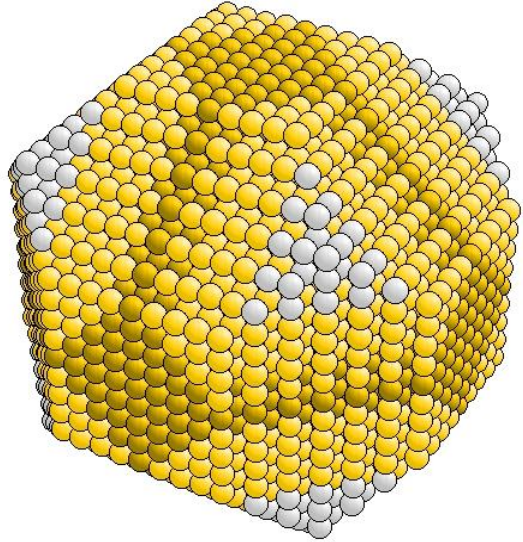

**Figure C.9.** Atom ball model of an atom centered cubo-rhombic NP, $sc(20, 14, 42)$ ($K = K_a$, all atom balls), with its cubo-rhombo-octahedral NP components, $sc(20, 14, 36)$ ($K = K_b$, dark and light yellow balls), and $sc(20, 14, 28)$ ($K = K_c$, dark yellow balls), see text.

**Outer *K* range** of sc(*N*, *M*, *K*) where with (C.45)

$K \geq K_a$ (C.53)

For these *K* values the NP becomes cubo-rhombic and does not exhibit any {111} facets (except for microfacets with three atoms). It is structurally identical with sc(*N*, *M*, $K_a$) as discussed above and in Sec. C.2.1.

**Upper central *K* range** of sc(*N*, *M*, *K*) where with (C.45), (C.51)

$K_b \leq K \leq K_a$ (C.54)

For these *K* values the initial sc(*N*, *M*, $K_a$) NP is capped at its <111> corners creating eight triangular {111} facets. Altogether, these NPs exhibit six {100} facets, twelve {110} facets, and eight {111} facets, see Fig. C.10.

The **{100} facets** are square with four <100> edges of length (2*M* - *N*) $a_o$.

The **{110} facets** are octagonal or rectangular ($K = K_b$) with two <110> edges of length (3*M* - *K*) $\sqrt{2}a_o$, two <100> edges of (2*M* - *N*) $a_o$, and four <111> edges of (*K* + *N* - 4*M*)/2 $\sqrt{3}a_o$.

The **{111} facets** are triangular with three <110> edges of length (3*M* - *K*) $\sqrt{2}a_o$.

The NP structure is illustrated in Fig. C.10 for the ac NP sc(24, 14, 36) ($K_a$ = 42, $K_b$ = 32, yellow atom balls), where white balls above the {111} facets are added to yield the corresponding cubo-rhombic sc(*N*, *M*, $K_a$) NP.

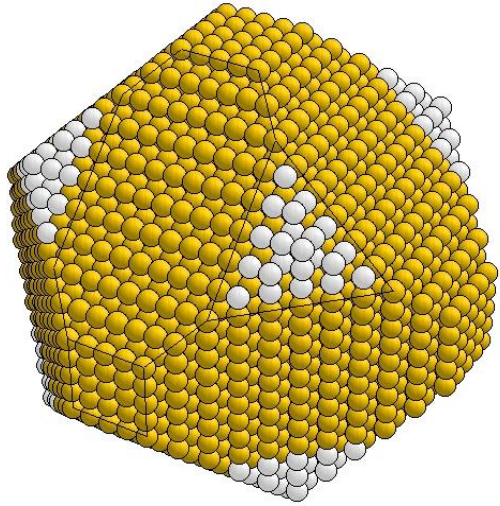

**Figure C.10.** Atom ball model of an atom centered cubo-rhombo-octahedral NP, sc(24, 14, 36), see text. Black lines sketch the square {100}, octagonal {110}, and triangular {111} facets.

The total number of NP atoms, $N_{vol}(N, M, K)$, and the number of facet atoms, $N_{facet}(N, M, K)$, (outer polyhedral shell) are given with (C.22), (C.23), (C.6) by

$$N_{vol}(N, M, K) = N_{vol}(N, M, -) - H(H+2)(2H-1)/3 + h \qquad (C.55)$$

$$N_{facet}(N, M, K) = N_{facet}(N, M, -) - 2H^2 + 2h, \qquad H = 3M - K \qquad (C.56)$$

For $K = K_b$, the sc($N, M, K$) NP assumes a particular shape where its twelve **{110} facets** are rectangular with two edges of length $(N - M) \sqrt{2} a_o$ and of $(2M - N) a_o$ while the **{100}** and **{111} facets** are square and triangular as described before. This is illustrated in Fig. C.11 for the vc NP sc(15, 9, 21) ($K_b = 21$).

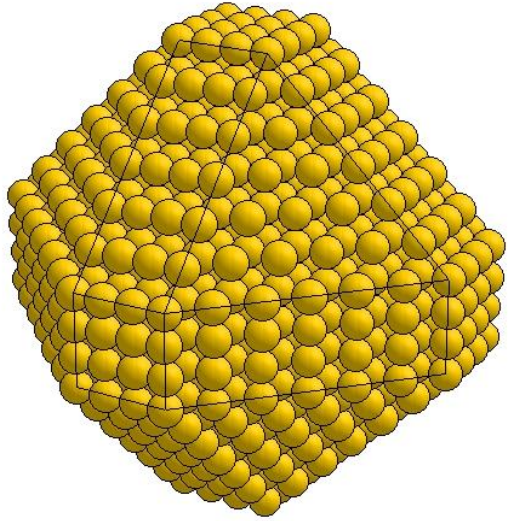

**Figure C.11.** Atom ball model of a void centered cubo-rhombo-octahedral NP, sc(15, 9, 21). Black lines sketch the square {100}, rectangular {110}, and triangular {111} facets.

**Lower central $K$ range** of sc($N, M, K$) where with (C.51), (C.52)

$$K_c \leq K \leq K_b \qquad (C.57)$$

For these $K$ values the capping of the initial sc($N, M, K_b$) along the <111> directions is continued to yield eight hexagonal {111} facets. As before, these NPs exhibit six {100} facets, twelve {110} facets, and eight {111} facets, see Fig. C.12.

The **{100} facets** are octagonal with alternating edges, four <100> of length $(K - 2M) a_o$ and four <110> of length $(4M - N - K)/2 \sqrt{2} a_o$.

The **{110} facets** are rectangular with two <110> edges of length $(N - M) \sqrt{2} a_o$ and two <100> edges of length $(K - 2M) a_o$.

The **{111} facets** are hexagonal with <110> edges of alternating lengths $(4M - N - K)/2 \sqrt{2} a_o$ and $(N - M) \sqrt{2} a_o$.

The NP structure is illustrated in Fig. C.12 for the NP sc(22, 14, 30) ($K_b$ = 34, $K_c$ = 28, yellow atom balls) where white balls above the {111} facets are added to sc($N$, $M$, $K$) to yield the corresponding sc($N$, $M$, $K_b$) NP.

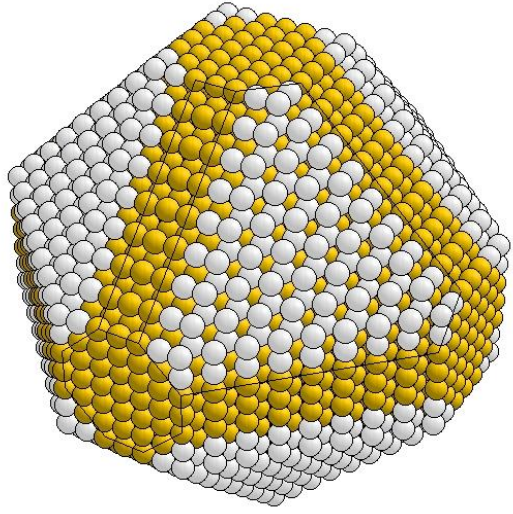

**Figure C.12.** Atom ball model of an atom centered cubo-rhombo-octahedral NP, sc(22, 14, 30), see text. Black lines sketch the octagonal {100}, rectangular {110}, and hexagonal/triangular {111} facets.

The total number of NP atoms, $N_{vol}(N, M, K)$, and the number of facet atoms, $N_{facet}(N, M, K)$, (outer polyhedral shell) are given with (C.55), (C.56), (C.6) by

$$N_{vol}(N, M, K) = N_{vol}(N, M, K_b) - H\,[G\,(G + 2) - 2/3\,(H^2 - 4)]/2 \qquad (C.58)$$

$$N_{facet}(N, M, K) = N_{facet}(N, M, -) - 2(N - M)^2 - 2G\,H + 2h \qquad (C.59)$$

$$H = 4M - N - K, \qquad G = 2M + N - K$$

**Inner $K$ range** of sc($N$, $M$, $K$) where with (C.52)

$$K \leq K_c \qquad (C.60)$$

For these $K$ values the NP becomes cubo-octahedral and does not exhibit {110} facets (except for possible microstrips). It is structurally identical with sc($N$, $M_a$, $K$) as discussed above and in Sec. C.2.2.

A classification of sc($N$, $M$, $K$) NP types for any $N$, $M$, $K$ combination is given by Table T.15 of the Supplement.

## D. Cubic Macroparticles

Compact particles with cubic lattices and of $O_h$ symmetry, discussed in Secs. A - C, are uniquely described by polyhedral parameters *N*, *M*, *K* referring to distances between parallel facets. This description becomes particularly simple for large nanoparticles (macroparticles, MP) when *N*, *M*, *K* assume very large values. As a consequence, {*hkl*} monolayer planes, describing the facets, can still be defined by their normal directions in Cartesian coordinates but their distribution inside the MP becomes quasi-continuous rather than discrete. As a result, {*hkl*} facets confining the MPs can vary in their distance by quite small amounts compared with the particle diameter, and corresponding structure parameters *N*, *M*, *K* may be considered quasi-continuous quantities and described approximately by real numbers *A*, *B*, *C*. Thus, assuming overall $O_h$ symmetry of the MPs and confinement by facets of the three cubic netplane families, {100}, {110}, and {111}, diameters of these macroparticles can be written as

$$D_{\{100\}} = A, \quad D_{\{110\}} = B/\sqrt{2}, \quad D_{\{111\}} = C/\sqrt{3} \tag{D.1}$$

with *A*, *B*, *C* real valued. As a consequence, the MPs can be denoted **cub(*A*, *B*, *C*)** in the most general case. If a facet type does not appear in the MP the corresponding parameter value *A*, *B*, or *C* may be ignored and is replaced by a minus sign. As an example, a cubic MP with only {100} and {110} facets is denoted cub(*A*, *B*, -). These notations will be used in the following discussion.

### D.1. Generic cub(*A*, -, -), (-, *B*, -), (-, -, *C*) Macroparticles

Generic cubic macroparticles (MPs) of $O_h$ symmetry are confined by facets with orientations of only one {*hkl*} netplane family. Here we consider {100}, {110}, and {111} facets derived from the densest monolayers of the cubic lattice allowing three different generic MP types.

**(a)** **Generic cubic** MPs, denoted **cub(*A*, -, -)** are confined by all six {100} monolayers with distances $D_{\{100\}} = A$ between parallel confining monolayers. This yields six {100} facets, see Fig. D.1.

The **{100} facets** are square with four <100> edges of length *A*.

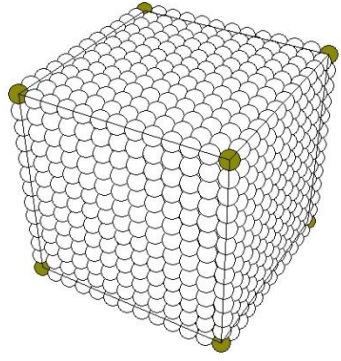

**Figure D.1.** Generic cubic MP filled with atom balls of an sc lattice. The corners are emphasized by dark color and black lines outline the MP.

The largest distance from the MP center to its surface along <hkl> directions, $s_{<hkl>}$, is given by

$$s_{<100>}(A, -, -) = A/2 \tag{D.2a}$$
$$s_{<110>}(A, -, -) = \sqrt{2}\,A/2 \tag{D.2b}$$
$$s_{<111>}(A, -, -) = \sqrt{3}\,A/2 \tag{D.2c}$$

These quantities will be used in Secs. D.2.

**(b)** **Generic rhombohedral** MPs, denoted **cub(-, B, -)**, are confined by all twelve {110} monolayers with distances $D_{\{110\}} = B/\sqrt{2}$ between parallel confining monolayers. This yields twelve {110} facets, see Fig. D.2.

The **{110} facets** are rhombic with <111> edges of length $\sqrt{3}\,B/4$. Thus, the MPs can be described as rhombic dodecahedra reminding of the shape of Wigner-Seitz cells of the fcc crystal lattice [20].

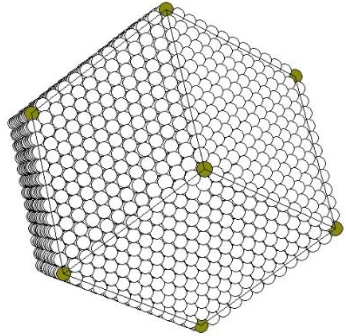

**Figure D.2.** Generic rhombohedral MP filled with atom balls of a bcc lattice. The corners are emphasized by dark color and black lines outline the MP.

The largest distance from the MP center to its surface along <hkl> directions, $s_{<hkl>}$, is given by

$$s_{<100>}(-, B, -) = B/2 \qquad (D.3a)$$
$$s_{<110>}(-, B, -) = \sqrt{2}\, B/4 \qquad (D.3b)$$
$$s_{<111>}(-, B, -) = \sqrt{3}\, B/4 \qquad (D.3c)$$

These quantities will be used in Secs. D.2.

**(c)** **Generic octahedral** MPs, denoted **cub(-, -, K)**, are confined by all eight {111} monolayers with distances $D_{\{111\}} = C/\sqrt{3}$ between parallel confining monolayers. This yields eight {111} facets, see Fig. D.3.

The **{111} facets** are triangular with three <110> edges of length $\sqrt{2}\, C/2$.

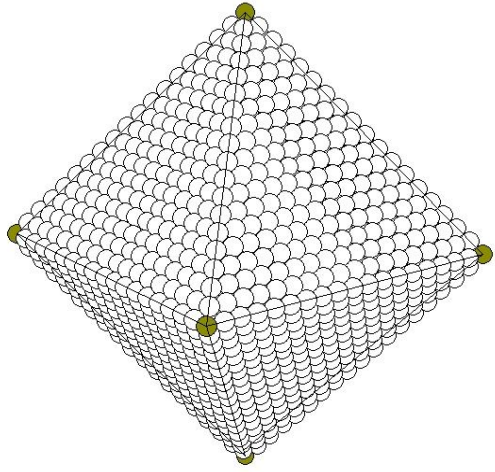

**Figure D.3.** Generic octahedral MP filled with atom balls of an fcc lattice. The corners are emphasized by dark color and black lines outline the MP.

The largest distance from the MP center to its surface along <hkl> directions, $s_{<hkl>}$, is given by

$$s_{<100>}(-, -, C) = C/2 \qquad (D.4a)$$
$$s_{<110>}(-, -, C) = \sqrt{2}\, C/4 \qquad (D.4b)$$
$$s_{<111>}(-, -, C) = \sqrt{3}\, C/6 \qquad (D.4c)$$

These quantities will be used in Secs. D.2.

Table T.16 of the Supplement collects types and shapes of all generic cub MPs.

## D.2. Non-generic Cubic Macroparticles

Non-generic cubic macroparticles of $O_h$ symmetry show facets with orientations of more than one {$hkl$} netplane family. This can be considered as combining confinements of the corresponding generic MPs, discussed in Sec. D.1, sharing their symmetry center. Thus, non-generic MPs are mutual intersections of more than one generic MP, where one MP cuts corners and edges from the other(s) to form additional facets. Here we discuss non-generic MPs **cub(*A*, *B*, *C*)**, which combine constraints of up to three generic MPs, cubic cub(*A*, -, -), rhombohedral cub(-, *B*, -), and octahedral cub(-, -, *C*). Thus, they allow {100}, {110}, as well as {111} facets. Clearly, the corresponding polyhedral parameters *A*, *B*, *C* depend on each other and determine the overall MP shape. In particular, if a participating generic MP encloses another participant it will not contribute to the overall MP shape. Thus, the respective {$hkl$} facets will not appear at the surface of the non-generic MP. In the following, we consider the three types of non-generic MPs, which combine constraints due to two generic MPs (Secs. D.2.1-3), before we discuss the most general case of cub(*A*, *B*, *C*) MPs in Sec. D.2.4.

### D.2.1. Truncated cub(*A*, *B*, -) Macroparticles

Non-generic **cubo-rhombic** MPs, denoted **cub(*A*, *B*, -)**, are confined by facets of the two generic MPs, cub(*A*, -, -) (cubic) and cub(-, *B*, -) (rhombohedral), see Fig. D.4. If the edges of the cubic MP cub(*A*, -, -) lie inside the rhombohedral MP cub(-, *B*, -), the resulting combination cub(*A*, *B*, -) will be generic cubic. This requires

$$s_{<110>}(A, -, -) \leq s_{<110>}(-, B, -) \tag{D.5}$$

and with (D.2), (D.3), leads to

$$B \geq 2A \tag{D.6}$$

On the other hand, if the corners of the rhombohedral MP cub(-, *B*, -) lie inside the cubic MP cub(*A*, -, -), the resulting combination cub(*A*, *B*, -) will be generic rhombohedral. This requires

$$s_{<100>}(-, B, -) \leq s_{<100>}(A, -, -) \tag{D.7}$$

and with (D.2), (D.3), leads to

$$B \leq A \tag{D.8}$$

Thus, the two generic MPs intersect and yield an MP cub(*A*, *B*, -) with both {100} and {110} facets only for polyhedral parameters *A*, *B* with

$$A < B < 2A \tag{D.9}$$

while cub($A$, $B$, -) is generic cubic for $B \geq 2A$ and generic rhombohedral for $B \leq A$. Further, generic cubic and rhombohedral MPs can be described by cub($N$, $M$, -) where

$$\text{cub}(A, -, -) = \text{cub}(A, B = 2A, -) \qquad \text{(cubic)} \qquad \text{(D.10a)}$$

$$\text{cub}(-, B, -) = \text{cub}(A = B, B, -) \qquad \text{(rhombohedral)} \qquad \text{(D.10b)}$$

The surfaces of cubo-rhombic MPs cub($A$, $B$, -) exhibit six {100} facets and twelve {110} facets, see Fig. D.4.

The **{100} facets** are square with four <100> edges of length ($B$ - $A$).

The **{110} facets** are hexagonal with four <111> edges of length $\sqrt{3}/4$ ($2A$ - $B$) and two <100> edges of length ($B$ - $A$).

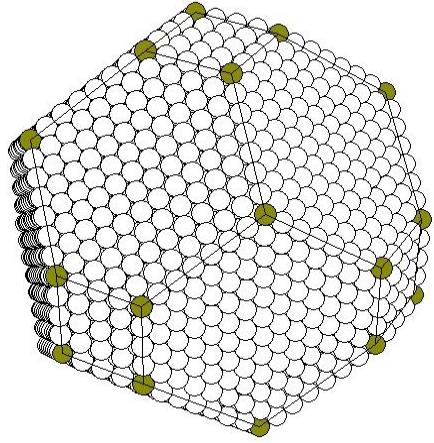

**Figure D.4.** Cubo-rhombic MP filled with atom balls of a bcc lattice. The corners are emphasized by dark color and black lines outline the MP.

A classification of cub($A$, $B$, -) MP types for any $A$, $B$ combination is given by Table T.17 of the Supplement.

### D.2.2. Truncated cub($A$, -, $C$) Macroparticles

Non-generic **cubo-octahedral** MPs, denoted **cub($A$, -, $C$)**, are confined by facets of the two generic MPs, cub($A$, -, -) (cubic) and cub(-, -, $C$) (octahedral). If the corners of the cubic MP cub($A$, -, -) lie inside the octahedral MP cub(-, -, $C$), the resulting combination cub($A$, -, $C$) will be generic cubic. This requires

$$s_{<111>}(A, -, -) \leq s_{<111>}(-, -, C) \qquad \text{(D.11)}$$

and with (D.2), (D.4), leads to

$$C \geq 3A \qquad \text{(D.12)}$$

On the other hand, if the corners of the octahedral MP cub(-, -, $C$) lie inside the cubic MP cub($A$, -, -), the resulting combination cub($A$, -, $C$) will be generic octahedral. This requires

$$s_{<100>}(-, -, C) \leq s_{<100>}(A, -, -) \tag{D.13}$$

and with (D.2), (D.4), leads to

$$C \leq A \tag{D.14}$$

Thus, the two generic MPs intersect and yield an MP cub($A$, -, $C$) with both {100} and {111} facets only for polyhedral parameters $A$, $C$ with

$$A < C < 3A \tag{D.15}$$

while cub($A$, -, $C$) is generic cubic for $C \geq 3A$ and generic octahedral for $C \leq A$. Further, generic cubic and octahedral cub MPs can be described by cub($A$, -, $C$) where

$$\text{cub}(A, -, -) = \text{cub}(A, -, C = 3A) \qquad \text{(cubic)} \tag{D.16a}$$

$$\text{cub}(-, -, C) = \text{cub}(A = C, -, C) \qquad \text{(octahedral)} \tag{D.16b}$$

The surfaces of cubo-octahedral MPs cub($A$, -, $C$) include six {100} and eight {111} facets, see Figs. D.5, D.6. Amongst the intersecting species according to (D.15) we can distinguish between **truncated octahedral** MPs where $C < 2A$ and **truncated cubic** MPs for $C > 2A$, with **cuboctahedral** MPs for $C = 2A$ separating. This will be discussed in the following.

**Truncated octahedral** MPs ($C < 2A$), Fig. D.5a, can be characterized by their facets as follows.

The **{100} facets** are square with four <110> edges of length $\sqrt{2}\,(C - A)/2$.

The **{111} facets** are hexagonal with <110> edges of alternating lengths $\sqrt{2}\,(C - A)/2$ and $\sqrt{2}\,(2A - C)/2$.

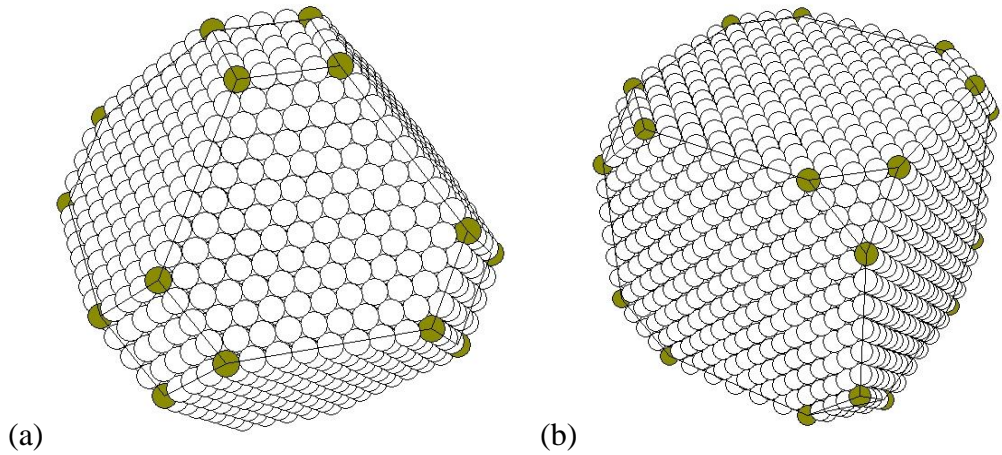

(a)          (b)

**Figure D.5.** Cubo-octahedral MPs filled with atom balls of an fcc lattice, (a) truncated octahedral, (b) truncated cubic type. The corners are emphasized by dark color and black lines outline the MPs.

**Truncated cubic** MPs ($C > 2A$), Fig. D.5b, can be characterized by their facets as follows.

The **{100} facets** are octagonal with alternating edges, four <100> of length ($C - 2A$) and four <110> of length $\sqrt{2}\,(3A - C)/2$.

The **{111} facets** are triangular with <110> edges of length $\sqrt{2}\,(3A - C)/2$.

There are MPs which can be assigned to both truncated cubic and truncated octahedral type, the **cuboctahedral** MPs cub($A$, -, $C$), defined by $C = 2A$ These MPs exhibit six {100} and eight {111} facets, see Fig. D.6. All **{100} facets** are square with four <110> edges of length $A/\sqrt{2}$ while all **{111} facets** are triangular with three <110> edges of length $A/\sqrt{2}$ shared with those of the {100} facets.

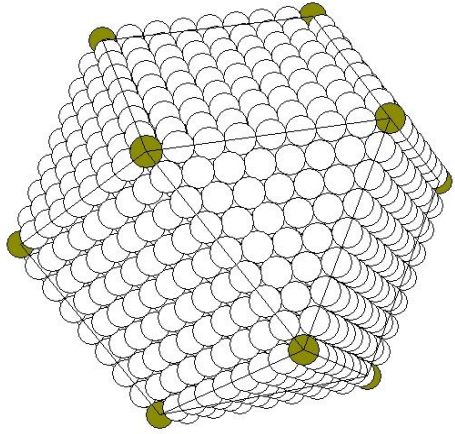

**Figure D.6.** Cuboctahedral MP filled with atom balls of an fcc lattice. The corners are emphasized by dark color and black lines outline the MPs.

A classification of cub($A$, -, $C$) MP types for any $A$, $C$ combination is given by Table T.18 of the Supplement.

### D.2.3. Truncated cub(-, $B$, $C$) Macroparticles

Non-generic **rhombo-octahedral** MPs, denoted **cub(-, $B$, $C$)**, are confined by facets of the two generic MPs, cub(-, $B$, -) (rhombohedral) and cub(-, -, $C$) (octahedral), see Fig. D.7. If the corners of the rhombohedral MP cub(-, $B$, -) lie inside the octahedral MP cub(-, -, $C$), the resulting combination cub(-, $B$, $C$) will be generic rhombohedral. This requires

$$s_{<111>}(-, B, -) \leq s_{<111>}(-, -, C) \tag{D.17}$$

and with (D.3), (D.4), leads to

$$C \geq 3/2\, B \tag{D.18}$$

On the other hand, if the corners of the octahedral MP cub(-, -, *C*) lie inside the rhombohedral MP cub(-, *B*, -), the resulting combination cub(-, *B*, *C*) will be generic octahedral. This requires

$$s_{<100>}(-, -, C) \leq s_{<100>}(-, B, -) \tag{D.19}$$

and with (D.3), (D.4), leads to

$$C \leq B \tag{D.20}$$

Thus, the two generic MPs intersect and yield an MP cub(-, *B*, *C*) with both {110} and {111} facets only for polyhedral parameters *B*, *C* with

$$B < C < 3/2\ B \tag{D.21}$$

while cub(-, *B*, *C*) is generic rhombohedral for $C \geq 3/2$ and generic octahedral for $C \leq B$. Further, generic rhombohedral and octahedral cub MPs can be described by cub(-, *B*, *C*) where

$$\text{cub}(-, B, -) = \text{cub}(-, B, C = 3/2\ B) \qquad \text{(rhombohedral)} \tag{D.22a}$$
$$\text{cub}(-, -, C) = \text{cub}(-, B = C, C) \qquad \text{(octahedral)} \tag{D.22b}$$

The surfaces of rhombo-octahedral MPs cub(-, *B*, *C*) exhibit twelve {110} and eight {111} facets, see Fig. D.7.

The **{110} facets** are hexagonal with four <111> edges of length $(C - B)/2\ \sqrt{3}$ and two <110> edges of $(3/2\ B - C)\ \sqrt{2}$.

The **{111} facets** are triangular with three <110> edges of length $(3/2\ B - C)\ \sqrt{2}$.

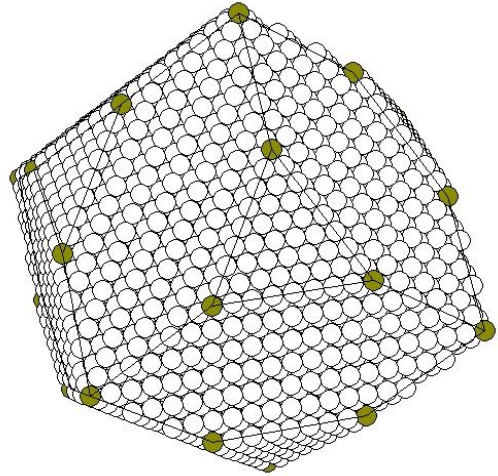

**Figure D.7.** Rhombo-octahedral MP filled with atom balls of an fcc lattice. The corners are emphasized by dark color and black lines outline the MP.

A classification of cub(-, $B$, $C$) MP types for any $B$, $C$ combination is given by Table T.19 of the Supplement.

### D.2.4. General cub($A$, $B$, $C$) Macroparticles

Non-generic **cubo-rhombo-octahedral** MPs, denoted **cub($A$, $B$, $C$)**, are confined by facets of all three generic MPs, cub($A$, -, -) (cubic), cub(-, $B$, -) (rhombohedral), and cub(-, -, $C$) (octahedral), see Fig. D.8. Thus, they can show {100}, {110}, and {111} facets. A general discussion of these MPs requires results for generic and non-generic MPs, see Secs. D.1, D.2.1-3, as will be detailed in the following.

First, we consider the general notation for generic MPs discussed in Sec. D.1. Cubic MPs cub($A$, -, -) are surrounded by smallest rhombohedral MPs cub(-, $B$, -) if $B = 2A$ and by smallest octahedral MPs cub(-, -, $K$) if $C = 3A$. This allows a notation

$$\text{cub}(A, -, -) = \text{cub}(A, B = 2A, C = 3A) \tag{D.23}$$

Rhombohedral MPs cub(-, $B$, -) are surrounded by smallest cubic MPs cub($A$, -, -) if $A = B$ and by smallest octahedral MPs cub(-, -, $C$) if $C = 3/2\ B$. This yields

$$\text{cub}(-, B, -) = \text{cub}(A = B, B, C = 3/2\ B) \tag{D.24}$$

Octahedral MPs cub(-, -, $C$) are surrounded by smallest cubic MPs cub($A$, -, -) if $A = C$ and by smallest rhombohedral MPs cub(-, $B$, -) if $B = C$. This yields

$$\text{cub}(-, -, C) = \text{cub}(A = C, B = C, C) \tag{D.25}$$

General notations for non-generic cub MPs with two facet types, discussed in Secs. D.2.1-3, are obtained by analogous arguments. Cubo-rhombic MPs cub($A$, $B$, -) are surrounded by smallest octahedral MPs cub(-, -, $C$) if $C = C_a$ with

$$C_a(A, B) = \min(3A, 3/2\ B) = 3/2\ B \tag{D.26}$$

which allows a notation

$$\text{cub}(A, B, -) = \text{cub}(A, B, C = C_a) \tag{D.27}$$

Cubo-octahedral MPs cub($A$, -, $C$) surrounded by smallest rhombohedral MPs cub(-, $B$, -) if $B = B_a$ with

$$\begin{aligned} B_a(A, C) &= \min(2A, C) & &\text{(D.28a)} \\ &= 2A & \text{(truncated cubic)} & &\text{(D.28b)} \\ &= C & \text{(truncated octahedral)} & &\text{(D.28c)} \end{aligned}$$

yielding

$$\mathrm{cub}(A, -, C) = \mathrm{cub}(A, B = B_a, C) \tag{D.29}$$

Rhombo-octahedral MPs cub(-, $B$, $C$) are surrounded by smallest cubic MPs cub($A$, -, -) if $A = A_a$ with

$$A_a(B, C) \quad = \min(B, C) = B \tag{D.30}$$

yielding

$$\mathrm{cub}(-, B, C) = \mathrm{cub}(A = A_a, B, C) \tag{D.31}$$

In the most general case of a cub($A$, $B$, $C$) MP, we start from a cubo-rhombic MP, cub($A$, $B$, -), with its constraints $A \leq B \leq 2A$. Then we add constraints of a generic octahedral MP, cub(-, -, $C$), to yield the cubo-rhombo-octahedral MP fcc($A$, $B$, $C$). This requires, according to the discussion above, $C$ values below $C_a$. Here we distinguish four different ranges of parameter $C$, defined by separating values $C_a \geq C_b \geq C_c$, where with $C_a$ from (D.26)

$$\mathrm{Cub}(A, B) = 2B - A \tag{D.32}$$
$$C_c(A, B) \quad = B \tag{D.33}$$

The ranges are illustrated in Fig. D.8 for the cubo-rhombic MP cub($A$, $B$, $C_a$).

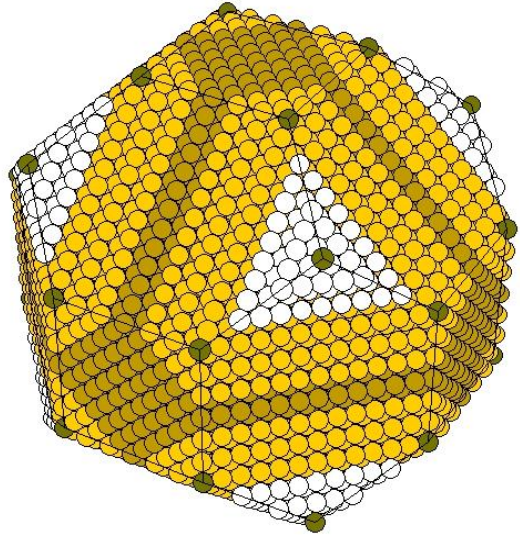

**Figure D.8.** Cubo-rhombic MP cub(A, B, $C_a$) filled with atom balls of an fcc lattice (all atom balls) with its cubo-rhombo-octahedral MP components cub(A, B, $C_b$) (dark and light yellow balls), and cub(A, B, $C_c$) (dark yellow balls). The corners are emphasized by dark color and black lines outline the boundaries of the MP.

**Outer *C* range** of cub(*A*, *B*, *C*) where with (D.26)

$$C \geq C_a \tag{D.34}$$

For these *C* values the MP becomes cubo-rhombic and does not exhibit any {111} facets. It is structurally identical with cub(*A*, *B*, -) = cub(*A*, *B*, $C_a$) as discussed above and in Sec. D.2.1.

**Upper central *C* range** of cub(*A*, *B*, *C*) where with (D.26), (D.32)

$$C_b \leq C \leq C_a \tag{D.35}$$

For these *C* values the initial cub(*A*, *B*, $C_a$) MP is capped at its <111> corners forming eight larger triangular {111} facets. Altogether, these MPs exhibit six {100} facets, twelve {110} facets, and eight {111} facets, see Fig. D.9.

The **{100} facets** are square with four <100> edges of length (*B* - *A*).

The **{110} facets** are octagonal or rectangular (*C* = $C_b$) with two <110> edges of length (3*B* - 2*C*)/√2, two <100> edges of (*B* - *A*), and four <111> edges of (*C* - $C_b$)/2 √3.

The **{111} facets** are triangular with three <110> edges of length (3*B* - 2*C*)/√2.

The MP structures are illustrated in Fig. D.9 for the MP filled by yellow atom balls where white balls above the {111} facets are added to yield the corresponding cubo-rhombic cub(*A*, *B*, $C_a$) MP.

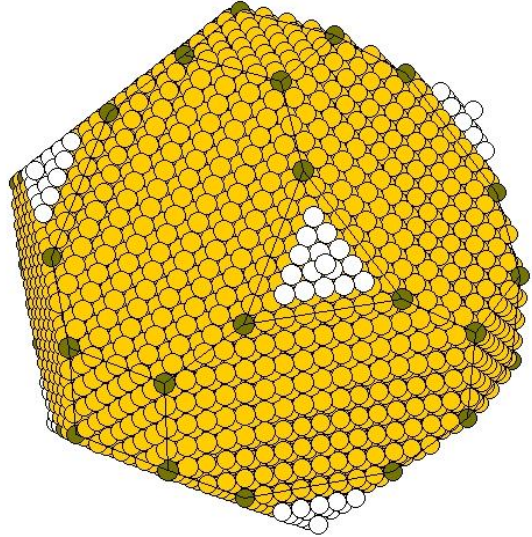

**Figure D.9.** Cubo-rhombo-octahedral MP cub(*A*, *B*, *C*) for $C_b$ < *C* < $C_a$ filled with atom balls of an fcc lattice, see text. The corners are emphasized by dark color and black lines outline the boundaries of the MP.

For $C = C_b$, the cub($A$, $B$, $C$) MP assumes a particular shape, see Fig. D.10. Its six **{100} facets** are square with four <100> edges of length ($B$ - $A$). Its twelve **{110} facets** are rectangular with two <110> edges of length ($2A$ - $B$)/$\sqrt{2}$ two <100> edges of ($B$ - $A$). Finally, its eight **{111} facets** are triangular with three <110> edges of length ($2A$ - $B$)/$\sqrt{2}$. This is illustrated in Fig. D.10.

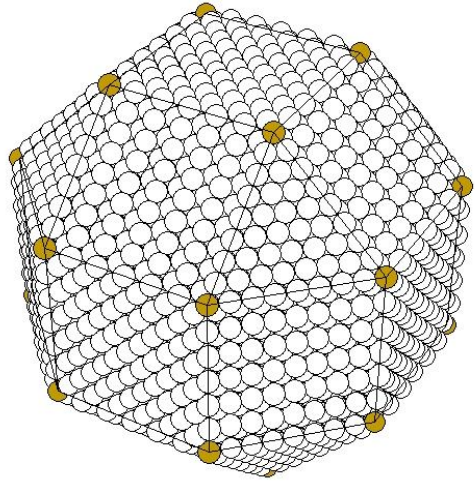

**Figure D.10.** Cubo-rhombo-octahedral MP cub($A$, $B$, $C_b$) filled with atom balls of an fcc lattice. The corners are emphasized by dark color and black lines outline the boundaries of the MP.

**Lower central $C$ range** of cub($A$, $B$, $C$) where with (D.32), (D.33)

$$C_c \leq C \leq C_b \tag{D.36}$$

For these $C$ values the capping of the initial cub($A$, $B$, $C_b$) along the <111> directions is continued to yield eight hexagonal {111} facets. As before, these MPs exhibit six {100} facets, twelve {110} facets, and eight {111} facets, see Fig. D.11.

The **{100} facets** are octagonal with alternating edges, four <100> of length ($C$ - $B$) and four <110> of length ($C_b$ - $C$)/$\sqrt{2}$.

The **{110} facets** are rectangular with two <110> edges of length ($2A$ - $B$)/$\sqrt{2}$ and two <100> edges of length ($C$ - $B$).

The **{111} facets** are hexagonal with <110> edges of alternating lengths ($C_b$ - $C$)/$\sqrt{2}$ and ($2A$ - $B$)/$\sqrt{2}$.

This is illustrated in Fig. D.11 for the MP filled by yellow atom balls where white balls above the {111} facets are added to yield the corresponding cub($A$, $B$, $C_b$) MP.

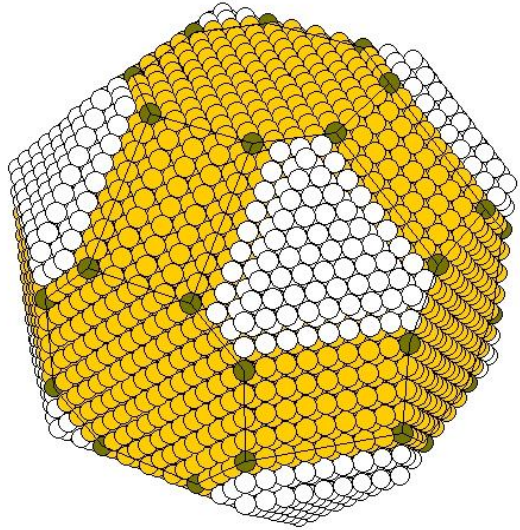

**Figure D.11.** Cubo-rhombo-octahedral MP cub($A$, $B$, $C$) for $C_c < C < C_b$ filled with atom balls of an fcc lattice, see text. The corners are emphasized by dark color and black lines outline the boundaries of the MP.

**Inner $C$ range** of cub($A$, $B$, $C$) where with (D.33)

$$C \leq C_c \tag{D.37}$$

For these $C$ values the MP becomes cubo-octahedral and does not exhibit any {110} facets. It is structurally identical with cub($A$, -, $C$) = cub($A$, $B_a$, $C$) as discussed above and in Sec. D.2.2.

A classification of cub($A$, $B$, $C$) MP types for any $A$, $B$, $C$ combination is given by Table T.20 of the Supplement.

## III. Conclusions

The present work gives a full theoretical account of the shape and structure of nanoparticles (NPs) forming compact polyhedral sections of ideal primitive cubic lattices, where face centered, body centered, and simple variants are considered. We focus on particles of $O_h$ symmetry which are confined by facets of highly dense monolayers, reflecting Miller index families {100}, {110}, and {111} of the corresponding lattice. The structure evaluation identifies different types of generic NPs, which serve for the definition of general polyhedral NPs as intersections of corresponding generic NPs. This allows a classification of polyhedral NPs according to three integer valued polyhedral parameters *N*, *M*, *K*. These are connected with particle diameters along corresponding facet normal directions, reflecting {*hkl*} monolayer families of the underlying lattice. Detailed structural properties of the general polyhedral NPs, such as shape, size, and facet surfaces, are discussed in analytical and numerical detail with visualization of characteristic examples. This illustrates the complexity of seemingly simple nanoparticles in a quantitative structure account. As a result of the multitude of shapes, a classification of NPs by counting facet edge atoms cannot be easily achieved [11], in contrast to the unique (*N*, *M*, *K*) notation.

While the overall NP shapes are quite similar between the different cubic lattice types, structural fine details differ, where we mention only three examples. First, NPs with internal fcc, bcc, and sc lattice structure can all be highly symmetric showing $O_h$ symmetry. However, fcc and sc NPs offer two high-symmetry centers, atom and void sites, whereas bcc NPs with $O_h$ symmetry allow only atom sites as centers. Second, the densest {*hkl*} monolayers differ between the three cubic lattice types, {111} for fcc, {110} for bcc, and {100} for sc. Thus, equally shaped NPs of the three lattice types must exhibit different atom density at their surfaces. For example, generic NPs of highest surface atom density are octahedral for fcc, rhombohedral for bcc, and cubic for sc. Third, {100}, {110}, and {111} monolayer planes of adjacent NP facets can join at corners and edges which are not occupied by atoms of the ideal lattice. This gives rise to microfacets and narrow facet strips depending on the lattice type. For example, {100} microfacets of four atoms and {111} microfacets of three atoms occur and are densest at fcc NP surfaces. In contrast, at bcc NP surfaces they are more open and can be topped by a lattice atom to form a corner. As a result, rhombohedral bcc NPs of any size do not exhibit {100} nor {111} microfacets, see Fig. B.2, whereas their fcc counterparts allow both, see Figs. A.2, A.3.

Clearly, the present results deal only with ideal cubic NPs and cannot account for all possible structures of the most general metal nanoparticles, which are observed, for example, by electron microscopy [13]. Realistic NPs may exhibit very different shapes, including less compact particles, local structural disorder, symmetry breaking, combined with deviations from (or incompatibility with) the crystal lattice structure in their inner NP core. This can be examined only in case-by-case studies where exact quantitative data may be difficult to obtain. However, the present results can be used for models to describe facet geometries at compact real metal NP surfaces, and help to estimate typical particle sizes and shapes. The results may also serve for a repository of possible structure details expected for general real metal NPs.

# V. Supplementary Information

## S.1. Classification Tables of Cubic (*N*, *M*, *K*) Nanoparticles

Here we present Tables giving a full classification of cubic (fcc, bcc, and sc) NPs for all combinations of polyhedral parameters *N*, *M*, *K*. This includes NPs where one or two parameters define the structure already uniquely and missing parameter values are replaced by "-". Further, isomorphs of a given NP are defined as NPs which are structurally identical but differ in their notation. As an example, all NPs fcc(*N*, *M*, *K*) with $M > 2N$ and $K > 3N$ are isomorphs of the generic cubic NP fcc(*N*, 2*N*, 3*N*).

### S.1.1. Face Centered Cubic NPs

| Generic type | Constraints | Facets | Corners |
|---|---|---|---|
| Cubic fcc(*N*, -, -) | ac, *N* even, vc, *N* odd | {100} 6<br>{110} 0<br>{111} 0 | <100> 0<br><110> 0<br><111> 8 |
| | ac, *N* odd, vc, *N* even | {100} 6<br>{110} 0<br>{111} 8 | <100> 0<br><110> 0<br><111> 8 [&] |
| Rhombohedral fcc(-, *M*, -) | ac, *M* = 4*p* | {100} 0<br>{110} 12<br>{111} 0 | <100> 6<br><110> 0<br><111> 8 |
| | ac, *M* = 4*p* + 1<br>    *M* = 4*p* + 3<br>vc, *M* = 4*p* | {100} 6<br>{110} 12<br>{111} 8 | <100> 6 [&]<br><110> 0<br><111> 8 [&] |
| | ac, *M* = 4*p* + 2<br>vc, *M* = 4*p* + 1<br>    *M* = 4*p* + 3 | {100} 0<br>{110} 12<br>{111} 8 | <100> 6<br><110> 0<br><111> 8 [&] |
| | vc, *M* = 4*p* + 2 | {100} 6<br>{110} 12<br>{111} 0 | <100> 6 [&]<br><110> 0<br><111> 8 |
| Octahedral fcc(-, -, *K*) | ac, *K* even | {100} 0<br>{110} 0<br>{111} 8 | <100> 6<br><110> 0<br><111> 0 |
| | vc, *K* odd | {100} 6 [+]<br>{110} 12 [+]<br>{111} 8 | <100> 6 [&]<br><110> 0<br><111> 0 |

**Table T.1.** Types and notations of all generic fcc NPs. Here "ac" denotes atom centered and "vc" void centered NPs. The superscript label "&" denotes corner quadruplets about <100> and corner triplets about <111>.

| Constraints | NP types | Isomorphs |
|---|---|---|
| $M \geq 2N$ | Generic cubic | $(N, -, -) =$ $(N, M = 2N, -)$ |
| $N \leq M \leq 2N$ | Cubo-rhombic | $(N, M, -)$ |
| $M \leq N$ | Generic rhombohedral | $(-, M, -) =$ $(N = M - h', M, -)$ |

**Table T.2.** Constraints and types including isomorphs of fcc($N$, $M$, -) NPs. Parameter $h'$ is defined by (A.6).

| Constraints | NP types | Isomorphs |
|---|---|---|
| $K \geq 3N - h$ | Generic cubic | $(N, -, -) =$ $(N, -, K = 3N - h)$ |
| $2N \leq K \leq 3N - h$ | Cubo-octahedral, truncated cubic | $(N, -, K)$ |
| $K = 2N$ | Cuboctahedral | $(N, -, K = 2N)$, $(N = K/2, -, K)$ |
| $N \leq K \leq 2N$ | Cubo-octahedral, truncated octahedral | $(N, -, K)$ |
| $K \leq N$ | Generic octahedral | $(-, -, K) =$ $(N = K, -, K)$ |

**Table T.3.** Constraints and types including isomorphs of fcc($N$, -, $K$) NPs. Parameter $h$ is defined by (A.5).

| Constraints | NP types | Isomorphs |
|---|---|---|
| $K \geq K_a$ | Generic rhombohedral | $(-, M, -) =$ $(-, M, K = K_a)$ |
| $M \leq K \leq K_a$ | Rhombo-octahedral | $(-, M, K)$ |
| $K \leq M$ | Generic octahedral | $(-, -, K) =$ $(-, M = K, K)$ |

**Table T.4.** Constraints and types including isomorphs of fcc(-, $M$, $K$) NPs. Parameter $K_a$ is defined by (A.43).

| Constraints 1 | Constraints 2 | NP types | Isomorphs |
|---|---|---|---|
| $M \geq 2N$ | $K \geq 3N$ | Generic cubic | $(N, -, -) =$ $(N, 2N, 3N)$ |
| | $2N \leq K \leq 3N$ | Cubo-octahedral, truncated cubic | $(N, -, K) =$ $(N, 2N, K)$ |
| | $K = 2N$    ($K$ even) | Cuboctahedral | $(N, -, K) =$ $(N, 2N, 2N)$ |
| | $N \leq K \leq 2N$ | Cubo-octahedral, truncated octahedral | $(N, -, K) =$ $(N, K, K)$ |
| | $K \leq N$ | Generic octahedral | $(-, -, K) =$ $(K, K, K)$ |
| $N + h' \leq M \leq 2N$ | $K \geq K_a$ | Cubo-rhombic | $(N, M, -) =$ $(N, M, K_a)$ |
| | $K_b \leq K \leq K_a$ | Cubo-rhombo-oct., upper central | $(N, M, K)$ |
| | $K_c \leq K \leq K_b$ | Cubo-rhombo-oct., lower central | $(N, M, K)$ |
| | $N \leq K \leq K_c$ | Cubo-octahedral, truncated octahedral | $(N, -, K) =$ $(N, K, K)$ |
| | $K \leq N$ | Generic octahedral | $(-, -, K) =$ $(K, K, K)$ |
| $M \leq N + h'$ | $K \geq K_a$ | Generic rhombohedral | $(-, M, -) =$ $(N_a, M, K_a)$ |
| | $M - h' \leq K \leq K_a$ | Octo-rhombohedral | $(-, M, K) =$ $(N_a, M, K)$ |
| | $K \leq M - h'$ | Generic octahedral | $(-, -, K) =$ $(N_a, M_a, K)$ |

**Table T.5.** Constraints and types including isomorphs of fcc($N$, $M$, $K$) NPs. Polyhedral parameters $N_a$, $M_a$, $K_a$, $K_b$, $K_c$ are defined by (A.50), (A.48), (A.43), (A.52), (A.53) and $h'$ by (A.6).

## S.1.2. Body Centered Cubic NPs

| Generic type | Constraints | Facets | Corners |
|---|---|---|---|
| Cubic<br>bcc($N$, -, -) | | {100} 6<br>{110} 0<br>{111} 0 | <100> 0<br><110> 0<br><111> 8 |
| Rhombohedral<br>bcc(-, $M$, -) | | {100} 0<br>{110} 12<br>{111} 0 | <100> 6<br><110> 0<br><111> 8 |
| Octahedral<br>bcc(-, -, $K$) | $K$ even | {100} 0<br>{110} 0<br>{111} 8 | <100> 6<br><110> 0<br><111> 0 |
| | $K$ odd | {100} 0<br>{110} 12<br>{111} 8 | <100> 6<br><110> 0<br><111> 0 |

**Table T.6.** Types and notations of all generic bcc NPs.

| Constraints | NP types | Isomorphs |
|---|---|---|
| $N \geq 2M$ | Generic rhombohedral | (-, $M$, -) =<br>($N = 2M$, $M$, -) |
| $M \leq N \leq 2M$ | Cubo-rhombic | ($N$, $M$, -) |
| $N \leq M$ | Generic cubic | ($N$, -, -) =<br>($N$, $M = N$, -) |

**Table T.7.** Constraints and types including isomorphs of bcc($N$, $M$, -) NPs.

| Constraints | NP types | Isomorphs |
|---|---|---|
| $K \geq 3N$ | Generic cubic | $(N, -, -) =$ $(N, -, K = 3N)$ |
| $2N \leq K \leq 3N$ | Cubo-octahedral, truncated cubic | $(N, -, K)$ |
| $K = 2N$ | Cuboctahedral | $(N, -, K = 2N)$, $(N = K/2, -, 2K)$ |
| $N \leq K \leq 2N$ | Cubo-octahedral, truncated octahedral | $(N, -, K)$ |
| $K \leq N$ | Generic octahedral | $(-, -, K) =$ $(N = K - g, -, K)$ |

**Table T.8.** Constraints and types including isomorphs of bcc($N$, -, $K$) NPs. Parameter $g$ is defined by (B.4).

| Constraints | NP types | Isomorphs |
|---|---|---|
| $K \geq 3M$ | Generic rhombohedral | $(-, M, -) =$ $(-, M, K = 3M)$ |
| $2M \leq K \leq 3M$ | Rhombo-octahedral | $(-, M, K)$ |
| $K \leq 2M$ | Generic octahedral | $(-, -, K) =$ $(-, M = (K - g)/2, K)$ |

**Table T.9.** Constraints and types including isomorphs of bcc(-, $M$, $K$) NPs. Parameter $g$ is defined by (B.4).

| Constraints 1 | Constraints 2 | NP types | Isomorphs |
|---|---|---|---|
| $N \geq 2M$ | $K \geq 3M$ | Generic rhombohedral | $(-, M, -) = (N_a, M, K_a)$ |
| | $2M + g \leq K \leq 3M$ | Rhombo-octahedral | $(-, M, K) = (N_a, M, K)$ |
| | $K \leq 2M + g$ | Generic octahedral | $(-, -, K) = (N_a, M_a, K)$ |
| $M \leq N \leq 2M$ | $K \geq 3M$ | Cubo-rhombic | $(N, M, -) = (N, M, K_a)$ |
| | $4M - N \leq K \leq 3M$ | Cubo-rhombo-oct., upper central | $(N, M, K)$ |
| | $2M \leq K \leq 4M - N$ | Cubo-rhombo-oct., lower central | $(N, M, K)$ |
| | $N + g \leq K \leq 2N$ | Cubo-octahedral, truncated octahedral | $(N, -, K) = (N, M_a, K)$ |
| | $K \leq N + g$ | Generic octahedral | $(-, -, K) = (N_a, M_a, K)$ |
| $N \leq M$ | $K \geq 3N$ | Generic cubic | $(N, -, -) = (N, M_a, K_a)$ |
| | $2N \leq K \leq 3N$ | Cubo-octahedral, truncated cubic | $(N, -, K) = (N, M_a, K)$ |
| | $K = 2N$   $K$ even | Cuboctahedral | $(N, M_a, K)$ |
| | $N + g \leq K \leq 2N$ | Cubo-octahedral, truncated octahedral | $(N, -, K) = (N, M_a, K)$ |
| | $K \leq N + g$ | Generic octahedral | $(-, -, K) = (N_a, M_a, K)$ |

**Table T.10.** Constraints and types including isomorphs of bcc($N, M, K$) NPs. Polyhedral parameters $N_a, M_a, K_a$ are defined by (B.48), (B.46), (B.44) and $g$ by (B.4).

## S.1.3. Simple Cubic NPs

| Generic type | Constraints | Facets | Corners |
|---|---|---|---|
| Cubic<br>sc($N$, -, -) | --- | {100} 6<br>{110} 0<br>{111} 0 | <100> 0<br><110> 0<br><111> 8 |
| Rhombohedral ac<br>sc(-, $M$, -) | $M$ even | {100} 0<br>{110} 12<br>{111} 0 | <100> 6<br><110> 0<br><111> 8 |
|  | $M$ odd | {100} 0<br>{110} 12<br>{111} 8 | <100> 6<br><110> 0<br><111> 8 [&] |
| Rhombohedral vc<br>sc(-, $M$, -) | $M$ even | {100} 6<br>{110} 12<br>{111} 8 | <100> 6 [&]<br><110> 0<br><111> 8 [&] |
|  | $M$ odd | {100} 6<br>{110} 12<br>{111} 0 | <100> 6 [&]<br><110> 0<br><111> 8 |
| Octahedral<br>sc(-, -, $K$) | $K$ even | {100} 0<br>{110} 0<br>{111} 8 | <100> 6<br><110> 0<br><111> 0 |
|  | $K$ odd | {100} 6<br>{110} 12<br>{111} 8 | <100> 6 [&]<br><110> 0<br><111> 0 |

**Table T.11.** Types and notations of all generic sc NPs. The superscript label "&" denotes corner quadruplets about <100> and corner triplets about <111>.

**(a) Atom centered**, $N$ even

| Constraints | NP types | Isomorphs |
|---|---|---|
| $N \geq 2M$ | Generic rhombohedral | $(-, M, -) =$ <br> $(N = 2M, M, -)$ |
| $M \leq N \leq 2M$ | Cubo-rhombic | $(N, M, -)$ |
| $N \leq M$ | Generic cubic | $(N, -, -) =$ <br> $(N, M = N, -)$ |

**(b) Void centered**, $N$ odd

| Constraints | NP types | Isomorphs |
|---|---|---|
| $N \geq 2M - 1$ | Generic rhombohedral | $(-, M, -) =$ <br> $(N = 2M - 1, M, -)$ |
| $M \leq N \leq 2M - 1$ | Cubo-rhombic | $(N, M, -)$ |
| $N \leq M$ | Generic cubic | $(N, -, -) =$ <br> $(N, M = N, -)$ |

**Table T.12.** Constraints and types including isomorphs of sc($N$, $M$, -) NPs, (a) atom centered and (b) void centered.

**(a) Atom centered**, $N$, $K$ even

| Constraints | NP types | Isomorphs |
|---|---|---|
| $K \geq 3N$ | Generic cubic | $(N, -, -) =$ <br> $(N, -, K = 3N)$ |
| $2N \leq K \leq 3N$ | Cubo-octahedral truncated cubic | $(N, -, K)$ |
| $K = 2N$ | Cuboctahedral | $(N, -, K = 2N)$, <br> $(N = K/2, -, K)$ |
| $N \leq K \leq 2N$ | Cubo-octahedral truncated octahedral | $(N, -, K)$ |
| $K \leq N$ | Generic octahedral | $(-, -, K) =$ <br> $(N = K, -, K)$ |

**(b) Void centered**, *N*, *K* odd

| Constraints | NP types | Isomorphs |
|---|---|---|
| $K \geq 3N$ | Generic cubic | $(N, -, -) =$ $(N, -, K = 3N)$ |
| $2N + 1 \leq K \leq 3N$ | Cubo-octahedral truncated cubic | $(N, -, K)$ |
| $N + 2 \leq K \leq 2N - 1$ | Cubo-octahedral truncated octahedral | $(N, -, K)$ |
| $K \leq N + 2$ | Generic octahedral | $(-, -, K) =$ $(N = K - 2, -, K)$ |

**Table T.13.** Constraints and types including isomorphs of sc(*N*, -, *K*) NPs, (a) atom centered and (b) void centered.

**(a) Atom centered**, *K* even

| Constraints | NP types | Isomorphs |
|---|---|---|
| $K \geq 3M - h$ | Generic rhombohedral | $(-, M, -) =$ $(-, M, K = 3M - h)$ |
| $2M \leq K \leq 3M - h$ | Rhombo-octahedral | $(-, M, K)$ |
| $K \leq 2M$ | Generic octahedral | $(-, -, K) =$ $(-, M = K/2, K)$ |

**(b) Void centered**, *K* odd

| Constraints | NP types | Isomorphs |
|---|---|---|
| $K \geq 3M - h$ | Generic rhombohedral | $(-, M, -) =$ $(-, M, K = 3M - h)$ |
| $2M + 1 \leq K \leq 3M - h$ | Rhombo-octahedral | $(-, M, K)$ |
| $K \leq 2M + 1$ | Generic octahedral | $(-, -, K) =$ $(-, M = (K - 1)/2, K)$ |

**Table T.14.** Constraints and types including isomorphs of sc(-, *M*, *K*) NPs, (a) atom centered and (b) void centered. Parameter *h* is defined by (C.6).

| Constraints 1 | Constraints 2 | NP types | Isomorphs |
|---|---|---|---|
| $N \geq 2M$ | $K \geq 3M$ | Generic rhombohedral | $(-, M, -) =$ $(N_a, M, K_a)$ |
| | $2M \leq K \leq 3M$ | Rhombo-octahedral | $(-, M, K) =$ $(N_a, M, K)$ |
| | $K \leq 2M$ | Generic octahedral | $(-, -, K) =$ $(N_a, M_a, K)$ |
| $M \leq N \leq 2M$ | $K \geq 3M$ | Cubo-rhombic | $(N, M, -) =$ $(N, M, K_a)$ |
| | $4M - N \leq K \leq 3M$ | Cubo-rhombo-oct., upper central | $(N, M, K)$ |
| | $2M \leq K \leq 4M - N$ | Cubo-rhombo-oct., lower central | $(N, M, K)$ |
| | $N + 2g \leq K \leq 2M$ | Cubo-octahedral, truncated octahedral | $(N, -, K) =$ $(N, M_a, K)$ |
| | $K \leq N + 2g$ | Generic octahedral | $(-, -, K) =$ $(N_a, M_a, K)$ |
| $N \leq M$ | $K \geq 3N$ | Generic cubic | $(N, -, -) =$ $(N, M_a, K_a)$ |
| | $2N \leq K \leq 3N$ | Cubo-octahedral, truncated cubic | $(N, -, K) =$ $(N, M_a, K)$ |
| | $K = 2N$ | Cuboctahedral | $(N, M_a, K)$ |
| | $N + 2g \leq K \leq 2N$ | Cubo-octahedral, truncated octahedral | $(N, -, K) =$ $(N, M_a, K)$ |
| | $K \leq N + 2g$ | Generic octahedral | $(-, -, K) =$ $(N_a, M_a, K)$ |

**Table T.15.** Constraints and types including isomorphs of sc($N$, $M$, $K$) NPs. Polyhedral parameters $N_a$, $M_a$, $K_a$ are defined by (C.49), (C.47), (C.45) and $g$ by (C.5).

## S.1.4. Cubic MPs

| Generic type | Facets | | Corners | |
|---|---|---|---|---|
| Cubic<br>cub($N$, -, -) | {100} | 6 | <100> | 0 |
| | {110} | 0 | <110> | 0 |
| | {111} | 0 | <111> | 8 |
| Rhombohedral<br>cub(-, $M$, -) | {100} | 0 | <100> | 6 |
| | {110} | 12 | <110> | 0 |
| | {111} | 0 | <111> | 8 |
| Octahedral<br>cub(-, -, $K$) | {100} | 0 | <100> | 6 |
| | {110} | 0 | <110> | 0 |
| | {111} | 8 | <111> | 0 |

**Table T.16.** Types and notations of all generic cub MPs.

| Constraints | MP types | Isomorphs |
|---|---|---|
| $B \geq 2A$ | Generic cubic | $(A, -, -) =$<br>$(A, B = 2A, -)$ |
| $A \leq B \leq 2A$ | Cubo-rhombic | $(A, B, -)$ |
| $B \leq A$ | Generic rhombohedral | $(-, B, -) =$<br>$(A = B, B, -)$ |

**Table T.17.** Constraints and types including isomorphs of cub($A$, $B$, -) MPs.

| Constraints | MP types | Isomorphs |
|---|---|---|
| $C \geq 3A$ | Generic cubic | $(A, -, -) =$<br>$(A, -, C = 3A)$ |
| $2A \leq C \leq 3A$ | Cubo-octahedral,<br>truncated cubic | $(A, -, C)$ |
| $C = 2A$ | Cuboctahedral | $(A, -, C = 2A)$,<br>$(A = C/2, -, C)$ |
| $A \leq C \leq 2A$ | Cubo-octahedral,<br>truncated octahedral | $(A, -, C)$ |
| $C \leq A$ | Generic octahedral | $(-, -, C) =$<br>$(A = C, -, C)$ |

**Table T.18.** Constraints and types including isomorphs of cub($A$, -, $C$) MPs.

| Constraints | MP types | Isomorphs |
|---|---|---|
| $C \geq 3/2\,B$ | Generic rhombohedral | $(-, B, -) =$ $(-, B, C = 3/2\,B)$ |
| $B \leq C \leq 3/2\,B$ | Rhombo-octahedral | $(-, B, C)$ |
| $C \leq B$ | Generic octahedral | $(-, -, C) =$ $(-, B = C, C)$ |

**Table T.19.** Constraints and types including isomorphs of cub(-, $B$, $C$) MPs.

| Constraints 1 | Constraints 2 | MP types | Isomorphs |
|---|---|---|---|
| $B \geq 2A$ | $C \geq 3A$ | Generic cubic | $(A, -, -) =$ $(A, B_a, C_a)$ |
| | $2A \leq C \leq 3A$ | Cubo-octahedral, truncated cubic | $(A, -, C) =$ $(A, B_a, C)$ |
| | $C = 2A$ | Cuboctahedral | $(A, B_a, C)$ |
| | $A \leq C \leq 2A$ | Cubo-octahedral, truncated octahedral | $(A, -, C) =$ $(A, B_a, C)$ |
| | $C \leq A$ | Generic octahedral | $(-, -, K) =$ $(A_a, B_a, C)$ |
| $A \leq B \leq 2A$ | $C \geq 3/2\,B$ | Cubo-rhombic | $(A, B, -) =$ $(A, B, 3/2\,B)$ |
| | $2B - A \leq C \leq 3/2\,B$ | Cubo-rhombo-oct., upper central | $(A, B, C)$ |
| | $B \leq C \leq 2B - A$ | Cubo-rhombo-oct., lower central | $(A, B, C)$ |
| | $A \leq C \leq B$ | Cubo-octahedral, truncated octahedral | $(A, -, C) =$ $(A, B_a, C)$ |
| | $C \leq A$ | Generic octahedral | $(-, -, C) =$ $(A_a, B_a, C)$ |
| $B \leq A$ | $C \geq 3/2\,B$ | Generic rhombohedral | $(-, B, -) =$ $(A_a, B, C_a)$ |
| | $B \leq C \leq 3/2\,B$ | Rhombo-octahedral | $(-, B, C) =$ $(A_a, B, C)$ |
| | $C \leq B$ | Generic octahedral | $(-, -, C) =$ $(A_a, B_a, C)$ |

**Table T.20.** Constraints and types including isomorphs of cub($A$, $B$, $C$) MPs. Polyhedral parameters $A_a$, $B_a$, $C_a$ are defined by (D.30), (D.28), (D.26).

## S.2. Alternative Descriptions of Cubic (*N*, *M*, *K*) Nanoparticles

There are two other strategies to describe each of the general fcc(*N*, *M*, *K*), bcc(*N*, *M*, *K*), and sc(*N*, *M*, *K*) NPs which differ from those discussed in Secs. A.2.4, B.2.4, and C.2.4. In each case, the two strategies yield the same (*N*, *M*, *K*) NP description as given before and will be mentioned briefly in the following.

### S.2.1. Face Centered Cubic NPs

Here we start from a cubo-octahedral NP, fcc(*N*, -, *K*), with its constraints $N \leq K \leq 3N - h$, see (A.28). Then we add constraints of a generic rhombohedral NP, fcc(-, *M*, -), to yield the cubo-rhombo-octahedral NP fcc(*N*, *M*, *K*). This requires, according to the discussion in Sec. A.2.4, *M* values below $M_a$ where

$$M_a(N, K) = \min(K, 2N) \tag{S.1}$$

Cubo-octahedral NPs, fcc(*N*, -, *K*), can exist as truncated octahedral, $K \leq 2N$, and as truncated cubic, $K \geq 2N$, species. Thus, there are two starting shapes. This is illustrated in Fig. S.1 for fcc(16, 26, 26) (truncated octahedral, $K \leq 2N$) and for fcc(16, 32, 36) (truncated cubic, $K \geq 2N$). In both cases, we can distinguish four different ranges of parameter *M*, defined by separating values $M_a \geq M_b \geq M_c$, where with (S.1), (A.4)

$$M_b(N, K) = (N + K)/2 \tag{S.2}$$

$$M_c(N, K) = 2K/3 \quad = (2K + 3)/3 \qquad K = 6p + 3g \tag{S.3a}$$
$$\qquad\qquad\quad = (2K + 2)/3 \qquad K = 6p + 2 + 3g \tag{S.3b}$$
$$\qquad\qquad\quad = (2K + 1)/3 \qquad K = 6p + 4 - 3g \tag{S.3c}$$

Note that $M_b(N, K)$, $M_c(N, K)$ may be fractional.

Analogous to the discussion in Sec. A.2.4, we define an **outer *M* range**, $M \geq M_a$, (all atom balls in Fig. S.1). Here the fcc(*N*, *M*, *K*) NP becomes cubo-octahedral and is structurally identical with fcc(*N*, $M_a$, *K*). Next, there is an **upper central *M* range**, $M_b \leq M \leq M_a$, (white atom balls in Fig. S.1) and a **lower central *M* range**, $M_c \leq M \leq M_b$, (light yellow atom balls in Fig. S.1). In both ranges, the NP is truly cubo-rhombo-octahedral. Finally, in the **inner *M* range**, $M \leq M_c$, (dark yellow atom balls in Fig. S.1) the NP becomes cubo-rhombic, structurally identical with fcc(*N*, *M*, $K_a$).

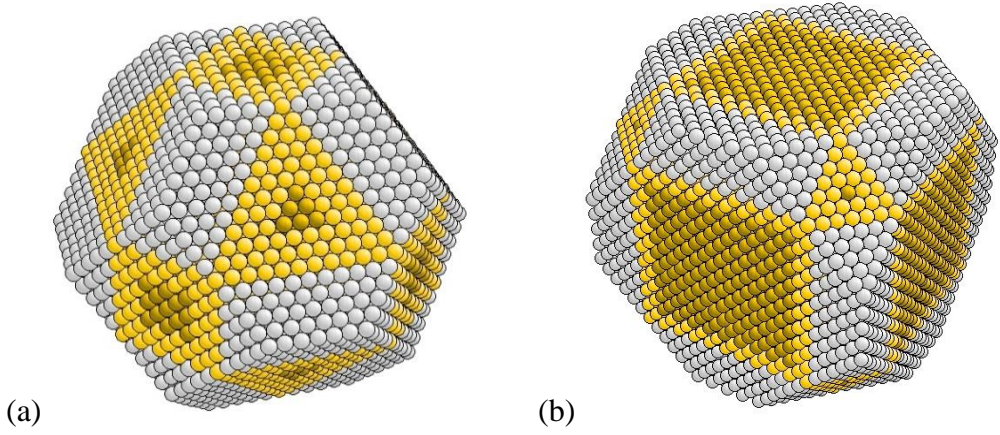

(a)                          (b)

**Figure S.1.** Atom ball models of atom centered cubo-octahedral NPs,
(a) fcc(16, 26, 26) ($M = M_a$, all atom balls), with its cubo-rhombo-octahedral NP components, fcc(16, 21, 26) ($M = M_b$, dark and light yellow balls), and fcc(16, 18, 26) ($M = M_c$, dark yellow balls);
(b) fcc(16, 32, 36) ($M = M_a$, all atom balls), with its cubo-rhombo-octahedral NP components, fcc(16, 26, 36) ($M = M_b$, dark and light yellow balls), and fcc(16, 24, 36) ($M = M_c$, dark yellow balls), see text.

The second strategy starts from a rhombo-octahedral NP, fcc(-, $M$, $K$), with its constraints $M \leq K \leq 3M/2$, etc., see (A.38). Then we add constraints of a generic cubic NP, fcc($N$, -, -), to yield the cubo-rhombo-octahedral NP fcc($N$, $M$, $K$). This requires, according to the discussion in Sec. A.2.4, $N$ values below $N_a$, where with (A.6)

$$N_a(M, K) \quad = M - h' \tag{S.4}$$

As with the first strategy, we can distinguish four different ranges of parameter $N$, defined by separating values $N_a \geq N_b \geq N_c$, where with (S.4)

$$N_b(M, K) \ = 2M - K \tag{S.5}$$
$$N_c(M, K) \ = M/2 \qquad\qquad (M \text{ even}) \tag{S.6a}$$
$$\qquad\qquad = (M - 1)/2 \qquad (M \text{ odd}) \tag{S.6b}$$

This is illustrated in Fig. S.2 for the atom centered rhombo-octahedral NP fcc(24, 24, 32).

Analogous to the discussion in Sec. A.2.4, we define an **outer $N$ range**, $N \geq N_a$, (all atom balls in Fig. S.2). Here the fcc($N$, $M$, $K$) NP becomes rhombo-octahedral and is structurally identical with fcc($N_a$, $M$, $K$). Next, there is an **upper central $N$ range**, $N_b \leq N \leq N_a$, (white atom balls in Fig. S.2) and a **lower central $N$ range**, $N_c \leq N \leq N_b$, (light yellow atom balls in Fig. S.2). In both ranges, the NP is truly cubo-rhombo-octahedral. Finally, in the **inner $N$ range**, $N \leq N_c$, (dark yellow atom balls in Fig. S.2) the NP becomes cubo-octahedral, structurally identical fcc($N$, $M_a$, $K$).

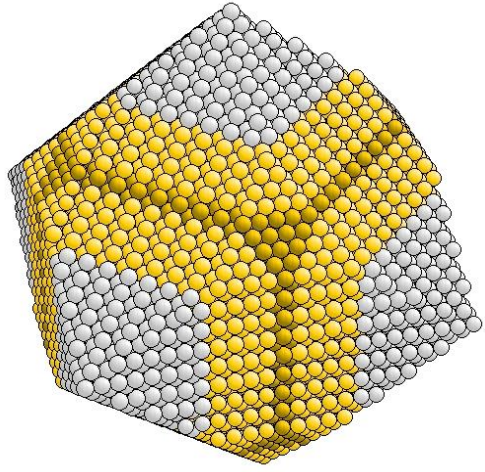

**Figure S.2.** Atom ball model of an atom centered rhombo-octahedral NP, fcc(24, 24, 32) ($N = N_a$, all atom balls), with its cubo-rhombo-octahedral NP components, fcc(16, 24, 32) ($N = N_b$, dark and light yellow balls), and fcc(12, 24, 32) ($N = N_c$, dark yellow balls), see text.

### S.2.2. Body Centered Cubic NPs

Here we start from a cubo-octahedral NP, bcc($N$, -, $K$), with its constraints $N \leq K \leq 3N$ ($K$ even) or $N + 1 \leq K \leq 3N$ ($K$ odd). Then we add constraints of a generic rhombohedral NP, bcc(-, $M$, -), to yield the cubo-rhombo-octahedral NP bcc($N$, $M$, $K$). This requires, according to the discussion in Sec. B.2.4, $M$ values below $M_a$ where

$$
\begin{aligned}
M_a(N, K) &= K/2 & (K \leq 2N, K \text{ even}) & \quad\text{(S.7a)} \\
&= (K - 1)/2 & (K \leq 2N + 1, K \text{ odd}) & \quad\text{(S.7b)} \\
&= N & (K \geq 2N) & \quad\text{(S.7c)}
\end{aligned}
$$

Cubo-octahedral NPs, bcc($N$, -, $K$), can exist as truncated octahedral, $K \leq 2N$, and as truncated cubic, $K \geq 2N$, species. Thus, there are two starting shapes. This is illustrated in Fig. S.3 for bcc(18, 15, 30) (truncated octahedral, $K \leq 2N$) and for bcc(17, 17, 39) (truncated cubic, $K \geq 2N$). In both cases, we can distinguish four different ranges of parameter $M$, defined by separating values $M_a \geq M_b \geq M_c$, where with (S.7)

$$M_b(N, K) = (N + K)/4 \quad\text{(S.8)}$$

$$M_c(N, K) = K/3 \quad\text{(S.9)}$$

Note that $M_b(N, K)$, $M_c(N, K)$ may be fractional.

Analogous to the discussion in Sec. B.2.4, we define an **outer $M$ range**, $M \geq M_a$, (all atom balls in Fig. S.3). Here the bcc($N$, $M$, $K$) NP becomes cubo-octahedral and is structurally identical with bcc($N$, $M_a$, $K$). Next, there is an **upper central $M$ range**, $M_b \leq M \leq M_a$, (white atom balls in Fig. S.3) and a **lower central $M$ range**, $M_c \leq M \leq M_b$, (light yellow atom balls in Fig. S.3). In both ranges, the NP is truly cubo-rhombo-octahedral. Finally, in the **inner $M$ range**, $M \leq M_c$, (dark yellow atom balls in Fig. S.3) the NP becomes cubo-rhombic, structurally identical with bcc($N$, $M$, $K_a$).

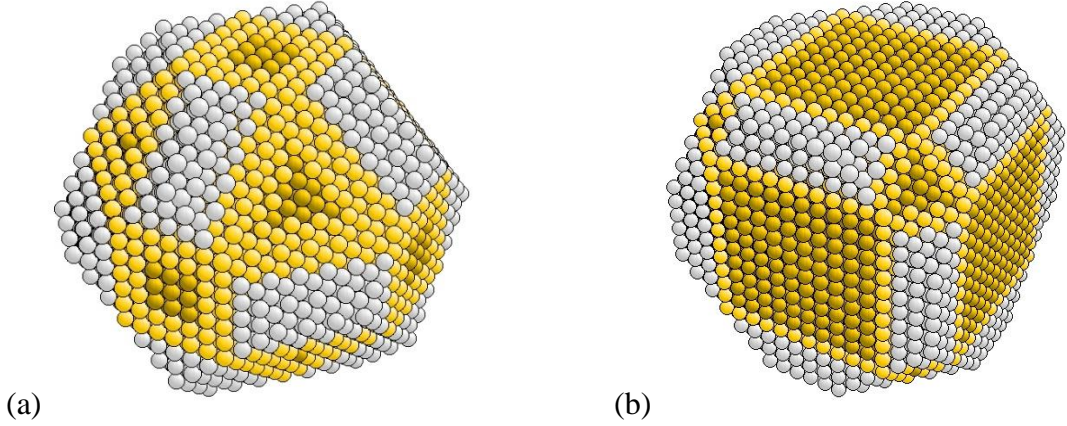

(a) (b)

**Figure S.3.** Atom ball models of cubo-octahedral NPs,
(a) bcc(18, 15, 30) ($M = M_a$, all atom balls), with its cubo-rhombo-octahedral NP components, bcc(18, 12, 30) ($M = M_b$, dark and light yellow balls), and bcc(18, 10, 30) ($M = M_c$, dark yellow balls);
(b) bcc(17, 17, 39) ($M = M_a$, all atom balls), with its cubo-rhombo-octahedral NP components, bcc(17, 14, 39) ($M = M_b$, dark and light yellow balls), and bcc(17, 13, 39) ($M = M_c$, dark yellow balls), see text.

The second strategy starts from a rhombo-octahedral NP, bcc(-, $M$, $K$), with its constraints $2M - g \leq K \leq 3M$, see (B.37). Then we add constraints of a generic cubic NP, bcc($N$, -, -), to yield the cubo-rhombo-octahedral NP bcc($N$, $M$, $K$). This requires, according to the discussion in Sec. B.2.4, $N$ values below $N_a$, where

$$N_a(M, K) = 2M \tag{S.10}$$

As with the first strategy, we can distinguish four different ranges of parameter $N$, defined by separating values $N_a \geq N_b \geq N_c$, where with (S.10)

$$N_b(M, K) = 4M - K \tag{S.11}$$
$$N_c(M, K) = M \tag{S.12}$$

This is illustrated in Fig. S.4 for the rhombo-octahedral NP bcc(26, 13, 33).

Analogous to the discussion in Sec. B.2.4, we define an **outer $N$ range**, $N \geq N_a$, (all atom balls in Fig. S.4). Here the bcc($N$, $M$, $K$) NP becomes rhombo-octahedral and is structurally identical with bcc($N_a$, $M$, $K$). Next, there is an **upper central $N$ range**, $N_b \leq N \leq N_a$, (white atom balls in Fig. S.4) and a **lower central $N$ range**, $N_c \leq N \leq N_b$, (light yellow atom balls in Fig. S.4). In both ranges, the NP is truly cubo-rhombo-octahedral. Finally, in the **inner $N$ range**, $N \leq N_c$, (dark yellow atom balls in Fig. S.4) the NP becomes cubo-octahedral, structurally identical bcc($N$, $M_a$, $K$).

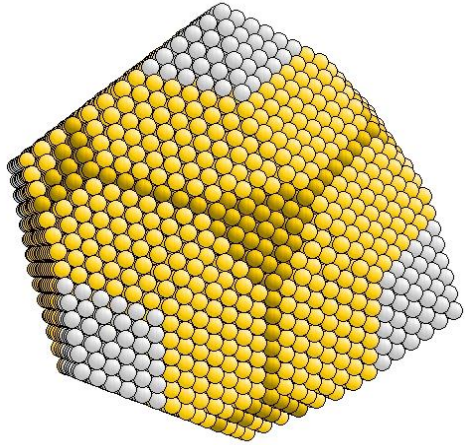

**Figure S.4.** Atom ball model of a rhombo-octahedral NP, bcc(26, 13, 33) ($N = N_a$, all atom balls), with its cubo-rhombo-octahedral NP components, bcc(20, 13, 33) ($N = N_b$, dark and light yellow balls), and bcc(13, 13, 33) ($N = N_c$, dark yellow balls), see text.

## S.2.3. Simple Cubic NPs

Here we start from a cubo-octahedral NP, sc($N$, -, $K$), with its constraints $N + 2g \leq K \leq 3N$, see (C.27). Then we add constraints of a generic rhombohedral NP, sc(-, $M$, -), to yield the cubo-rhombo-octahedral NP sc($N$, $M$, $K$). This requires, according to the discussion in Sec. C.2.4, $M$ values below $M_a$ where with (C.5)

$$M_a(N, K) = (K - g)/2 \qquad (K \leq 2N) \qquad \text{(S.13a)}$$
$$= N \qquad (K \geq 2N) \qquad \text{(S.13b)}$$

Cubo-octahedral NPs, sc($N$, -, $K$), can exist as truncated octahedral, $K \leq 2N$, and as truncated cubic, $K \geq 2N$, species. Thus, there are two starting shapes. This is illustrated in Fig. S.5 for sc(20, 18, 36) (truncated octahedral) and for sc(20, 20, 44) (truncated cubic). In both cases, we can distinguish four different ranges of parameter $M$, defined by separating values $M_a \geq M_b \geq M_c$, where with (S.13)

$$M_b(N, K) = (N + K)/4 \qquad (S.14)$$

$$M_c(N, K) = K/3 \qquad (S.15)$$

Note that $M_b(N, K)$, $M_c(N, K)$ may be fractional.

Analogous to the discussion in Sec. C.2.4, we define an **outer $M$ range**, $M \geq M_a$, (all atom balls in Fig. S.5). Here the sc($N, M, K$) NP becomes cubo-octahedral and is structurally identical with sc($N, M_a, K$). Next, there is an **upper central $M$ range**, $M_b \leq M \leq M_a$, (white atom balls in Fig. S.5) and a **lower central $M$ range**, $M_c \leq M \leq M_b$, (light yellow atom balls in Fig. S.5). In both ranges, the NP is truly cubo-rhombo-octahedral. Finally, in the **inner $M$ range**, $M \leq M_c$, (dark yellow atom balls in Fig. S.5) the NP becomes cubo-rhombic, structurally identical with sc($N, M, K_a$).

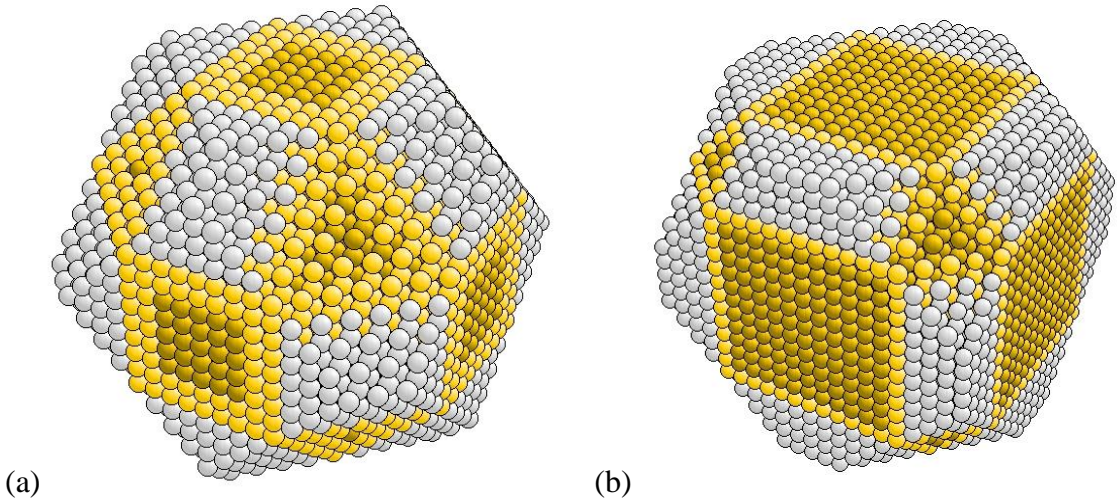

(a)          (b)

**Figure S.5.** Atom ball models of atom centered cubo-octahedral NPs, (a) truncated octahedral sc(20, 18, 36) ($M = M_a$, all atom balls), with its cubo-rhombo-octahedral NP components, sc(20, 14, 36) ($M = M_b$, dark and light yellow balls), and sc(20, 12, 36) ($M = M_c$, dark yellow balls); (b) truncated cubic sc(20, 20, 44) ($M = M_a$, all atom balls), with its cubo-rhombo-octahedral NP components, sc(20, 16, 44) ($M = M_b$, dark and light yellow balls), and sc(20, 15, 44) ($M = M_c$, dark yellow balls), see text.

The second strategy starts from a rhombo-octahedral NP, sc(-, $M, K$), with its constraints $2M + g \leq K \leq 3M - h$, see (C.38). Then we add constraints of a generic cubic NP, sc($N$, -, -), to yield the cubo-rhombo-octahedral NP sc($N, M, K$). This requires, according to the discussion in Sec. C.2.4, $N$ values below $N_a$. where with (C.5)

$$N_a(M, K) = 2M - g \tag{S.16}$$

As with the first strategy, we can distinguish four different ranges of parameter $N$, defined by separating values $N_a \geq N_b \geq N_c$, where with (S.16)

$$N_b(M, K) = 4M - K \tag{S.17}$$

$$N_c(M, K) = M \tag{S.18}$$

This is illustrated in Fig. S.6 for the atom centered rhombo-octahedral NP sc(28, 14, 34).

Analogous to the discussion in Sec. C.2.4, we define an **outer $N$ range**, $N \geq N_a$, (all atom balls in Fig. S.6). Here the sc($N$, $M$, $K$) NP becomes rhombo-octahedral and is structurally identical with sc($N_a$, $M$, $K$). Next, there is an **upper central $N$ range**, $N_b \leq N \leq N_a$, (white atom balls in Fig. S.6) and a **lower central $N$ range**, $N_c \leq N \leq N_b$, (light yellow atom balls in Fig. S.6). In both ranges, the NP is truly cubo-rhombo-octahedral. Finally, in the **inner $N$ range**, $N \leq N_c$, (dark yellow atom balls in Fig. S.6) the NP becomes cubo-octahedral, structurally identical sc($N$, $M_a$, $K$).

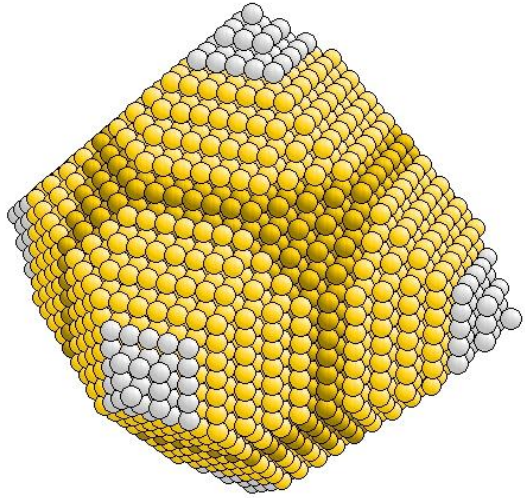

**Figure S.6.** Atom ball model of an atom centered rhombo-octahedral NP, sc(28, 14, 34) ($N = N_a$, all atom balls), with its cubo-rhombo-octahedral NP components, sc(22, 14, 34) ($N = N_b$, dark and light yellow balls), and sc(14, 14, 34) ($N = N_c$, dark yellow balls), see text.

## S.3. Symmetry Centers

The cubic NPs, sc($N$, $M$, $K$), bcc($N$, $M$, $K$), and fcc($N$, $M$, $K$) of $O_h$ symmetry contain atoms or high symmetry voids at their center depending on the lattice type and on parameters $N$, $M$, $K$.

The **face centered cubic** (fcc) lattice offers two different centers of $O_h$ symmetry, an atom site and a high symmetry void, as shown in Fig. S.7.

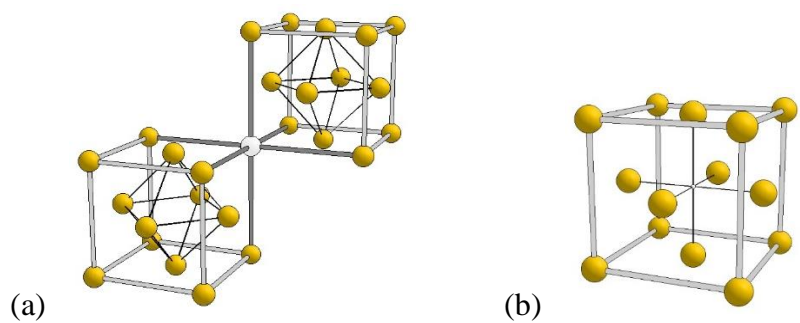

    (a)                    (b)

**Figure S.7.** $O_h$ symmetry centers of the fcc lattice, (a) atom site and (b) high symmetry void site. The centers are emphasized by white color and connected with their nearest neighbor atoms (void site) and next nearest neighbors (atom site).

This discriminates between two symmetry types of octahedral and cubic fcc($N$, $M$, $K$) NPs, about an atom center and about a high symmetry void, depending on the parities (even, odd, any) of parameters $N$, $M$, $K$ as spelled out in the following Table.

| NP center type | $N$ | $M$ | $K$ |
|---|---|---|---|
| atom | any | any | even |
| void | any | any | odd |

The **body centered cubic** (bcc) lattice offers only one center of $O_h$ symmetry which coincides with an atom site, as shown in Fig. S.8.

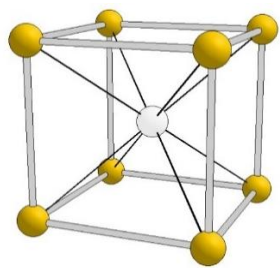

**Figure S.8.** $O_h$ symmetry center of the bcc lattice at atom site. The center is emphasized by white color and connected with its nearest neighbor atoms.

This allows for one symmetry type of bcc($N$, $M$, $K$) NPs, about an atom center, independent of the parities of parameters $N$, $M$, $K$.

The **simple cubic** (sc) lattice offers two different centers of $O_h$ symmetry, an atom site and a high symmetry void site, as shown in Fig. S.9.

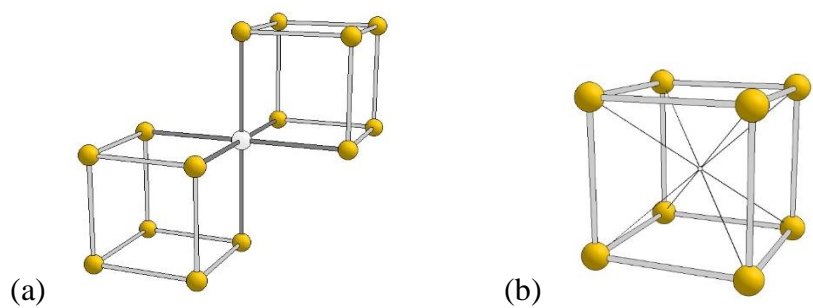

(a)        (b)

**Figure S.9.** $O_h$ symmetry centers of the sc lattice, (a) atom site and (b) high symmetry void site. The symmetry centers are emphasized by white color and connected with their nearest neighbor atoms.

This discriminates between two types of octahedral and cubic sc($N$, $M$, $K$) NPs, about an atom center and about a high symmetry void, depending on the parities (even, odd, any) of parameters $N$, $M$, $K$ as spelled out in the following Table.

| NP center type | $N$ | $M$ | $K$ |
|---|---|---|---|
| atom | even | any | even |
| void | odd | any | odd |